\documentclass{aa}
\usepackage[utf8]{inputenc}
\usepackage[flushleft]{threeparttable}
\usepackage{breqn}
\usepackage{array}
\usepackage{rotating}
\usepackage{natbib}
\bibpunct{(}{)}{;}{a}{}{,} % to follow the A&A style

\title{X-ray spectra of the Fe-L complex II: atomic data constraints from EBIT experiment and X-ray grating observations of Capella}
\author{Liyi Gu \inst{1,2} 
\and 
Chintan Shah \inst{3,4}
\and 
Junjie Mao \inst{5,2}
\and 
A.J.J. Raassen \inst{2,6}
\and
Jelle de Plaa \inst{2}
\and 
Ciro Pinto \inst{7}
\and 
Hiroki Akamatsu \inst{2} 
\and 
Norbert Werner \inst{8,9,10}
\and 
Aurora Simionescu \inst{2,11,12}
\and 
François Mernier \inst{13,2}
\and 
Makoto Sawada \inst{1}
\and
Pranav Mohanty \inst{11}
\and 
Pedro Amaro \inst{14}
\and 
Ming Feng Gu \inst{15}
\and 
F. Scott Porter \inst{3} 
\and 
José R. Crespo López-Urrutia \inst{4}
\and 
Jelle S. Kaastra \inst{2,11} 
}
\institute{
RIKEN High Energy Astrophysics Laboratory, 2-1 Hirosawa, Wako, Saitama 351-0198, Japan
\and
SRON Netherlands Institute for Space Research, Sorbonnelaan 2, 3584 CA Utrecht, the Netherlands 
\and 
NASA/Goddard Space Flight Center, 8800 Greenbelt Rd, Greenbelt, MD 20771, USA
\and 
Max-Planck-Institut f$\rm \ddot{u}$r Kernphysik, Heidelberg, D-69117 Heidelberg, Germany
\and 
Department of Physics, University of Strathclyde, Glasgow, G4 0NG, UK
\and 
Astronomical Institute ``Anton Pannekoek'', Science Park 904, 1098 XH Amsterdam, University of Amsterdam, The Netherlands
\and 
INAF - IASF Palermo, Via U. La Malfa 153, 90146 Palermo PA, Italy
\and 
MTA-E$\rm \ddot{o}$tv$\rm \ddot{o}$s University Lend$\rm \ddot{u}$let Hot Universe Research Group, P$\rm \acute{a}$zm$\rm \acute{a}$ny P$\rm \acute{e}$ter 
s$\rm \acute{e}$t$\rm \acute{a}$ny 1/A, Budapest, 1117, Hungary
\and 
Department of Theoretical Physics and Astrophysics, Faculty of Science, Masaryk University, Kotl$\rm \acute{a}$$\rm \breve{r}$sk$\rm \acute{a}$ 2, Brno, 611 37, Czech Republic
\and 
School of Science, Hiroshima University, 1-3-1 Kagamiyama, Higashi-Hiroshima 739-8526, Japan
\and
Leiden Observatory, Leiden University, PO Box 9513, 2300 RA Leiden, the Netherlands 
\and
Kavli Institute for the Physics and Mathematics of the Universe (WPI), University of Tokyo, Kashiwa 277-8583, Japan  
\and 
ESTEC/ESA, Keplerlaan 1, 2201AZ Noordwijk, The Netherlands
\and 
Laboratory of Instrumentation, Biomedical Engineering and Radiation Physics (LIBPhys-UNL), Department of Physics, NOVA School of Science and Technology, NOVA University Lisbon, 2829-516 Caparica, Portugal
\and 
Space Science Laboratory, University of California, Berkeley, CA 94720, USA
}
\date{March 2020}

\abstract{The {\it Hitomi} results for the Perseus cluster have shown that accurate atomic models are essential to the success of X-ray spectroscopic missions, 
and just as important as knowledge on instrumental calibration and astrophysical modeling. Preparing the models requires a multifaceted approach, including
theoretical calculations, laboratory measurements, and calibration using real observations. In a previous paper, we presented a calculation
of the electron impact cross sections on the transitions forming the Fe-L complex. In the present work, we systematically test the calculation against cross sections
of ions measured in an electron beam ion trap experiment. A two-dimensional analysis in the electron beam energies and X-ray photon energies 
is utilized to disentangle radiative channels following dielectronic
recombination, direct electron-impact excitation, and resonant excitation processes in the experimental data. The data calibrated through laboratory measurements are further fed 
into global modeling of the {\it Chandra} grating spectrum of Capella. We investigate and compare the fit quality, as well as sensitivity 
of the derived physical parameters to the underlying atomic data and the astrophysical plasma modeling. We further list the potential areas of disagreement
between the observation and the present calculations, which in turn calls for renewed efforts in theoretical calculations and targeted laboratory measurements. 
}

\keywords{Atomic data -- Methods: laboratory: atomic  -- Techniques: spectroscopic -- Stars: coronae  }
\titlerunning{Fe-L complex part II }
\authorrunning{L. Gu}

\begin{document}

\maketitle

\section{Introduction}

High resolution X-ray spectroscopy provides unique opportunities for the exploration of both the microscopic physics 
of celestial bodies and the fundamental laws of cosmology. The observational window of X-ray spectroscopy, unlocked
by the spectrometers onboard {\it Chandra}, {\it XMM-Newton}, and {\it Hitomi}, will be fully open with the micro-calorimeters
on the future {\it XRISM} and {\it Athena} missions. These telescopes will enable well-resolved spectroscopy of all kinds of cosmic
X-ray sources, advancing the understanding into a broad range of physical conditions, from the heating source in the corona of a star 
to the formation of the largest scale baryonic structure. 

The increasing sensitivity and resolving power of X-ray spectrometers require accurate modeling of the X-ray spectra,
which is in turn built on a range of fundamental atomic data, mainly including the wavelengths and cross sections of radiative and collisional
processes. The connection between X-ray astrophysics and atomic physics has never been so tight. The existing spectra already
revealed the limits of the available atomic data, which are mostly obtained through theoretical calculations. The first {\it Hitomi} 
results on the Perseus cluster showed surprising discrepancies
between the measurements using different atomic codes, for instance, 15\% for the derived iron metallicity, while
the statistical uncertainties from the observation are only 1\% \citep{hitomiatomic2018}. These discrepancies represent systematic uncertainties
that are even larger than the uncertainties due to the calibration of the {\it Hitomi} instruments. Another example showed that
the laboratory measurements of the transition energies and cross sections of the oxygen innershell photoionization lines might be in
tension with the results using the {\it Chandra} grating data, casting doubt on the interpretation from some of the observations \citep{mclaughlin2017}.

The Fe-L complex is a dominant feature in the X-ray spectra from many collisional plasma sources, such as stars, interstellar medium, and groups of galaxies.
The Fe-L complex is composed of a range of radiative transitions onto $n=2$ states of Na-like to Li-like Fe ions, 
powered by multiple channels of collisional excitation, recombination, and ionization mechanisms. These Fe lines 
are very bright, making them key diagnostics of electron temperature, density, chemical abundances, gas motion, and photon scattering opacity
of the sources \citep{p1996, xu2002, werner2006}. Nevertheless, we lack adequate accuracy in the atomic data for
the Fe-L complex \citep{gu2006b, beiersdorfer2002, brown2006, bernitt2012, bei2018, mernier2018, shah2019, shah2020}, where the line formation is complicated. Unless the issue is addressed, 
the large errors in the data might potentially lead to unacceptable uncertainties 
in the future {\it XRISM}~\citep{xrism2018} and {\it Athena}~\citep{barret2016} analysis. 

In \citet{gu2019} (hereafter paper I), we performed a distorted-wave calculation of the electron impact cross sections on the Fe-L ions,
paying special attention to the resonant populating processes, including resonant excitation and dielectronic recombination. 
These calculations are systematically compared with the available $R$-matrix results, on both the collisional rates and the 
model spectra based on line formation calculation. We found that the two calculations agree within 20\% on most of the main transitions,
while the discrepancies become much larger for the weaker lines. 
Thus, the new Fe-L calculations must be verified before delivery to the community. This can be done by the systematic testing
of the atomic data against (1) laboratory measurements using e.g., electron beam ion traps (EBITs) and (2) deep astrophysical
observations of standard objects. In this paper, we first put forward an experimental benchmark using a recent EBIT measurement 
of the Na-like and Ne-like Fe lines (Section~\ref{sec:ebit}), focusing on the dielectronic recombination, direct and resonant excitation processes. 
In the second half, we implement the atomic data obtained through calculations and EBIT experiments into the analysis of the Capella
grating data (Section~\ref{sec:capella}). Similar work was done in \citet{hitomiatomic2018}, in which the K-shell lines are directly calibrated against
the {\it Hitomi} data of the Perseus cluster. Based on the above tests, we discuss potential areas of disagreement, 
as well as feasible corrections to the atomic data.

Throughout the paper, the errors are given at a 68\% confidence level.

\section{Laboratory benchmark}
\label{sec:ebit}
\subsection{Fe-L data measured with FLASH-EBIT}

\begin{table*}[!htbp]
\centering
\caption{Terminology of the relevant \ion{Fe}{XVII} lines }
\label{tab:fexvii}
\begin{threeparttable}
\begin{tabular}{ccccccccccc}
\hline
Name   & wavelength ({\AA})$^a$           & transition  & channel & formation$^b$   \\
\hline
3$C$    &   15.014  & 2$s^2$2$p^6$ ($^1S_0$) - 2$s^2$2$p^5$3$d$ ($^1P_1$) & $3d-2p$ & mainly DE \\
3$D$    &   15.261  & 2$s^2$2$p^6$ ($^1S_0$) - 2$s^2$2$p^5$3$d$ ($^3D_1$) & $3d-2p$ & mainly DE \\
3$E$    &   15.453  & 2$s^2$2$p^6$ ($^1S_0$) - 2$s^2$2$p^5$3$d$ ($^3P_1$) & $3d-2p$ & DE and DR cascades \\
3$F$    &   16.780  & 2$s^2$2$p^6$ ($^1S_0$) - 2$s^2$2$p^5$3$s$ ($^3P_1$) & $3s-2p$ & DE, RE, DR, II, RR\\
3$G$    &   17.051  & 2$s^2$2$p^6$ ($^1S_0$) - 2$s^2$2$p^5$3$s$ ($^1P_1$) & $3s-2p$ & DE, RE, DR, II, RR\\
$M$2    &   17.096  & 2$s^2$2$p^6$ ($^1S_0$) - 2$s^2$2$p^5$3$s$ ($^3P_2$) & $3s-2p$ & DE, RE, DR, II, RR\\ 
\hline
\end{tabular}
\begin{tablenotes}
\item[$(a)$] Measured wavelengths by \citet{brown1998}.
\item[$(b)$] DE: direct excitation; RE: resonant excitation; DR: dielectronic recombination; II: innershell ionization; RR: radiative recombination.
\end{tablenotes}
\end{threeparttable}
\end{table*}

The experimental data used in the present analysis were reported by \citet{shah2019} (hereafter S19) using the FLASH-EBIT \citep{epp2010} facility located at the Max-Planck-Institut f$\rm \ddot{u}$r Kernphysik in Heidelberg, Germany. 
In the experiment, a monoenergetic electron beam emitted from the hot cathode is compressed by a strong magnetic field of 6~T generated by superconducting Helmholtz coils. The electron beam collisionally ionizes Fe neutral atoms to the desired charge state, which are radially confined by the negative space charge potential of the electron beam and axially by electrostatic potentials applied to the set of drift tubes. The charge state distributions of the trapped ions are driven by electron-impact ionization, recombination, and excitation processes.
The spontaneous radiative decay of excited states generates X-ray photons, which are collected at 90 degrees to the electron beam axis with a silicon-drift detector (resolution $\approx 120$~eV FWHM at 1~keV). 

The line emission cross sections were measured for $3d-2p$ and $3s-2p$ channels of Fe~\textsc{xvii} ions formed through dielectronic recombination, direct electron-impact excitation, resonant excitation, and radiative cascades (see terminology in Table~\ref{tab:fexvii}). The experiment (S19) improves previous works by reducing the collision-energy spread to only 5~eV FWHM at 800 eV, which is six-to-ten times improvement compared to~\citet{brown2006}, \citet{gillaspy2011}, and \citet{beiersdorfer2017}. Moreover, three orders of magnitude higher counting statistics allowed to distinguish narrow resonant excitation and dielectronic recombination features from direct excitation ones in the experiment.
By comparing the laboratory line fluxes with a theoretical calculation tailored to match experimental conditions, S19 found good agreement with various theories for the $3s-2p$ transition, while accurately confirming known discrepancies in the $3d-2p$ transition. 
The latter was found to be overestimated by $9-20$\% by state-of-the-art-theories. 
This result is well in line with earlier lab measurements~\citep{beiersdorfer2002,brown2006}. 

\begin{figure*}[!htbp]
\center
\resizebox{1.0\hsize}{!}{\includegraphics[angle=0]{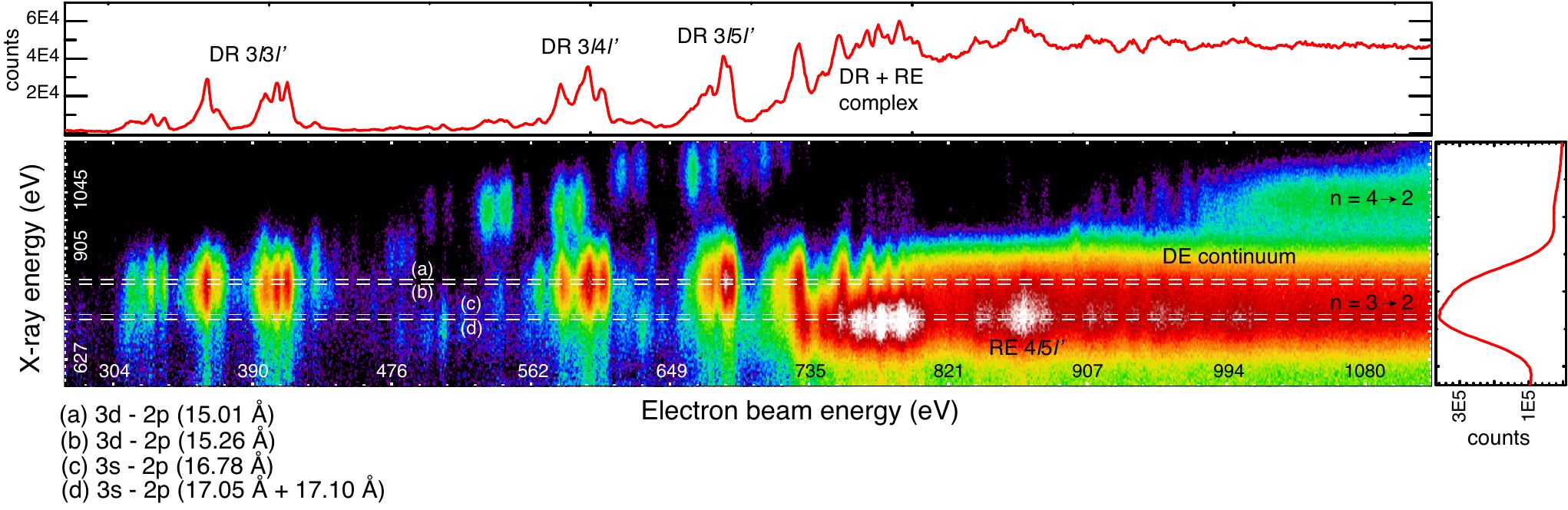}}
\caption{EBIT X-ray data as a function of the electron beam energy (abscissa) and the photon energy (ordinate),
corrected for the carbon foil transmission and the varying beam current. Color shows the number of counts.
Photon energies of the main 3$d$ - 2$p$ and 3$s$ - 2$p$ transitions of Ne-like Fe are marked with the dashed lines. The upper and
right panels show the count profiles projected along the beam energy and the photon energy, respectively. }
\label{fig:ebitdata}
\end{figure*}

It remains unclear whether the same consistency and discrepancy can also be expected for the general spectral codes (e.g., AtomDB, Chianti, and SPEX)
used in X-ray astronomy. In fact, these codes might behave differently from a dedicated calculation, 
as the database in the codes are often a combination of multiple sources of calculations, and are subject to various approximations. The main caveats for a direct EBIT-code benchmark are (1) cross sections in astronomical 
codes are often folded with the Maxwellian distribution, while these EBIT data were taken with a linear energy
weighting; (2) conventional analyzing techniques cannot fully disentangle different radiative
components (e.g., dielectronic recombination, radiative recombination, resonant excitation, direct excitation) 
when they overlap in beam energy in the EBIT data. The most obvious example is the complex around the beam 
energy $\sim 700-800$~eV as shown in Fig.~\ref{fig:ebitdata}, where the strong dielectronic recombination transitions 
of the Na-like ions are mixed up with the resonant excitation and direct excitation transitions of the Ne-like ions; (3) most 
of the lines are blended in photon energy due to
the poor X-ray resolution of the current data (Fig.~\ref{fig:ebitdata}). An accurate component study based on data decomposition
and Maxwellianization (\S~\ref{sec:2d}) is the key to solve these issues.

Here we calibrate the EBIT data following the procedure in S19. First, the X-ray detector was shielded from the  
UV light by a 1~$\mu$m carbon foil, which also blocks a part of the X-ray radiation from the trap. The transmission of the thin
foil was corrected based on the tabulated data from \citet{henke1993}, which gives 15\% for 600~eV and 61\% for
1000~eV. Second, the beam current $I_{e}$ was adjusted as a function of beam energy $E$ to keep the electron density
$n_e \propto I_e / \sqrt{E}$ constant in the experiment, thus the space charge potential of the electron beam. The EBIT data was 
therefore corrected for the energy-dependent electron
beam current density. 

Finally, the X-ray polarizations from the radiative cascade and the cyclotron motion of electrons inside the electron beam
were modeled based on the measurement and the theoretical procedure presented in~\citet{shah2018}. 
The polarization calculations were performed using with the FAC code \citep{gu2008} for each individual component, including dielectronic recombination, resonant excitation, and direct excitation, and presented in S19 as the machine-readable table (see data behind the Fig.~2 of S19). 
They also agree well with an independent experiment performed at LLNL EBIT, measuring X-ray line polarizations for Fe-L lines using the two-crystal technique~\citep{chen2004}.
Moreover, in S19, a comprehensive agreement for various line emission cross sections at 90 degrees was achieved using the FAC predictions~(\textit{cf.}~Figs.~2, 3, and 4 of S19). 
Two other completely independent experiments measuring the total dielectronic recombination cross sections the Test Storage Ring (TSR) at Heidelberg~\citep{schmidt2009} and the total direct excitation cross sections at LLNL EBIT~\citep{brown2006} have also shown a good agreement with the S19 experimental data. 
Such extensive comparisons between independent theories and experiments in S19 provided us good confidence over the validity of the polarization predictions. 
We also note that the FAC polarization predictions are also thoroughly benchmarked in previous experiments at FLASH-EBIT~\citep{shah2015,shah2016,amaro2017,shah2018}. 
Here, we inferred the total flux ($F_{\rm tot}$) from the observed flux in the experiment ($F_{\rm 90}$) through $F_{\rm tot} = 4 \pi F_{\rm 90} (3-P)/3 $, where $P$ is the degree of polarization. 
Based on FAC, the maximal effect of polarization is $\sim$40\% for the dielectronic recombination and direct excitation at the 3$d$ - 2$p$ manifold, and $\sim$5\% for the direct excitation at the 3$s$ - 2$p$ lines. The latter line complex is completely dominated by the radiative cascades thus depolarized~(see~\citealt{chen2004} and S19 for details). 
%

%As mentioned above, the X-ray photons are observed at 90 degrees with respect to the electron beam propagation axis. 
%
%Therefore, the total flux ($F_{\rm tot}$) can be recovered from the observed flux ($F_{\rm 90}$) through $F_{\rm tot} = 4 \pi F_{\rm 90} (3-P)/3 $, where $P$ is the degree of polarization. 
%

%\textbf{The polarization correction has an accuracy of ...
%Chintan: could you please add a few words here for the referee's first comment?}

%
\subsection{Two dimensional fits}
\label{sec:2d}

\begin{table*}[!htbp]
\centering
\caption{Fit quality of the EBIT data}
\label{tab:fits2d}
\begin{threeparttable}
\begin{tabular}{lccccccccccc}
\hline
Region of interest & beam energy$^{a}$ & photon energy$^{a}$ & number of peaks & fit goodness   \\
       & (keV)       & (keV)         &                 & $\chi^2$/dof   \\
\hline
1      & 0.254$-$0.295 & 0.547$-$0.915 & 3 & 5445/4862 \\
2      & 0.295$-$0.341 & 0.515$-$0.983 & 4 & 8632/6906 \\
3      & 0.341$-$0.377 & 0.516$-$1.053 & 5 & 7001/6149 \\ 
4      & 0.377$-$0.423 & 0.514$-$1.089 & 6 & 9610/8536 \\
5      & 0.423$-$0.459 & 0.514$-$1.099 & 4 & 5433/6731 \\
6      & 0.459$-$0.491 & 0.512$-$1.106 & 5 & 5961/6100 \\
7      & 0.491$-$0.524 & 0.512$-$1.106 & 5 & 6301/6436 \\
8      & 0.524$-$0.554 & 0.576$-$1.234 & 5 & 5589/6100 \\
9      & 0.554$-$0.612 & 0.571$-$1.245 & 10 & 12503/12648 \\
10     & 0.612$-$0.648 & 0.567$-$1.256 & 8 & 7589/8047 \\
11     & 0.648$-$0.699 & 0.566$-$1.298 & 11 & 11296/12072 \\
12     & 0.699$-$0.737 & 0.564$-$1.402 & 10 & 12039/10302 \\
13     & 0.737$-$0.807 & 0.564$-$1.102 & 19 & 12157/9971 \\
14     & 0.807$-$0.835 & 0.539$-$1.400 & 4 & 9646/7678 \\
15     & 0.835$-$0.888 & 0.555$-$1.403 & 11 & 18258/14666 \\
16     & 0.888$-$0.934 & 0.554$-$1.396 & 12 & 19347/15070 \\
17     & 0.934$-$0.987 & 0.552$-$1.408 & 10 & 16110/12158 \\
18     & 0.987$-$1.110 & 0.548$-$0.962 & 10 & 22098/16390 \\
\hline
\end{tabular}
\begin{tablenotes}
\item[$(a)$] Lower and upper boundaries of the regions of interest.
\end{tablenotes}
\end{threeparttable}
\end{table*}

As shown in Fig.~\ref{fig:ebitdata}, the X-ray line emission of Ne-like and Na-like Fe is made up of two main components: the 
resonant X-ray peaks (including dielectronic recombination and resonant excitation) and the continuum (direct
excitation and radiative recombination). An accurate determination of the transition intensities rests upon precise separation and extraction of each individual 
component, which is most related to two issues: (1) how to model the resonant component; and (2) how to separate the resonant
and the continuum components.

\begin{figure}[!htbp]
\centering
\resizebox{0.7\hsize}{!}{\includegraphics[angle=0]{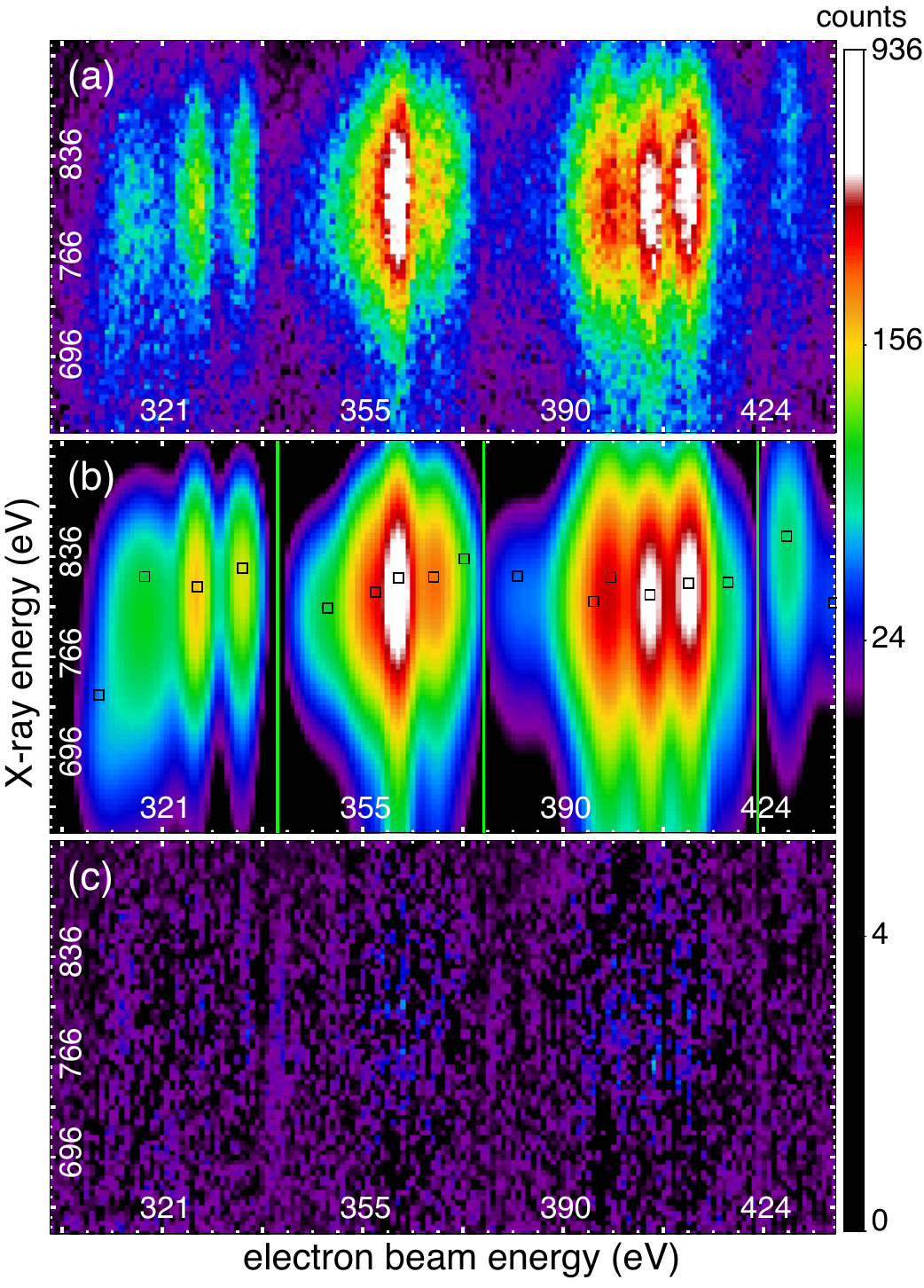}}
\caption{(a) EBIT data of the dielectronic recombination from $3l3l'$ states (where two electrons are excited to $n=3$). 
(b) Best-fit model, the central position of each peak is marked by a black box. 
The boundaries of the regions of interest are shown by green lines. (c) Background noise and residual of the fit.}
\label{fig:drn3}
\end{figure}

To solve the first issue, we model the spectrum at the resonant X-ray peaks in two dimensions and 
extract each peak from its adjacent transitions. Following, e.g., \citet{li2013}, the instrumental response is approximated using a set of semi-empirical 
equations,
\begin{equation}
\label{eq:eq1}
 F(x,y) = F_{x} (x) \left( F_{y} (y) + F^{\ast}_{y} (y) \right),
\end{equation}
where the response functions of the full energy peak along the electron beam energy $x$ and X-ray photon energy $y$ are
\begin{equation}
\label{eq:eq2}
F_{x} (x) = N_{1} e^{-\frac{(x-x_{0})^{2}}{2\sigma^{2}_{x}}},
\end{equation}
and 
\begin{equation}
\label{eq:eq3}
F_{y} (y) = N_{2} e^{-\frac{(y-y_{0})^{2}}{2\sigma^{2}_{y}}}.
\end{equation}
The exponential tail function towards the low energy from the full peak is 
\begin{dmath}
\label{eq:eq4}
F^{\ast}_{y} (y) = N_{3} e^{-\frac{y-y_{0}}{\sigma^{\ast}_{y}}} 
    Erfc\left( \frac{y-y_{0}}{N_{4}\sigma_{y}} + \frac{\sigma_{y}}{N_{4}\sigma^{\ast}_{y}} \right).
\end{dmath}
In the above equations, $N_{1}$, $N_{2}$, $N_{3}$, and $N_{4}$ are the normalization parameters, $x_0$ and $y_0$ are the excitation
and photon energies of the transition, $\sigma_{x}$, $\sigma_{y}$, and $\sigma^{\ast}_{y}$ are the 
line spreads of the instrument, and {\tt Erfc} is a complementary error function. All the parameters are left free for each X-ray peak.
The {\tt Sherpa} package \footnote{http://cxc.harvard.edu/sherpa/} \citep{freeman2001} in the {\tt CIAO} software \citep{fru2006}
is used to fit the EBIT data. To localize the 
noise analysis, the entire image is divided into several regions of interest (Fig.~\ref{fig:drn3}), and the fit is carried
out independently for each region. The number of peaks are set by a visual inspection, unless an F-test determines
that an additional resonant component is required. The quality of the fit to each region of interest is summarized in 
Table~\ref{tab:fits2d}. Generally, the fit is robust, thanks to the very high
count statistics and the excellent signal-to-noise level of the data. 

As shown in Fig.~\ref{fig:drn3}, the dielectronic recombination of Na-like Fe from $n=3$ levels are modeled
using Eqs.~(1-4). The overall reduced-$\chi^2$ is 1.2, for degrees of freedom of $\sim 22000$. The statistical uncertainties on the peak intensities 
are determined to be $1-4$\%, and the estimation of the relevant systematic uncertainties is described in \S~\ref{sec:uncert}. 
The average position accuracy is 0.5\% on the beam energy and 3.5\%
on the photon energy, relative to the absolute values obtained from the fits. 

A remaining issue is to separate the resonant and continuum components, especially for the 3$s$ - 2$p$ 
line at beam energy $>720$~eV where the resonant excitation, dielectronic recombination, and direct excitation overlap.
It is challenging to determine the resonant and continuum components simultaneously through fit, as their variations on the
beam energy are both substantial. To overcome the degeneracy, we first fix the continuum spectrum to the theoretical value
with the many-body perturbation theory (MBPT) method in the FAC code \citep{gu2006, gu2009}. The calculation has been described in S19 and benchmarked within the same experiment. 
A second-order MBPT algorithm and configuration interaction are used to enhance the accuracy of the atomic structure, which
in the end has a direct effect on the collision strength of the allowed transitions. To minimize the discrepancy between
theory and lab data, we scale the theoretical continuum spectra by the measurement data at beam energies of 
$1100-1115$~eV, where the direct excitation dominates \citep{brown2006}. As reported in S19~Tab.~1, the excitation continuum
for the 3$s$ - 2$p$ transition from the MBPT theory agrees with the EBIT data, while for the 3$d$ - 2$p$ transition, the 
MBPT overestimates the continuum by $\sim9$\%. Therefore, the scaling applied to the theoretical continuum is done independently
for the two transitions. After further correcting for polarization and the instrumental
broadening, the continuum model is subtracted from the original data, and the resonant peaks are modeled based on
Eqs.~(1-4). The continuum model can be updated by taking into account the residual from the resonant fits. Through
$4-5$ iterations, the procedure converges, and the variation on the continuum flux between two runs becomes less than $10^{-4}$.

\begin{figure}[!htbp]
\centering
\resizebox{0.7\hsize}{!}{\includegraphics[angle=0]{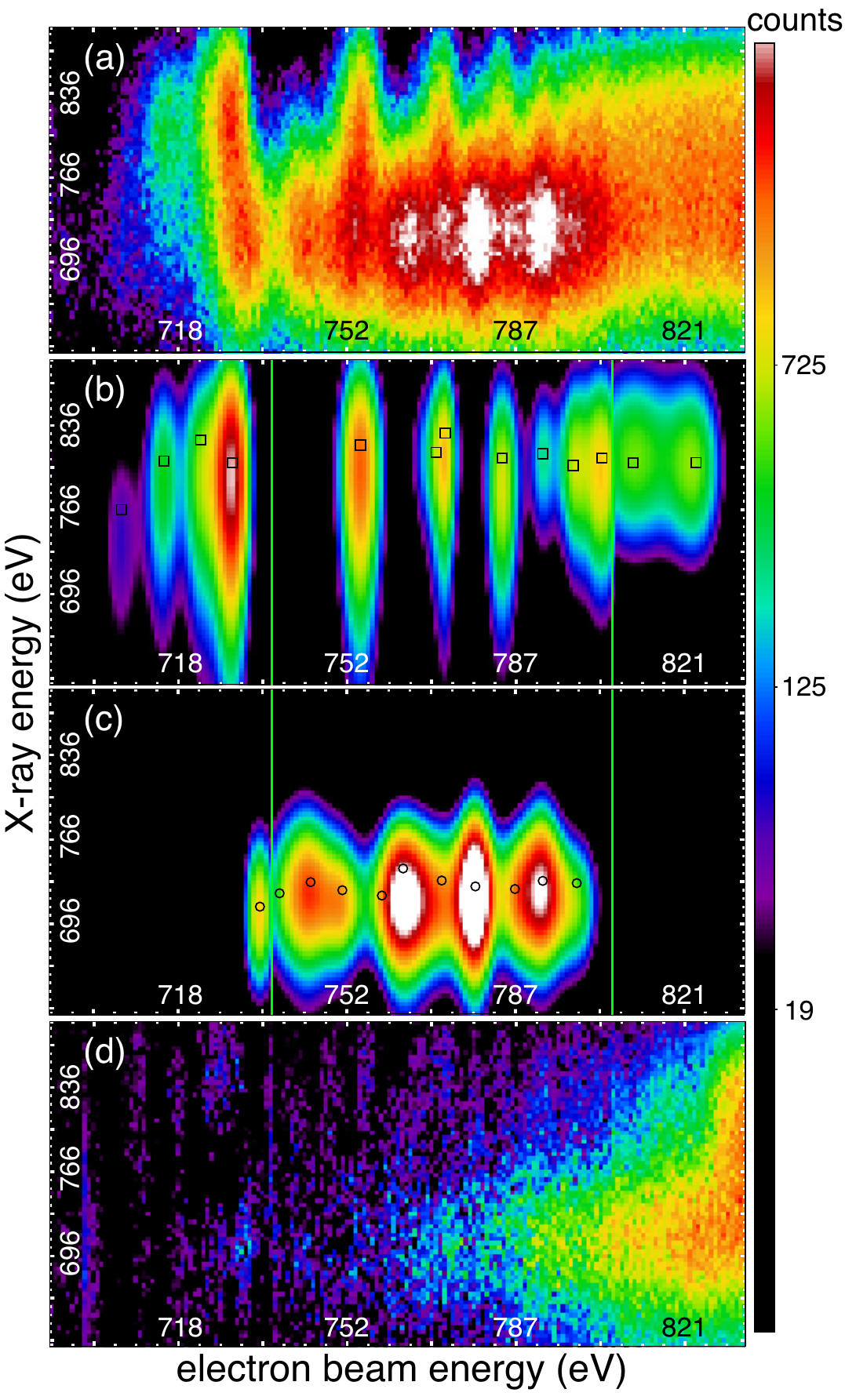}}
\caption{(a) EBIT data of the complex region (labeled ``DR + RE complex'' in Fig.~\ref{fig:ebitdata}). 
(b) Model of dielectronic recombination, the central positions are marked by black boxes. The boundaries of the fit 
windows are shown by green lines. (c) Model of resonant excitation, the central positions are marked by black circles.
(d) Background noise, residual of the fit, and the direct excitation continuum. }
\label{fig:drrecom}
\end{figure}

The continuum component is further decomposed into the 3$s$ - 2$p$ lines and the 3$d$ - 2$p$ lines. This is done in
two steps, following a similar procedure to the one described in S19. First, we project the continuum data in the $700-1120$~eV beam-energy band along the photon energy, and
fit it with a two-component model, each with a form of $F_{y} (y) + F^{\ast}_{y} (y)$. The photon energies $y_0$ of the two components
are fixed at 0.826~keV (3$d$ - 2$p$) and 0.726~keV (3$s$ - 2$p$). Second, the $700-1120$~eV electron beam band has further been divided into
14 segments, an independent two-component fit is done in each segment. Both components are treated as scale-down
of the broad-band continuum, $c[F_{y} (y) + F^{\ast}_{y} (y)]$, where $c$ is the normalization factor. The parameters on the detector response ($N_{2}$, $N_{3}$, $N_{4}$, $\sigma_{y}$, and $\sigma^{\ast}_{y}$) are fully fixed
to the results obtained in the first step. The line fluxes can thus be determined from the only free parameter $c$,
which is obtained through a fit at each beam energy segment.

The decomposition of the EBIT data at the complex region ($\sim 690-835$~eV) is shown in Fig.~\ref{fig:drrecom}. 
The dielectronic recombination from $n=6$ and higher, and the resonant excitation from $n=3$ and 4, are separated from 
the direct excitation continuum. The reduced-$\chi^2$ is $\sim$1.5 at the resonant peaks, where the systematic uncertainties
are higher than those at the dielectronic recombination peaks (\S~\ref{sec:uncert}). Thanks to very high counting statistics 
achieved in the experiment, the average position accuracy
for the resonant components is 0.6\% on the beam energy axis and 4\% on the photon energy axis, relative to the absolute values obtained in the fits.

To enable direct comparison with the astrophysical codes, we convolve the models of the resonant and the continuum components 
with a Maxwellian beam energy distribution. The line emissivities are calculated at four different temperatures, 0.25~keV, 0.5~keV, 
0.75~keV, and 1.0~keV. Obviously, the Maxwellian distribution covers a broader energy range than the EBIT data. This would not
affect the accuracy of the resonant component, because most of the dielectronic recombination and resonant excitation transitions 
can be found in the EBIT scan range. However, the collision strength of direct excitation has a continuous distribution towards
high energy (Paper I). To model the contribution beyond the EBIT scan range, we calculate the direct excitation cross section
up to 10~keV based on the distorted-wave method implemented in the FAC code. To connect the calculation smoothly to the lab data, 
the FAC model is scaled to the polarization-corrected EBIT data at 1.1~keV. The line emissivities of direct excitation are
obtained based on the combined EBIT data and the theoretical calculation. A direct calibration of the codes based on the Maxwellianized data 
is reported as follows.

\subsection{Code calibration}

The SPEX code calculates the ionization, recombination, and excitation processes on-the-fly from a database of fundamental atomic data,
including the level energies, radiative transition probabilities, Auger rates, and recombination/ionization/excitation rate coefficients.
The atomic data used to calculate Ne-like Fe has been described in Paper I. In essence, the dielectronic recombination rates are
obtained with a distorted wave calculation with the FAC code, while two parallel sets of excitation rate coefficients are available; one
with the distorted wave method (SPEX-FAC hereafter) and the other one based on the $R$-matrix calculation reported in \citet{liang2010} (SPEX-ADAS). For the Na-like Fe,
the current atomic constants in the SPEX code are mostly computed with the FAC code. It includes complete 2$s^2$2$p^6nl$, 2$s^2$2$p^5nln'l'$, 
and 2$s$2$p^6nln'l'$ configurations with principal quantum number $n$ and $n'$ $\leq 12$. The full-order configuration mixing and relativistic Breit interaction are taken into
account. The radiative relaxation routes and Auger routes for all the doubly-excited levels are calculated. The dataset is fed into the 
numerical solver for the rate equation in SPEX to calculate the steady-state level population, as well as the line emissivities from the
individual (or a set of) atomic processes. We calculate the model spectra of dielectronic recombination into Na-like Fe and the resonant and
direct excitation of the Ne-like Fe. A detailed comparison of the SPEX-FAC and SPEX-ADAS calculations is given in \S~\ref{sec:specmodel}.

The experimental data are normalized to the model spectra at the dielectronic recombination transition from the intermediate state 
2$p^5$3$d^2$ ($^2$F$_{7/2}$). This line has excitation energy of 0.412~keV and photon energy of 0.816~keV. It is selected because
this line is not only the strongest but also the simplest transition with the lowest allowed principal quantum number ($n=3$) of all the
related dielectronic recombination channels. The theoretical calculation is therefore expected to be reliable. The same normalization
technique was applied in S19, in which it has been further verified through a set of consistency checks on the resonant excitation peak at 
0.734~keV and on the radiative recombination continuum at 0.964~keV of beam energies. Moreover, a complete set of normalized experimental data has been also compared with the independent test storage ring experiment, both experiments agree very well (see Fig. 2 of S19). This shows the robustness of the calibration. The systematic uncertainty on the normalization factor has two main
sources: the statistical uncertainty of 2\% from the two-dimensional fit, and the systematic error of 3\% from the transmission curve
of the carbon filter. 

%As is discussed in detail in \S~\ref{sec:ebitde}, application of above normalization factor brings agreement between the direct excitation
%component and the theoretical calculation at the 3$s$ region. From previous reports (e.g., \citealt{brown2006} and S19), this agreement was
%well known to hold, which in turn  verifying the normalization method.

To have a crosscheck, we compare the dielectronic satellite intensities obtained from the two-dimensional fits with the data reported in S19, which were extracted from a strip region
with a width of 60~eV around the photon energy of the 3$d$ - 2$p$ line. The S19 data are reprocessed to have the same Maxwellian distribution and same beam energy resolution, and 
are normalized to our data at the 
2$p^5$3$d^2$ ($^2$F$_{7/2}$) transition. As seen in Fig.~\ref{fig:drfirst}, the two data show roughly consistent relative heights of the $n=3-5$ peaks, while our data show higher
peaks at $n=6$ and 7. Since the two data are taken from a same experiment, this discrepancy must solely come from the data analysis. We speculate that, due to
the poor resolution on photon energy, the conventional method adopted in S19 might miss a portion of flux, and the fraction included in the extraction region might vary among
different dielectronic peaks. Furthermore, the heights of the peaks would look different with the conventional method if the resolution of beam energy varies during the energy scan. Our reconstruction 
from the two-dimensional fits should naturally be more accurate, as it could handle both the detector response and the possible beam energy width variation.

\subsubsection{Dielectronic recombination}
\label{sec:dr}

\begin{figure}[!htbp]
\centering
\resizebox{\hsize}{!}{\includegraphics[angle=0]{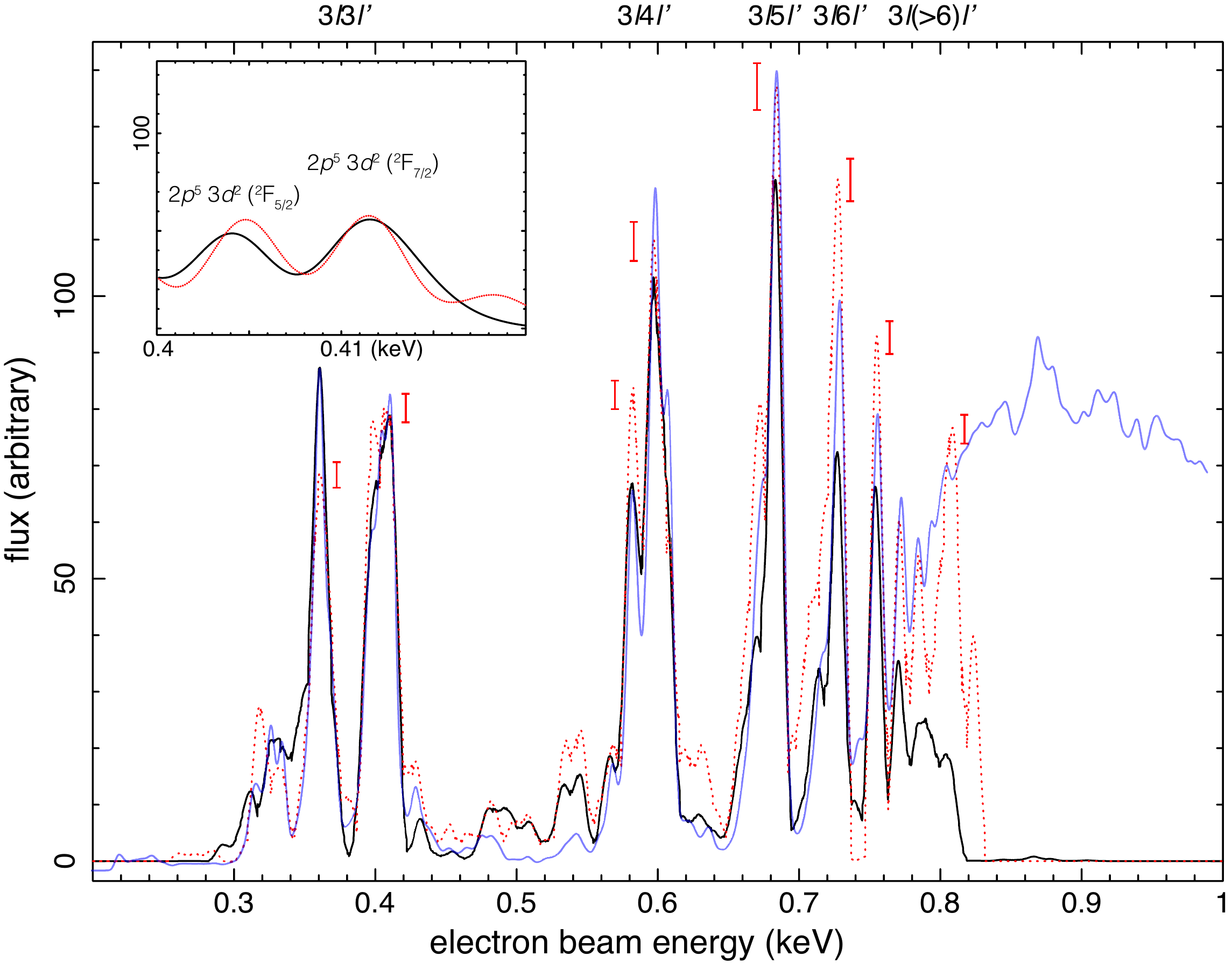}}
\caption{Dielectronic recombination line intensities as a function of electron beam energy. The original SPEX model and the EBIT data
are shown in black and red, respectively. Data uncertainties on the strong lines are marked as error bars alongside the peaks. 
Inset: normalization of the data and model at 2$p^5$3$d^2$ ($^2$F$_{7/2}$). The resolutions of the spectra are 4~eV in
the main plot, and 2~eV in the inset. For a comparison, the dielectronic recombination + direct excitation intensities reported in 
S19 are plotted in blue. The S19 data were extracted from a region of interest, 
with a width of 60~eV and centered on the 3$d$ - 2$p$ transition energy. We convolve the S19 data with a Maxwellian beam energy distribution at 0.5~keV,
and normalize the data to our curve shown in red at 2$p^5$3$d^2$ ($^2$F$_{7/2}$). Direct excitation becomes gradually dominant at beam energy $\geq 0.8$~keV in the S19 curve.}
\label{fig:drfirst}
\end{figure}

Dielectronic recombination satellites are known to contribute significantly to the 3$d$ - 2$p$ lines of Ne-like Fe \citep{beiersdorfer2017}. 
Moreover, \citet{bei2018} found that the satellites provide an accurate diagnostic of the electron temperature. However, the scientific potential
of the dielectronic satellites might be limited by the code accuracy; as shown in Fig.~\ref{fig:drfirst}, when the SPEX model and experimental data 
(both with a Maxwellian temperature of 0.5~keV) are 
normalized at quantum number $n=3$, the model clearly underestimates the line emissivities for $n=5$ and higher. For $n=4$ the model gives a reasonable fit to the
data. To calibrate these
satellites, we introduce a $n$-dependent tweak factor on the model spectra. First, the dielectronic satellites from the same $n$ are grouped in both the data and the model. For each group with quantum number $n$, the Auger rates of the SPEX code are multiplied by a constant $C_{n}$ to ``fit'' the observed EBIT 
spectrum. For $C_{n} > 1$, the dielectronic recombination rates increase, while the radiative branching ratios would decrease as more cascades become non-radiative. 
Taking these into account, the dielectronic recombination
line strengths would increase by a factor of $C_{n} (A_i + R_{ij}) / ( R_{ij} + C_{n}A_i)$, where $A_i$ is the original Auger rate from level $i$ and $R_{ij}$ is the radiative decay
rate from level $i$ to $j$. The best-fit $C_{n}$ factors are
$\sim 1.5-2$ for $n=5-8$. For $n=9$ and higher, the original Auger rates are multiplied by a factor of $\sim 3-4$. The large tweak factor at high $n$
would not only lift the $n=9-12$ peaks to the level of the EBIT data, but also can take into account the emission from $n>12$ components, which are missing in 
the current code. For convenience, the missing fluxes from $n>12$ are added to $n=9$. Practically, the migration is valid 
because the $n\geq 9$ peaks have nearly the same photon energy ($\delta E/E \leq 1 \times 10^{-4}$), making it almost impossible to disentangle the $n>12$ blend from the $n=9$ transitions
in astrophysical observations.

\begin{figure}[!htbp]
\centering
\resizebox{\hsize}{!}{\includegraphics[angle=0]{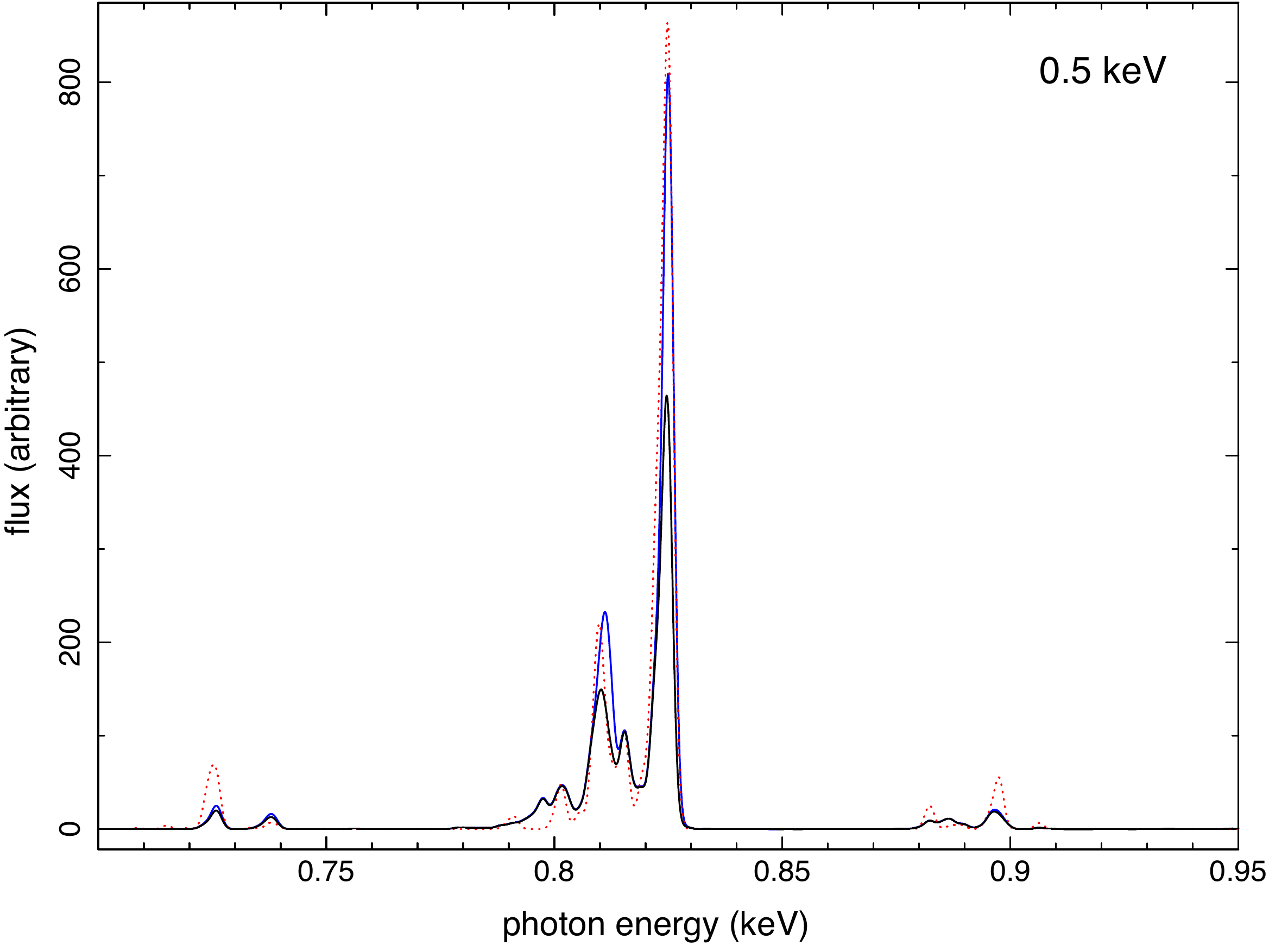}}
\caption{X-ray spectra of Na-like Fe dielectronic recombination from SPEX with the original (black) and the corrected (blue) Auger rates, compared with
the EBIT data (red), for a Maxwellian temperature of 0.5~keV. The spectral resolution is set to 2~eV. Note that the experimental photon energies obtained from 2D fits are corrected to the SPEX values, because the energy accuracy due to the detector resolution is poor compared to the theoretical values.}
\label{fig:drspec}
\end{figure}

\begin{figure*}[!htbp]
\center
\resizebox{0.8\hsize}{!}{\includegraphics[angle=0]{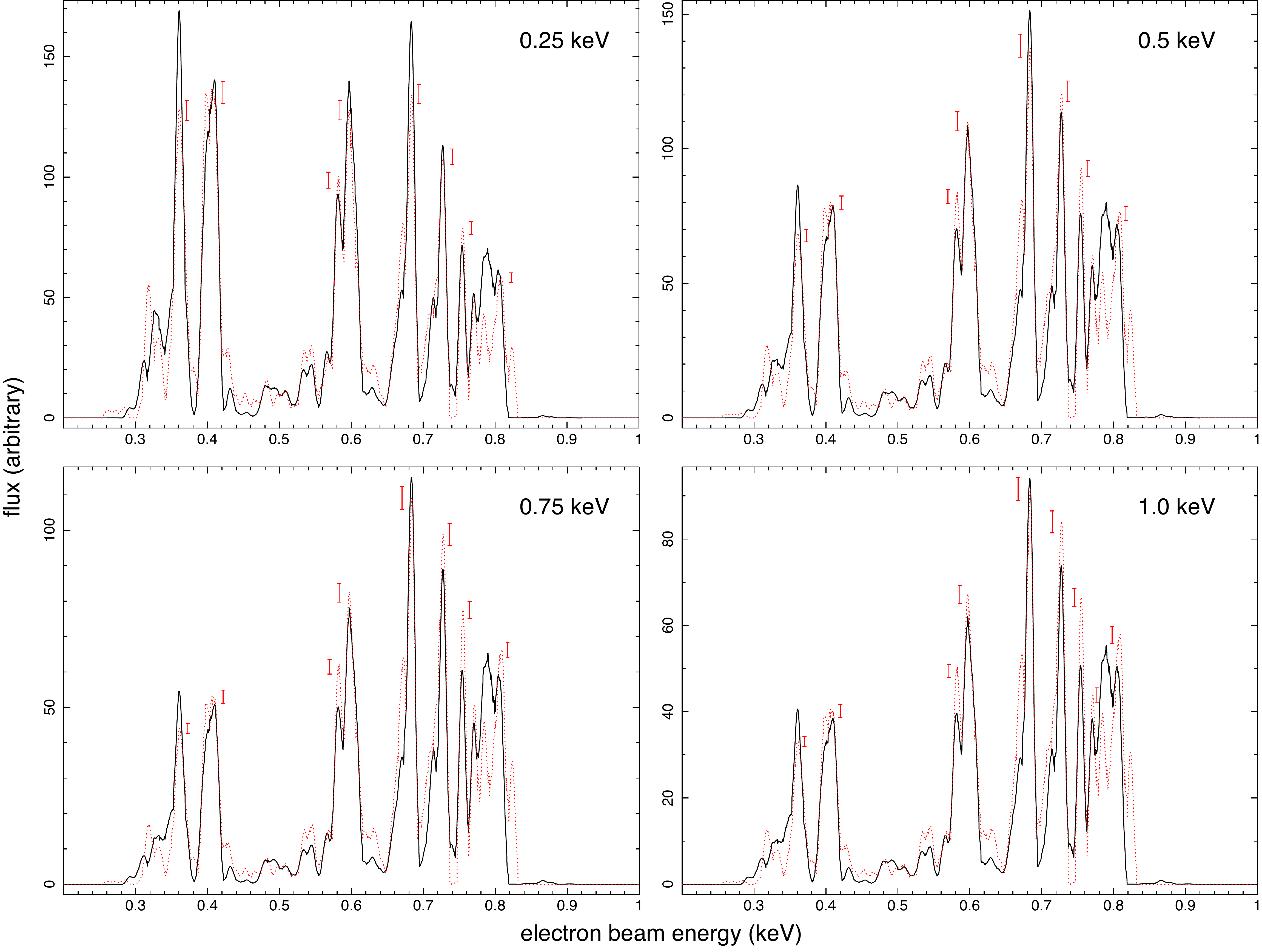}}
\caption{Electron beam spectra of dielectronic recombination from SPEX (black) and EBIT (red), after the tweak factor $C_n$ is applied to the SPEX code. Errors
are plotted in the same way as in Fig.~\ref{fig:drfirst}. In general
the code agrees well with the EBIT data for the four temperatures shown. }
\label{fig:drall}
\end{figure*}

As shown in Fig.~\ref{fig:drspec}, the tweaked X-ray spectrum now matches well with the EBIT data. The discrepancy between model and data on the total flux has been
reduced from 40\% to $\leq 5$\%. Although
this exercise is done only at a temperature of 0.5~keV, it is expected that the model spectra at other temperatures and ionization states are automatically
corrected, as the underlying atomic data (i.e., Auger rate and transition probability) are similar. This is confirmed in Fig.~\ref{fig:drall}, where the new SPEX model is compared with 
the EBIT data at 0.25, 0.5, 0.75, and 1.0~keV. The remaining discrepancies on the total fluxes are 4\%, 4\%, and 3\% for 0.25~keV, 0.75~keV, and 1.0~keV. 
A minor caveat is that the line profiles of the SPEX and EBIT resonances might sometimes appear to be slightly different (Fig.~\ref{fig:drall}). This is probably 
because even with the two-dimensional fits, it might still be difficult to fully separate the neighbouring resonances blended below the beam energy resolution.

%We note that the theoretical dielectronic satellite intensities in S19 appear to be larger than the measured ones for the high $n$ series, while in the present work with SPEX, 
%the theoretical values are instead smaller than the experimental ones. It is possible that the different limits on quantum number $n$, 60 in S19 and 12 in our work, can
%partially explain the discrepancies at the very high, unresolved dielectronic recombination lines. Nevertheless, cause of the disagreement at the $n = 4-12$ peaks is still unclear.

\subsubsection{Direct and resonant excitations}
\label{sec:ebitde}

\begin{figure*}[!htbp]
\centering
\resizebox{0.8\hsize}{!}{\includegraphics[angle=0]{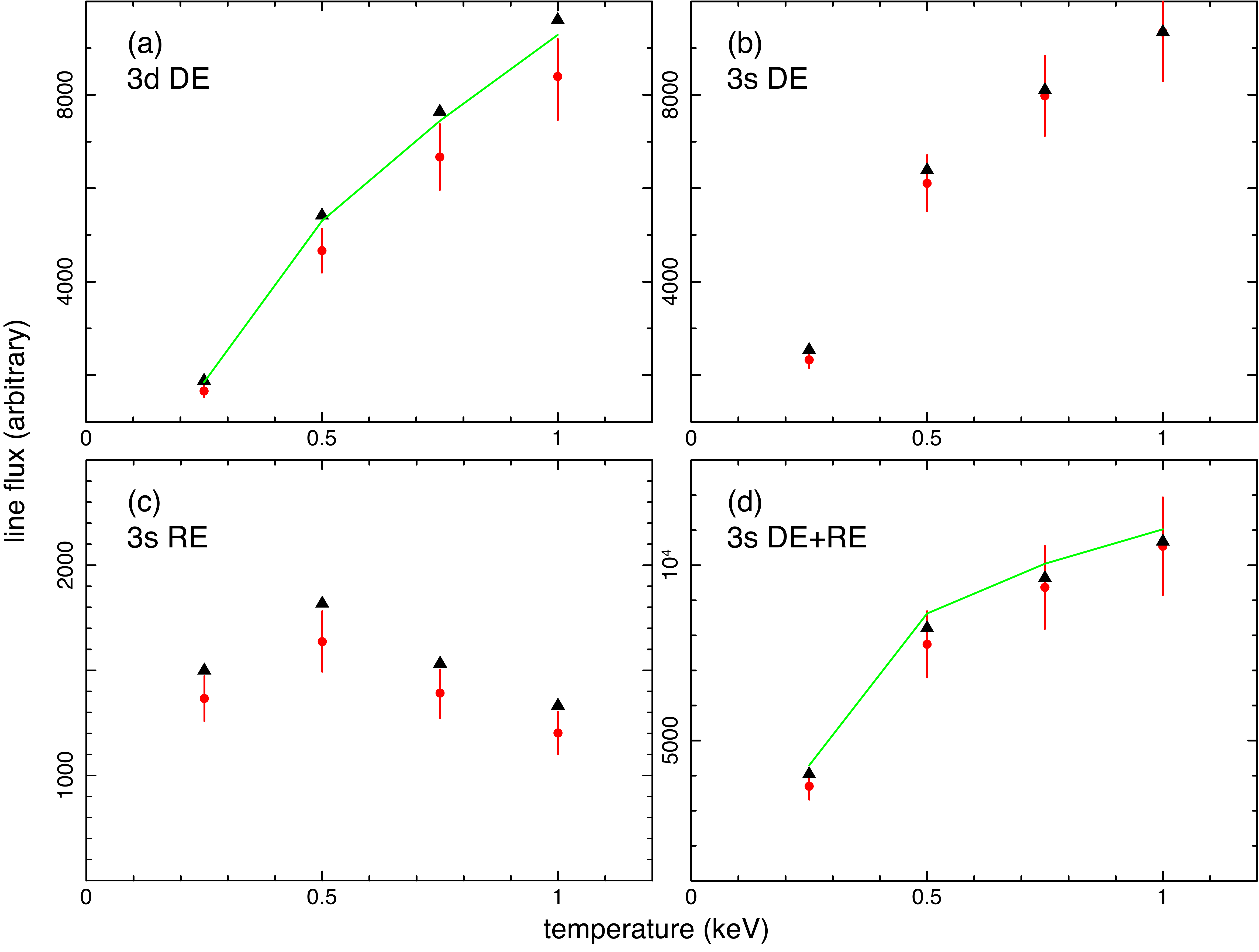}}
\caption{Comparison of SPEX calculation and EBIT measurement on the Ne-like Fe excitation. (a) Direct excitation of the 3$d$ manifold,
(b) direct excitation of the 3$s$ manifold, (c) resonant excitation onto the 3$s$ levels, and (d) combined direct and resonant excitation
of the 3$s$, from the original SPEX-FAC (black triangle), SPEX-ADAS (green line), and the EBIT experiment (red data). The SPEX-ADAS calculation does not 
separate direct and resonant excitation channels.}
\label{fig:dere}
\end{figure*}

Next, we compare the measurements of the direct and resonant excitation with the SPEX calculation on the 3$d$ and 3$s$ manifolds. It is well known
that direct excitation and the following cascade are essential ingredients for both manifolds, while resonant excitation contributes mainly to the 3$s$ levels
\citep{doron2002, chen2002, brown2006, shah2019}. As found in Paper I, both 3$lnl'$ ($n>5$) and 4$lnl'$ channels contribute
to the resonant excitation; the former can be found at the complex region shown in Fig.~\ref{fig:drrecom}, and the latter overlaps with the direct 
excitation continuum on the 3$s$ band. In the SPEX code, these resonant channels are pre-calculated into an analytic form, which is combined on-the-fly
with the direct excitation to determine the total rate coefficient. 

Figure~\ref{fig:dere} plots the SPEX-EBIT comparison on the total fluxes for direct and resonant excitation. The 3$d$ flux in the direct excitation
continuum reveals an overestimation of the SPEX flux with distorted wave ($R$-matrix) calculations by 16(13)\%. The difference is marginally significant
with respect to the uncertainty of the EBIT measurement (\S~\ref{sec:uncert}). Similar discrepancies are reported in 
\citet{loch2006,chen2011} with $R$-matrix approaches. As discussed in \citet{gu2009}, \citet{santana2015}, and S19, the issue is likely rooted
on the theoretical treatment of the high-order electron correlation, which mainly affects the transition probability and cross section of the 2$p^5$3$d$ ($^1$P$_{1}$) level,
and thus the 3$C$ line emission at 15.01~{\AA}. The other component in the 3$d$ manifold, the 2$p^5$3$d$ ($^3$D$_{1}$, so-called 3$D$) line seems to be better calculated (see \citet{brown2006}).

The source composition of the 3$s$ manifold is more complex than 3$d$. As reported in Paper I, cascades through the 3$s$ - 3$p$, 3$s$ - 3$s$, and 2$s$ - 2$p$
transitions are the major routes. Direct excitation is the main process (70\% at 0.8~keV) populating the upper states, while the rest is shared among the resonant 
excitation, dielectronic recombination, radiative recombination, and innershell ionization. Despite the intricate nature of the 3$s$ manifold, the EBIT spectrum reveals
a general agreement with the model for the direct and resonant excitation components (Fig.~\ref{fig:dere}). The SPEX flux of direct excitation is consistent with the 
data within 7\%. The resonant excitation based on the distorted wave calculation (Paper I)
appears to be slightly overestimated, though the difference (11\%) is marginal. The discrepancy becomes more subtle when the resonant excitation and direct excitation 
are combined in Fig.~\ref{fig:dere}(d), where a reasonable agreement within 8\% is found.

\begin{figure}[!htbp]
\centering
\resizebox{\hsize}{!}{\includegraphics[angle=0]{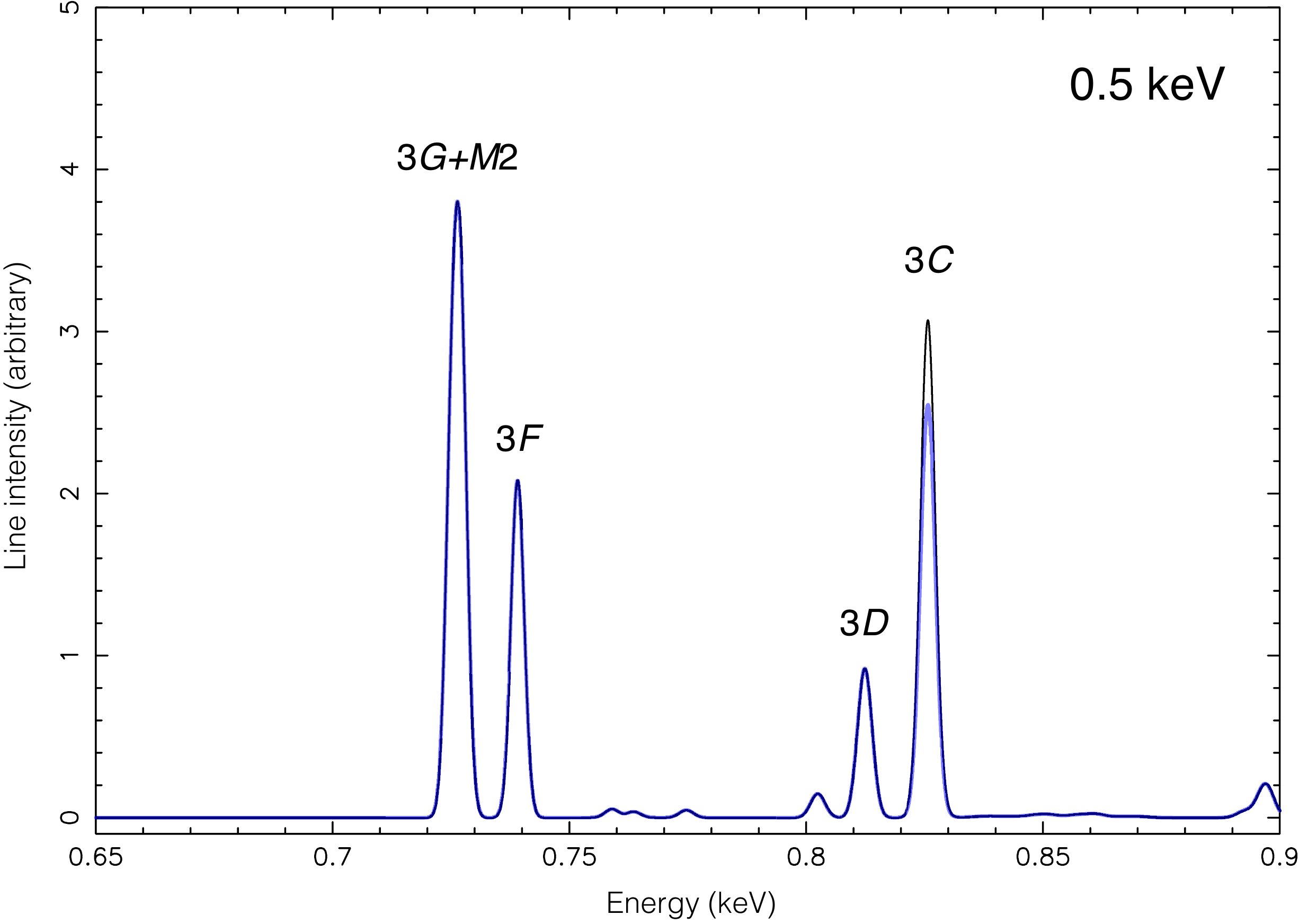}}
\caption{X-ray spectrum of Ne-like Fe with the original SPEX-FAC rates (black) and the rates corrected by the EBIT data (blue), calculated for the
equilibrium temperature of 0.5~keV. The spectral resolution is set to 2~eV.}
\label{fig:fexvii}
\end{figure}

For the code calibration, obviously, the primary issue would be to correct the atomic data on the 3$C$ line at 15.01~{\AA}. According to a previous EBIT result reported in 
\citet{brown2001}, the 3$C$/3$D$ ratio is observed to be $3.04 \pm 0.12$ at high temperature where Ne-like Fe contributes
100\% of the ion population. As shown in their Figure~2, the observed intensity ratio can be interpreted by a relative 
cross section of the 3$C$ and 3$D$ continua equal to 3.0, which agrees well with the ratios found in other experiments (e.g.,  \citealt{brown2006} and \citealt{shah2020}). This cross section ratio is then incorporated in the code calibration, combined with the results from the present EBIT experiments on the total 3$C$+3$D$ line intensity. As shown in Fig.~\ref{fig:fexvii}, the 3$C$ 
line emissivity of the Ne-like Fe is reduced by $\sim 20$\% from the original SPEX value. The new 3$C$+3$D$ intensity and
the 3$C$/3$D$ ratio are both in good agreement with the experimental values.

\begin{figure}[!htbp]
\centering
\resizebox{\hsize}{!}{\includegraphics[angle=0]{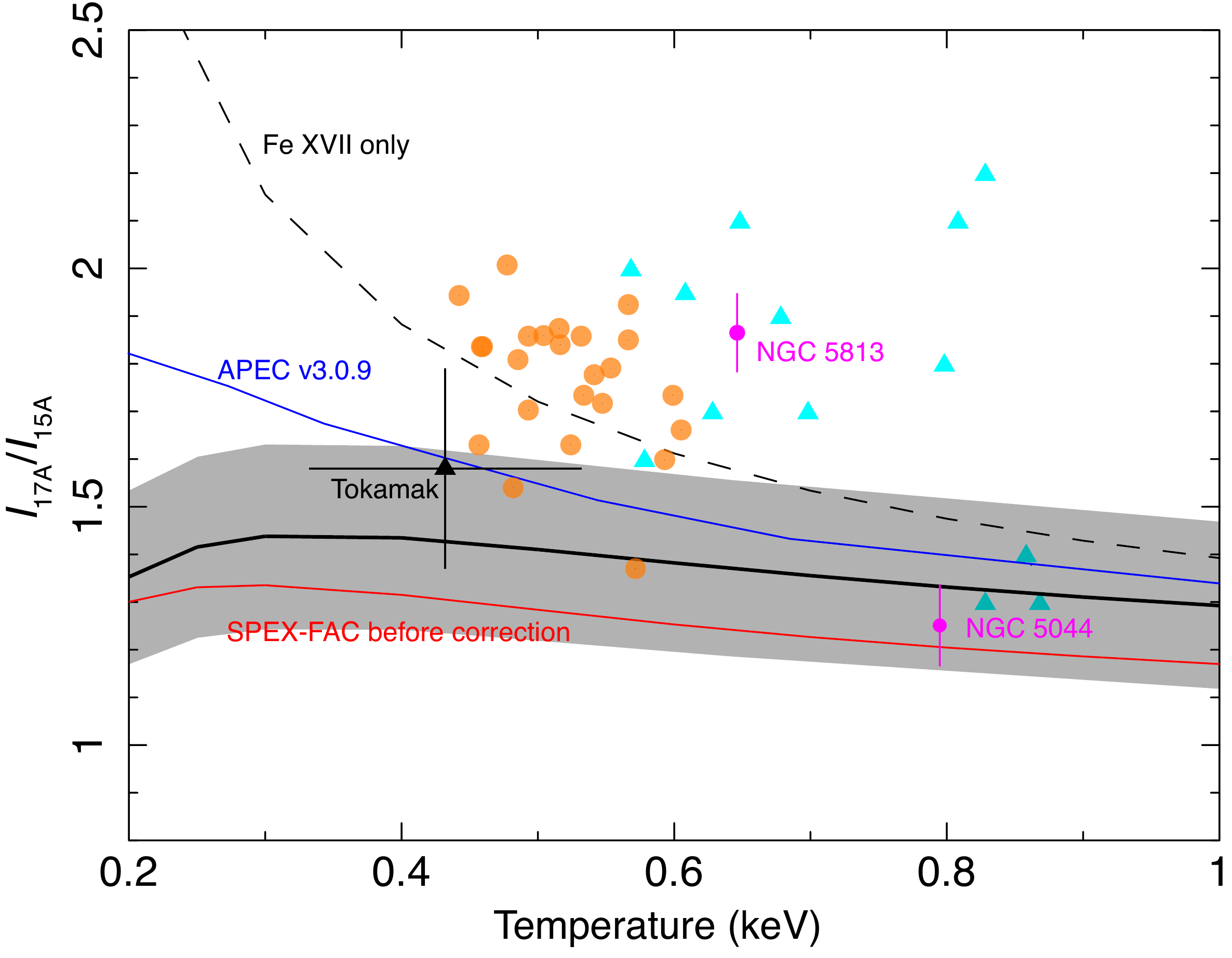}}
\caption{Observed and calculated values of the 17~{\AA}-to-15~{\AA} line ratio as a function of temperature. The SPEX model calibrated by the EBIT data is plotted by a thick black solid line, and the uncertainty is shown by the grey band. The original SPEX-FAC calculation, the APEC~v3.09, and the calibrated ratio with Ne-like Fe only are shown by the red solid, blue solid, and black dashed lines. The black triangle shows the laboratory measurement result taken from \citet{bei2004}, and the magenta data points
are the astrophysical measurements of elliptical galaxies taken from \citet{dp2012}. For NGC~5813, the deviation from the model curve is caused by the resonant scattering 
in the interstellar medium. The line ratios measured from the {\it Chandra} High-Energy Transmission Grating data of a sample of stellar coronae \citep{gu2009} are shown in orange, and those measured with the {\it XMM-Newton} Reflection Grating Spectrometer
for a sample of galaxies \citep{anna2017} are shown in cyan.}
\label{fig:lineratio}
\end{figure}

Combining the EBIT calibration to the Na-like and Ne-like transitions, we now obtain an update to the 17~{\AA}-to-15~{\AA} line ratio. This line ratio has broad astrophysical
interests, in particular in the search for photon resonant scattering by ionized plasma \citep{xu2002, beiersdorfer2002, bei2004, anna2017}. The 17~{\AA} line consists of two 3$s$ transitions: 
2$s^{2}$2$p^{5}$3$s$ $^1$P$_1$ (3$G$) and 2$s^{2}$2$p^{5}$3$s$ $^3$P$_2$ ($M$2) to the ground, and the 15~{\AA} flux is contributed by the Ne-like 3$C$ line and Na-like dielectronic
recombination lines near $15.01$~{\AA}. As plotted in Fig.~\ref{fig:lineratio}, the EBIT spectrum indicates a 17~{\AA}-to-15~{\AA} line ratio of 1.44 at 0.3~keV, and 
1.29 at 1.0~keV. The line ratio with the original SPEX-FAC calculation is lower by $4-10$\%, within the uncertainty of the EBIT data. The small discrepancy between the two
comes from the fact that the Na-like dielectronic recombination intensity increase (\S~\ref{sec:dr}) has partially cancelled out the correction on the direct excitation 
for the Ne-like 3$C$ line.

We have further compared the EBIT line ratio with that from the APEC~v3.0.9 database \citep{smith2001, foster2012}. The APEC model is systematically higher than the EBIT value, by 35\%
at 0.2~keV and 4\% at 1.0~keV. This apparent discrepancy at low temperature likely comes from the dielectronic recombination calculation. The EBIT result is in agreement with the 
Tokamak measurement \citep{bei2004} within their reported uncertainties.

It should be noted that the current EBIT experiment does not calibrate all the spectral components forming the 15~{\AA} and 17~{\AA} lines. An 
important missing component is the dielectronic recombination from F-like Fe, which
contributes through radiative cascade to the 3$s$ manifold. As shown in Paper I, the dielectronic recombination cascade contributes $\sim 35$\% of the $M$2 line at 1~keV,
and is expected to be more important at higher energies. Besides, the current work does not cover the radiative recombination and innershell ionization components.

\subsubsection{Experimental uncertainties}
\label{sec:uncert}

The overall uncertainties on the EBIT data are estimated as follows. Based on the two-dimensional fit, a statistical uncertainty on the normalization 
is calculated for each dielectronic recombination and resonant excitation peak (Tables~\ref{tab:ebit_de} and \ref{tab:ebit_re}). The statistical error is combined in quadrature with the 
systematic error on the transmission curve of the filter $\sim 3$\%, as well as the systematic uncertainty on the normalization factor calculated at 
the transition 2$p^5$3$d^2$ ($^2$F$_{7/2}$). As shown in Fig.~\ref{fig:drfirst}, the errors on the dielectronic recombination resonances are $5-8$\% 
on the peaks with $n=3 - 10$. For the resonant excitation, the average uncertainties on the total Maxwellianized fluxes are $\sim 9$\%.

Error sources on the direct excitation are (1) the statistical errors, in which the errors from the coupling of the 3$d$ - 2$p$ and 3$s$ - 2$p$
transitions are also taken into account, (2) 3\% from the carbon foil transmission, (3) error on the normalization factor at 2$p^5$3$d^2$ ($^2$F$_{7/2}$), 
and (4) the uncertainty on the theoretical calculation in the $1.1-10.0$~keV band (\S~\ref{sec:2d}). For the last component, we conservatively add a 
12\% uncertainty to the high beam energy band, which is determined by comparing the excitation cross sections obtained with the distorted wave and the MBPT methods at low beam energy band.
Combining these components in quadrature gives a total error of $\sim 8$\% at 0.25~keV, and $\sim 11$\% at 1~keV (Fig.~\ref{fig:dere}).

\section{Test on Capella grating data}
\label{sec:capella}

An important aspect of the present work is implementing the new atomic calculations and the EBIT line measurement on the spectroscopy
of representative astrophysical objects. To this end, we choose the bright non-degenerate stellar corona
of the G1$+$G8 binary Capella as the test target. The primary X-ray source is the quiescent coronal plasma, 
which is known to be in a quasi-collisional equilibrium state, heated to a temperature distribution over 
the range of $10^5 - 10^7$~K \citep{dupree1993, brickhouse2000, p2001, behar2001, desai2005, gu2006b, gu2009}. 
The spectrum of the source shows a huge amount of emission lines from $1.5-175$~{\AA}, including 
\ion{Fe}{XVI} and \ion{Fe}{XVII} lines. As a calibration source for instrumental performances (resolving
power, response matrix, line spread function), Capella has been observed many times with X-ray grating spectrometers \citep{canizares2005}.
The very bright Fe-L lines, appropriate temperature, and extremely high-quality grating data in the archive make it
the most suited target for the test. 

The work is based on the {\it Chandra} High-Energy Transmission Grating (HETG) data, covering a 
wavelength range of $1.2-32$~{\AA} with a spectral resolution of 0.023~{\AA} ($\sim 1.2$~eV at 800~eV, Medium Energy Grating) and 
0.012~{\AA} ($\sim 34.7$~eV at 6000~eV, High Energy Gratings). We do not include data from other instruments to minimize uncertainties from 
the cross-instrument calibration. A global, self-consistent fit is carried out for the broad HETG bandwidth, 
since the local fit (for instance, only in the $14-18$~{\AA} for \ion{Fe}{XVII}) might miss features from 
complex astrophysical conditions (e.g., multi-temperature). As described below, the accurate calibration, as well as the completeness
of the spectral modeling, are therefore essential to yield a reasonable description of the Capella spectrum using
a global fit.

This section is arranged as follows: first we describe the fit with the codes based on theoretical cross sections (\S~\ref{sec:calib}, \S~\ref{sec:specmodel}),
then we present the results with the EBIT-calibrated rate coefficients (\S~\ref{sec:ebitfit}). We will also overview the atomic data 
constraints obtained for the most used plasma codes.

\subsection{Data preparation}

\begin{table}[!htbp]
\centering
\caption{\textit{Chandra} ACIS/HETG observations of Capella}
\label{tab:capella}
\begin{threeparttable}
\begin{tabular}{ccccccccccc}
\hline
Observation ID                     & Start time              & Exposure (ks)      \\
\hline
57                                 & 2000-03-03              & 29.2               \\
1010                               & 2001-02-11              & 29.9               \\
1099                               & 1999-08-28              & 14.8               \\
1100                               & 1999-08-28              & 14.8               \\
1101                               & 1999-08-29              & 14.8               \\
1103                               & 1999-09-24              & 41.0               \\
1199                               & 1999-08-30              & 2.1                \\
1235                               & 1999-08-28              & 14.8               \\
1236                               & 1999-08-28              & 14.8               \\
1237                               & 1999-08-29              & 14.8               \\
1318                               & 1999-09-25              & 27.0               \\
2583                               & 2002-04-29              & 28.1               \\
3674                               & 2003-09-27              & 29.2               \\
5040                               & 2004-09-10              & 29.1               \\
5955                               & 2005-03-28              & 29.2               \\
6471                               & 2006-04-18              & 30.1               \\
9638                               & 2008-04-19              & 31.5               \\
10599                              & 2009-04-22              & 29.7               \\
11931                              & 2009-11-18              & 30.0               \\
13089                              & 2010-12-01              & 30.1               \\
14239                              & 2011-12-29              & 30.0               \\
16418                              & 2013-12-23              & 30.1               \\
17324                              & 2014-12-01              & 29.1               \\
18357                              & 2016-07-26              & 15.0               \\
18364                              & 2016-07-27              & 15.1               \\
\hline
\end{tabular}
\end{threeparttable}
\end{table}

Archival {\it Chandra} HETG observations of Capella are summarized in Table~\ref{tab:capella}.
The total clean exposure time is 594.9~ks. A subset of the data has been reported in
\citet{p2001}, \citet{behar2001}, \citet{desai2005}, and \citet{gu2006b}. The data were reduced using the CIAO v4.10 and
calibration database (CALDB) v4.8. The {\it chandra\_repro} script is used for the data screening and
production of spectral files for each observation. The spectra and the associated response files
are combined using the CIAO {\it combine\_grating\_spectra} tool. The Medium Energy Grating (MEG)
spectrum in the wavelength range of $3-32$~{\AA} and the High Energy Gratings (HEG) spectrum in $1.5-3$~{\AA} are
fit jointly.

\subsection{Calibration and background}
\label{sec:calib}

\begin{figure}[!htbp]
\centering
\resizebox{\hsize}{!}{\includegraphics[angle=0]{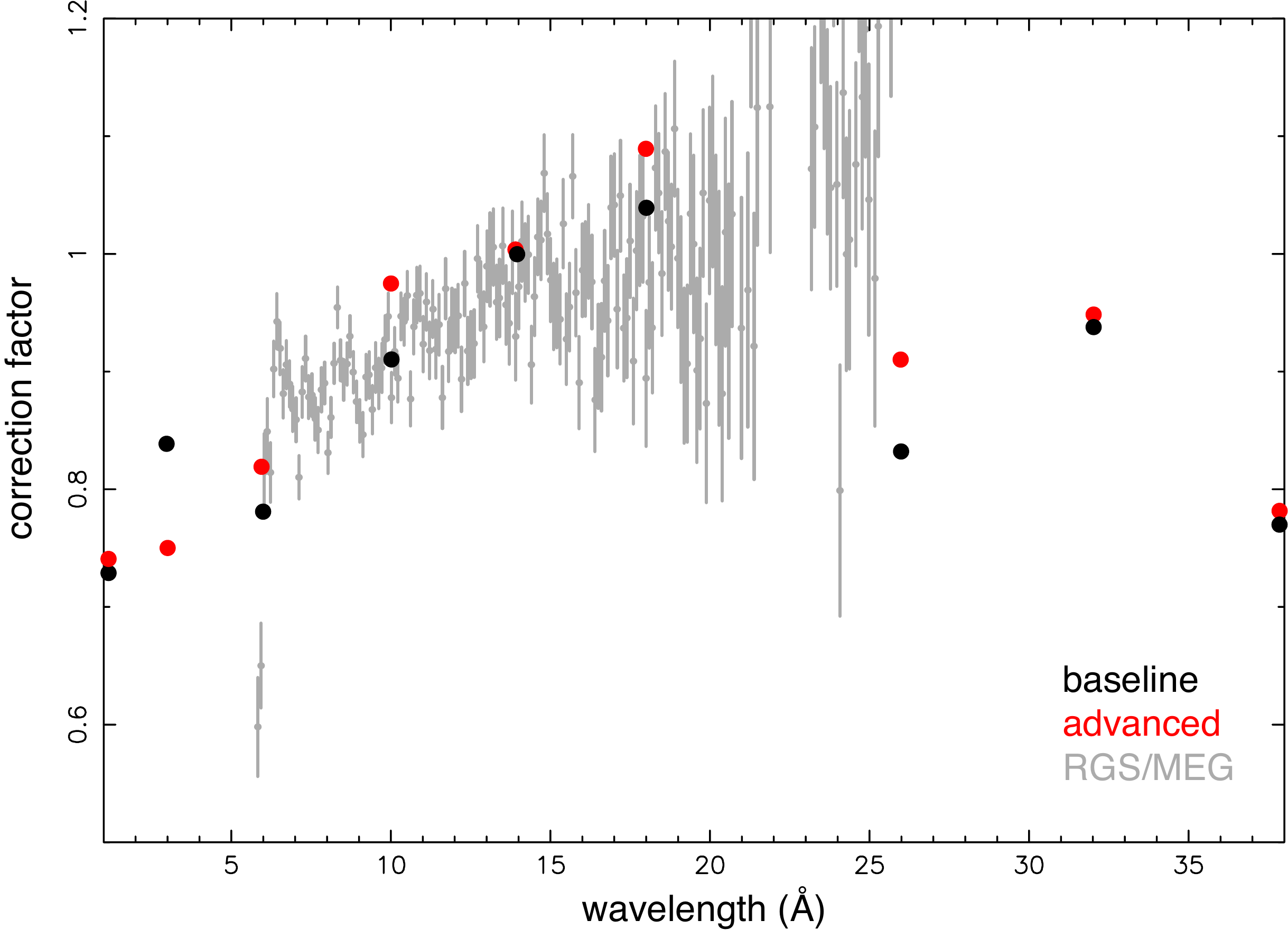}}
\caption{Effective area correction factors from the baseline and the advanced fits shown in black and red data points, respectively.
They are compared with the average effective area ratios between the {\it XMM-Newton} Reflection Grating Spectrometer and the
{\it Chandra} Medium Energy Grating for a sample of AGN sources (Kaastra, private communication). }
\label{fig:aeff}
\end{figure}

To remove possible residual calibration errors on the HETG effective area, we incorporate two
correction functions in the spectral analysis. One represents uncertainty in the \ion{O}{I} edge
($22.6 - 22.9$~{\AA}), another is the possible error in the broad-band effective area of the X-ray mirrors.
The edge correction is added as an extra neutral-O absorption component ({\it hot} model) in the spectral 
analysis. A positive (negative) absorption column density would mean that the actual edge is deeper (shallower) 
than the standard calibration. For the second component, we incorporate a {\it knak} component 
which determines the continuum correction function using piecewise power laws in the energy-correction factor
space. The grid points are set at 1, 3, 6, 10, 14, 18, 26, 32, and 38~{\AA}, skipping the location 
of \ion{O}{I} edge that would be taken into account by the first component. By making several iterations between a fit with
100~eV-wide bins and a fit with the optimal binning, the best-fit $\it knak$ and \ion{O}{I} edge models are 
determined. The two components are fixed throughout the fits. The same approach was also used
in the analysis of the Hitomi data of the Perseus cluster \citep{hitomiatomic2018}.

Shown in Fig.~\ref{fig:aeff}, the effective area correction factors derived with the $\it knak$ model are compared with the empirical values
obtained by a joint analysis of {\it XMM-Newton} and {\it Chandra} grating data for a sample of AGN sources (Kaastra, private communication).
Our models agree well with the data in 5-20~{\AA}. At the longer wavelengths, the models seem to fall below the data,
though the uncertainty of the data becomes larger.

The shape of the lines are determined by the instrumental line spread function and the relevant astrophysical 
effects such as random motion, both contain higher-order systematic uncertainties. The instrumental line spread function 
calibration has been cross-checked with the {\it XMM-Newton} Reflection Grating Spectrometer, and a good match has been
found between the two instruments (Kaastra, private communication). Then, to model accurately the additional broadening
from mainly the astrophysical effects,  
we incorporate the arbitrary line broadening model {\it vpro}, with a realistic 
profile shape calculated from the observed \ion{O}{VIII} Ly$\alpha$ line at $\sim 19$~{\AA}. This model further allows
fine-tuning with a Doppler broadening parameter to eliminate any residual biases from the uncertainties of the instrumental and astrophysical effects.

The standard pipeline instrumental background has been reprocessed in the following way. A Wiener filter is applied
to smooth out the noisy features in the background continuum, with the noise level determined by a Fourier transform.
This process is needed for utilizing the C-statistic on the fits of spectra with low count numbers~\citep{kaastra2017}.

\subsection{Spectral modeling with theoretical rates}
\label{sec:specmodel}

\noindent{\bf Atomic dataset}
\smallskip

Here we describe the source of the fundamental atomic data used in the Capella work. The SPEX-ADAS database is utilized for the 
fit with the baseline model. It contains the same atomic data as the SPEX-FAC database except for the collisional excitation
data on the Fe-L: the SPEX-ADAS incorporates the recent $R$-matrix results (\ion{Fe}{XVII} from \citealt{liang2010}, \ion{Fe}{XVIII}
from \citealt{witt2006}, \ion{Fe}{XIX} from \citealt{butler2008}, \ion{Fe}{XX} from \citealt{witt2007}, \ion{Fe}{XXI} from 
\citealt{badnell2001}, \ion{Fe}{XXII} from \citealt{liang2012}, \ion{Fe}{XXIII} from \citealt{fern2014}, and \ion{Fe}{XXIV}
from \citealt{liang2011}) for the low-lying levels (mostly up to $n=4$), while the SPEX-FAC utilizes the uniform distorted wave calculation 
with isolated resonances from Paper I. The $R$-matrix and distorted wave calculations are found to be consistent within 20\%
on the main Fe-L lines, although the discrepancies become significantly larger for the weaker transitions, in particular for 
\ion{Fe}{XVIII}, \ion{Fe}{XIX}, and \ion{Fe}{XX}. Generally speaking, the accuracy of $R$-matrix calculation is expected to be
superior to that of the distorted wave with isolated resonances. At the high levels, the SPEX-ADAS and SPEX-FAC converge to the same distorted wave
calculation, as the $R$-matrix data become gradually sparse with increasing $n$. 
The rates and wavelengths of non-Fe-L transitions in SPEX-ADAS and SPEX-FAC are the same as the public SPEX version 3.05.
We also run the analysis with the standard SPEX version 3.05.

Though the atomic data incorporated in the APEC code are in general consistent with SPEX, significant differences on 
several individual transitions have been reported \citep{hitomiatomic2018}.
We, therefore, include the latest APEC version 3.0.9 in the Capella fit, serving as an independent reference.

\begin{figure*}[!htbp]
\centering
\resizebox{0.8\hsize}{!}{\includegraphics[angle=0]{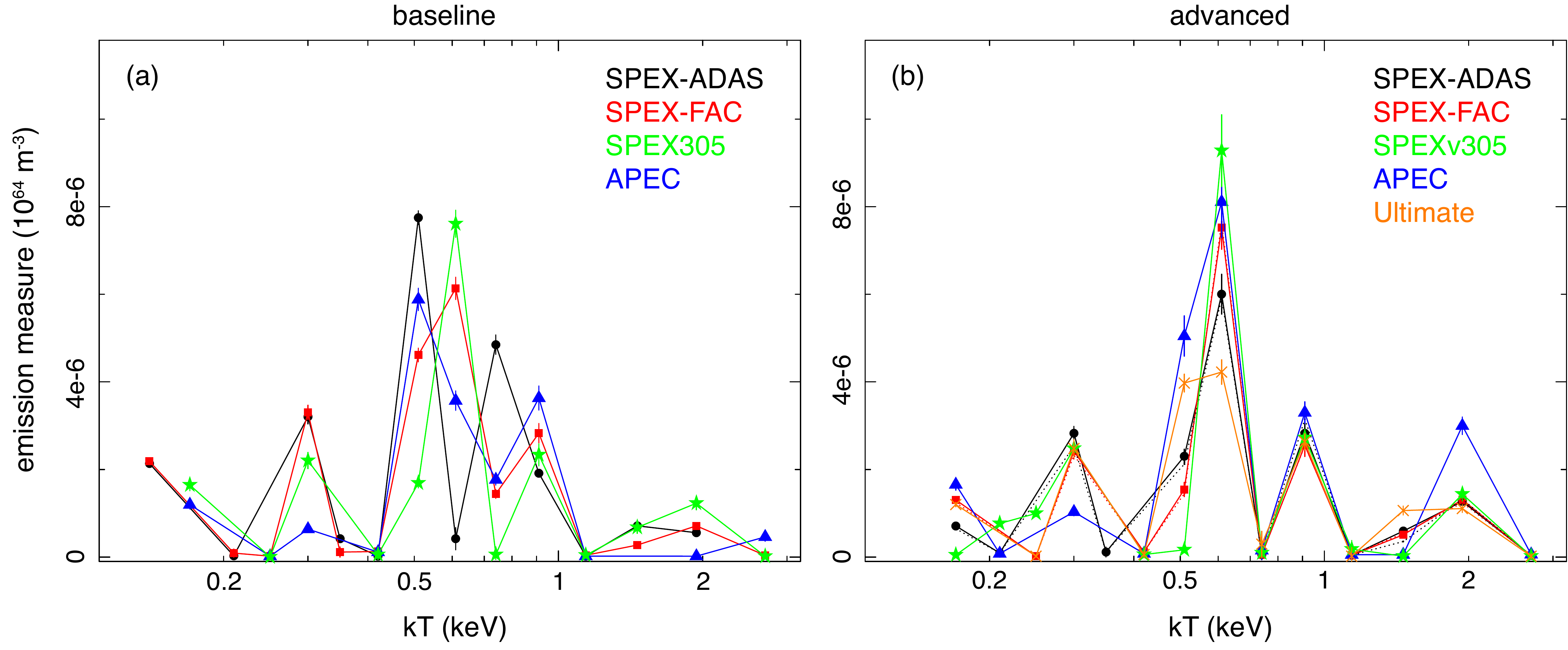}}
\caption{Differential emission measure distributions with the baseline fits (a) and the advanced fits (b), using the atomic data from 
the SPEX-ADAS calculation (black), SPEX-FAC calculation (red), SPEX version 3.05 (green), and APEC (blue). The advanced fits with the EBIT correction are shown in
dotted lines (they appear to nearly overlap with the non-EBIT-correction counterparts). The ultimate fit (see Apprendix B for detail) is plotted in orange. }
\label{fig:dem}
\end{figure*}

\begin{table*}[!htbp]
%\begin{sidewaystable}
\caption{Parameters of the reference model and sensitivity to model assumptions.
The first two lines give the best-fit values with their 1$\sigma$ statistical uncertainty. The next lines show the parameter
differences of the tested models relative to the baseline model (boldface for $>$3$\sigma$ differences).}
\label{tab:parameters}
\small
\centerline{
\begin{tabular}{@{}l@{\hskip 0.05in}c@{\hskip 0.05in}c@{\hskip 0.05in}c@{\hskip 0.05in}c@{\hskip 0.05in}c@{\hskip 0.05in}c@{\hskip 0.05in}c@{\hskip 0.05in}c@{\hskip 0.05in}c@{\hskip 0.05in}c@{\hskip 0.05in}c@{\hskip 0.05in}c@{\hskip 0.05in}c@{\hskip 0.05in}c@{\hskip 0.05in}c@{\hskip 0.05in}c@{\hskip 0.05in}c@{\hskip 0.05in}c@{\hskip 0.05in}c@{}}
\hline\hline
Model & $C_{\rm stat}^{\rm a}$ & C & N & O & Ne & Na & Mg & Al & Si & S    & Ar   & Ca  & Cr & Fe & Ni  & $N_{\rm {H,h}}^{\rm b}$ & $T_{\rm abs}^{\rm b}$ & $N_{\rm {H,c}}^{\rm c}$ & $v^{\rm d}$  \\
%          & & & & &  & & & & & & & & & & & {\scriptsize $10^{24}$ m$^{-2}$} & {\scriptsize keV} & {\scriptsize $10^{24}$ m$^{-2}$} & {\scriptsize km s$^{-1}$}\\     
\hline
Baseline               & 59597.8 & 0.77 & 1.06 & 0.41 & 0.45 & 1.10 & 0.75 & 0.79 & 0.84 & 0.61 & 0.43 & 0.48 & 0.85 & 0.587 & 0.59 & 0.65 & 0.37 & 0.06 & 108.4 \\
Error                  &          & 0.11 & 0.03 & 0.01 & 0.01 & 0.04 & 0.01 & 0.03 & 0.01 & 0.13 & 0.03 & 0.02 & 0.15 & 0.002 & 0.01 & 0.05 & 0.02 & 0.03 &  0.7  \\
\hline\noalign{\smallskip}
\multicolumn{15}{l}{\textsl{Plasma codes:}} \\
SPEX-FAC               & -1216.2 & 0.04 & 0.01 & 0.0  & 0.01 &-0.08 & 0.0  & 0.0 & 0.01 & 0.02 & 0.03 & 0.04 & 0.19 & {\bf 0.029} &-0.02 & {\bf 3.79} & {\bf 0.25} &-0.03 & 1.0 \\
SPEXv305               & 11148.8 &-0.09 &-0.07 &-0.03 &-0.02 &-0.10 & 0.0  & 0.02 & {\bf 0.04} & 0.03 &-0.06 &-0.06 & {\bf 0.44} & {\bf 0.124} &-0.02 & {\bf 0.47} & {\bf 0.21} &-0.04 & {\bf 24.5} \\
APECv309               & 17132.9 & {\bf 1.13} & {\bf 0.54} & {\bf 0.25} & {\bf 0.15} & {\bf 0.32} & {\bf 0.04} & 0.0 & 0.03 & 0.02 & {\bf 0.10} & {\bf 0.12} & {\bf -0.76} & {\bf 0.064} & {\bf 0.13} & {\bf 17.16} & {\bf 0.62} &-0.05 & {\bf-6.0}  \\
\hline\noalign{\smallskip}
\multicolumn{15}{l}{\textsl{Advanced model:}} \\
SPEX-ADAS              & -2684.8 & 0.25 & 0.09 & {\bf 0.08} & {\bf 0.06} &-0.06 & {\bf 0.11} & 0.09 & $-$ & $-$ & $-$ & $-$ & 0.03 & $-$  & {\bf 0.10} & {\bf 1.03}  & 0.03  &-0.05 & $-$  \\
SPEX-FAC               & -3741.8 & 0.10 & 0.0  & 0.01 & {\bf 0.04} & {\bf-0.20} & {\bf 0.08} & 0.08 & $-$ & $-$ & $-$ & $-$ & 0.17 & $-$ & {\bf 0.04} & {\bf 0.79} & {\bf 0.11} &-0.05 & $-$ \\
SPEXv305               & 9104.9  &-0.05 &-0.06 & {\bf -0.04} & 0.01 & {\bf -0.25} & {\bf 0.04} & 0.10 & $-$ & $-$ & $-$ & $-$ & 0.30 & $-$ & 0.01 & {\bf 0.93} & {\bf 0.15} &-0.05 & $-$ \\
APECv309               & 13363.4 & {\bf 1.22} & {\bf 0.40} & {\bf 0.22} & {\bf 0.15} & {\bf 0.39} & {\bf 0.18} & {\bf 0.26} & $-$ & $-$ & $-$ & $-$ & {\bf -0.80} & $-$ & {\bf 0.20} & {\bf -0.22} & -0.05 &-0.05 & $-$ \\
\hline\noalign{\smallskip}
\multicolumn{15}{l}{\textsl{Advanced model with EBIT calibration:}} \\
SPEX-ADAS              & -2139.4 & {\bf 0.36} & {\bf 0.16} & {\bf 0.13} & {\bf 0.08} &-0.06 & {\bf 0.13} & {\bf 0.14} & $-$ & $-$ & $-$ & $-$ & 0.03 & $-$ & {\bf 0.09} & {\bf 0.35} & -0.02 &-0.05 & $-$  \\
SPEX-FAC               & -3259.7 & 0.11 & 0.03 & 0.03 & {\bf 0.05} & {\bf-0.20} & {\bf 0.08} & {\bf 0.11} & $-$ & $-$ & $-$ & $-$ & 0.14 & $-$ & {\bf 0.04} & {\bf 0.85} & {\bf 0.09} &-0.05 & $-$  \\
\hline\noalign{\smallskip}
\multicolumn{15}{l}{\textsl{Ultimate fit$^{\rm e}$:}} \\
SPEX-ADAS              & -21695.9 & 0.08 & {\bf 0.12} & {\bf 0.11} & {\bf 0.11} & {\bf 0.33} & {\bf 0.21} & {\bf 0.22} & $-$ & $-$ & $-$ & $-$ & 0.08 & $-$ & {\bf 0.17} & 0.15 & -0.03 &-0.05 & $-$  \\
\hline
\end{tabular}
}
\begin{itemize}
\item[$a:$] Expected $C_{\rm stat} = 7137.40$. 
\item[$b:$] Absorption column density (in unit of $10^{24}$ m$^{-2}$) and temperature (keV) of the ionized $hot$ component.
\item[$c:$] Absorption column density (in unit of $10^{24}$ m$^{-2}$) of the neutral $hot$ component.
\item[$d:$] Doppler broadening (in unit of km s$^{-1}$) of the $vpro$ component for the line spread function.
\item[$e:$] See details in Appendix B. EBIT rates incorporated.
\end{itemize}
%\end{sidewaystable}
\end{table*}

\smallskip
\noindent{\bf Baseline model}
\smallskip

It is known that the X-ray spectra of stellar coronae require a differential emission measure
modeling \citep{mewe2001}. The multi-temperature structure of the coronal plasma can be well approximated 
by a combination of collisional ionization equilibrium (CIE) components. We define an optimal temperature grid of the model,
derived from a pre-calculation of the ionic charge state as a function of equilibrium temperature. First, the average
charge state $C$ of each astronomically abundant element is calculated for a fine mesh of temperature $T$.
We then obtain d$T$/d$C$ at each temperature, and the minimal temperature change (in unit of keV) over all the abundant elements
that brings one charge state further ($\delta C = 1$) can be approximated by
\begin{equation}
\label{eq:opt}
\delta T = 0.05 T^{0.7} + 0.08 T^2.
\end{equation}
This defines the most efficient temperature step size regarding the dependence on charge state. 
Based on Eq.~\ref{eq:opt}, a set of 18 CIE components are defined within the temperature range 
of $0.1-10.0$~keV. The emission measure of each component is free to vary, and the metal abundances 
of C, N, O, Ne, Mg, Al, Si, S, Ar, Ca, Cr, Fe, and Ni are also set as free parameters. The abundances
of other elements (with weak lines) are set to the Solar ratio. All the CIE components are 
assumed to have the same set of abundances. 

The interstellar
absorption by neutral and ionized material is modeled using two {\it hot} components. The
column densities of the two absorbers, and the temperature of the ionized component, are set free
in the fits. The abundances of the absorbing materials are fixed to the Solar ratio.

We further apply a redshift component to the CIE components, and leave it as a free parameter to allow
any residual uncertainties in the energy scale calibration, either of instrumental or astrophysical origin.
Effective area correction components (\S~\ref{sec:calib}) are also incorporated.

We use optimally binned spectra with the C-statistic. All abundances are relative to \citet{lodders2009} 
proto-solar abundances with free values relative to those abundances for the relevant elements. The ionization 
balance is set to the one in \citet{u2017}.

The baseline model provides a reasonable fit to the main transitions in the Capella spectrum. 
For the weaker transitions, the fit becomes worse, probably due to the remaining uncertainties in the
instrumental calibration and astrophysical effects, coupled with the unsolved issues with atomic data in the code.
The total C-statistic value is 59896 for an expected value of 7137, mostly due to the residuals in the weak lines (see detail in Appendix B). 
The fit is formally unacceptable, indicating that the current modeling of the complexity in the Capella spectrum is far from sufficient.
Even so, it is still useful to discuss the relative changes of the C-statistic values, as well as the changes in plasma parameters,
with respect to the baseline fit.

The differential emission measure distribution obtained with the baseline fit shows a primary peak at 0.51~keV, as well
as secondary ones at 0.3~keV and 0.74~keV (Fig.~\ref{fig:dem}). In general, it agrees with the previous measurements 
using the HETG instrument (e.g., \citealt{gu2006b}). The elemental abundances also agree within the uncertainties with 
the values reported in \citet{gu2006b}, except for N and Si, which are derived to be $\sim 40$\% higher in our work.

As shown in Table~\ref{tab:parameters}, replacing the SPEX-ADAS calculation with the SPEX-FAC calculation in the 
baseline model yields a slightly better fit. The peak in emission measure distribution is shifted to 0.61~keV, which 
is likely a merge between the adjacent peaks at 0.51~keV and 0.74~keV (Fig.~\ref{fig:dem}). The abundances remain 
nearly intact, except for Fe. It should be noted that any changes in non-Fe abundances are caused indirectly, as the
rate coefficients of non-Fe species are the same in SPEX-ADAS and SPEX-FAC calculations. 

The quality of fit with both SPEX-ADAS and SPEX-FAC improves significantly from the one with SPEX version 3.05, proving
that the new theoretical calculations are indeed better than those in the existing codes. There are multiple places where
the parameters derived with SPEX-ADAS/FAC significantly differ from those with SPEX version 3.05. In particular, the 
Fe abundance with SPEX version 3.05 is 21\% higher than the abundance with SPEX-ADAS, or 15\% higher than the SPEX-FAC value, while the
statistical uncertainty from the instrument is only 0.3\%. This reconfirms the conclusion
of Paper I that the new calculation tends to give lower Fe abundance than the standard SPEX code. The baseline
fit with the latest APEC code is found worse than those with the calculations in SPEX. The emission measure 
distribution and elemental abundances are both significantly changed with APEC, reflecting the systematic uncertainties
associated with the existing atomic codes.

\smallskip
\noindent{\bf Advanced model}
\smallskip

As shown above, the baseline model should be regarded as a simple approximation to the astrophysical condition of the
stellar corona. By properly incorporating several more degrees of freedom into the baseline model, we are able to achieve a more advanced physical description of the source.

We construct the advanced model as follows. First, the Fe abundance, which was tied among components in the baseline model, is now 
set free for all the CIE components with non-zero normalizations. The metallicities of a few components are tied to those of their 
adjacent components, as they cannot be determined independently with the current data. The C, N, O, Ne, Na, Mg, Al, Cr, and Ni abundances remain coupled, while
the Si, S, Ar, and Ca abundances are tied for two groups of components, those with temperatures $< 1$~keV and those above 1~keV. The model becomes 
quasi- multi-abundance, providing a better approximation to the possible metallicity gradient in the corona.

We further decouple the temperature used for ionization balance calculations, $T_{\rm bal}$, from the temperature $T_{\rm spec}$
used for the evaluation of the emitted spectrum for the set of ionic abundances determined using $T_{\rm bal}$. This is done by setting
the $rt = T_{\rm bal}/T_{\rm spec}$ a free parameter in the {\it cie} model. The $rt$ parameter could take into account any possible minor 
non-equilibrium ionization effect, as well as the systematic uncertainties on the ionization and recombination rates and on the 
calibration of the broadband continuum \citep{hitomiatomic2018}. 

Density diagnostics based on He-like triplets have been widely used for stellar sources \citep{ness2001, mewe2001}. As the efficiency of 
collisional deexcitation increases with density, the forbidden-to-intercombination line ratio is expected to be density-sensitive. Besides, the 
low-lying metastable levels might become significantly populated at high density, and the excitation, recombination, and radiative relaxation from/onto
these levels are therefore required to reproduce the line emission \citep{badnell2006}. Spectra of C-like, B-like, and Be-like species are 
in particular sensitive to the electron density \citep{mao2017, gu2019}. These effects are taken into account in the advanced model by
setting the plasma density as a free parameter. 

The line broadening parameter $v$ of the {\it vpro} model is set free for different temperature components. This allows extra uncertainties
in the line spread function calibration, as well as in the modeling of the astrophysical turbulence in the multi-zone corona. The redshift
component used for energy-scale calibration is also set free for different thermal components.

As shown in Table~\ref{tab:parameters}, the advanced model indeed improves the overall spectral fits. The different plasma codes 
converge to a similar form of the emission measure distribution, with a primary peak at 0.61~keV (Fig.~\ref{fig:dem}). The advanced fit further leads to 
the changes of coronal abundances, by $\leq 30$\%, with respect to the baseline fit with the corresponding code.

\begin{figure}[!htbp]
\centering
\resizebox{\hsize}{!}{\includegraphics[angle=0]{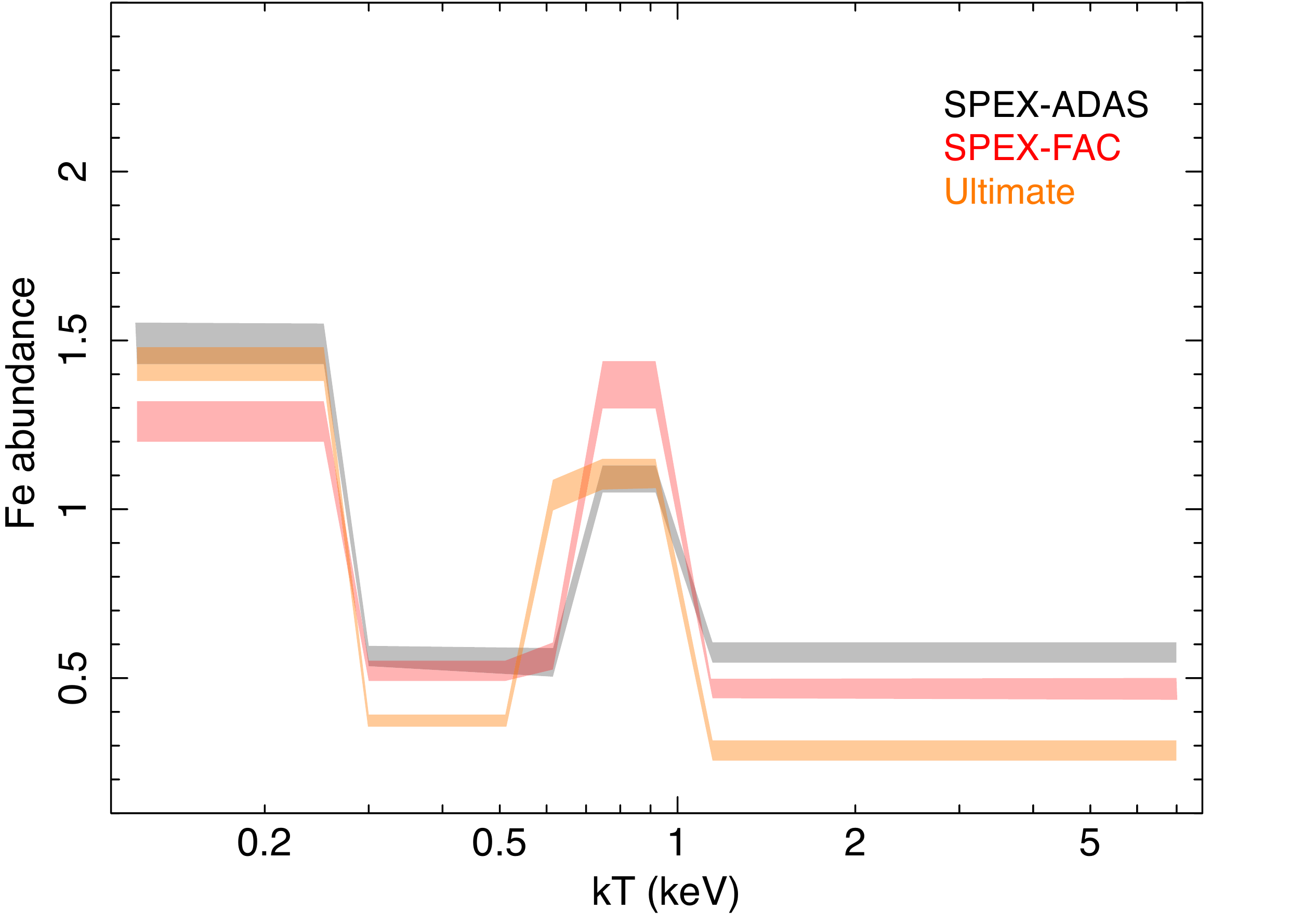}}
\caption{Fe abundances as a function of temperature obtained with the advanced modeling, using the atomic data of SPEX-ADAS (black) and SPEX-FAC (red).
The ultimate fit (see Appendix B) is shown in orange. }
\label{fig:fepro}
\end{figure}

\begin{figure*}[!htbp]
\centering
\resizebox{0.7\hsize}{!}{\includegraphics[angle=0]{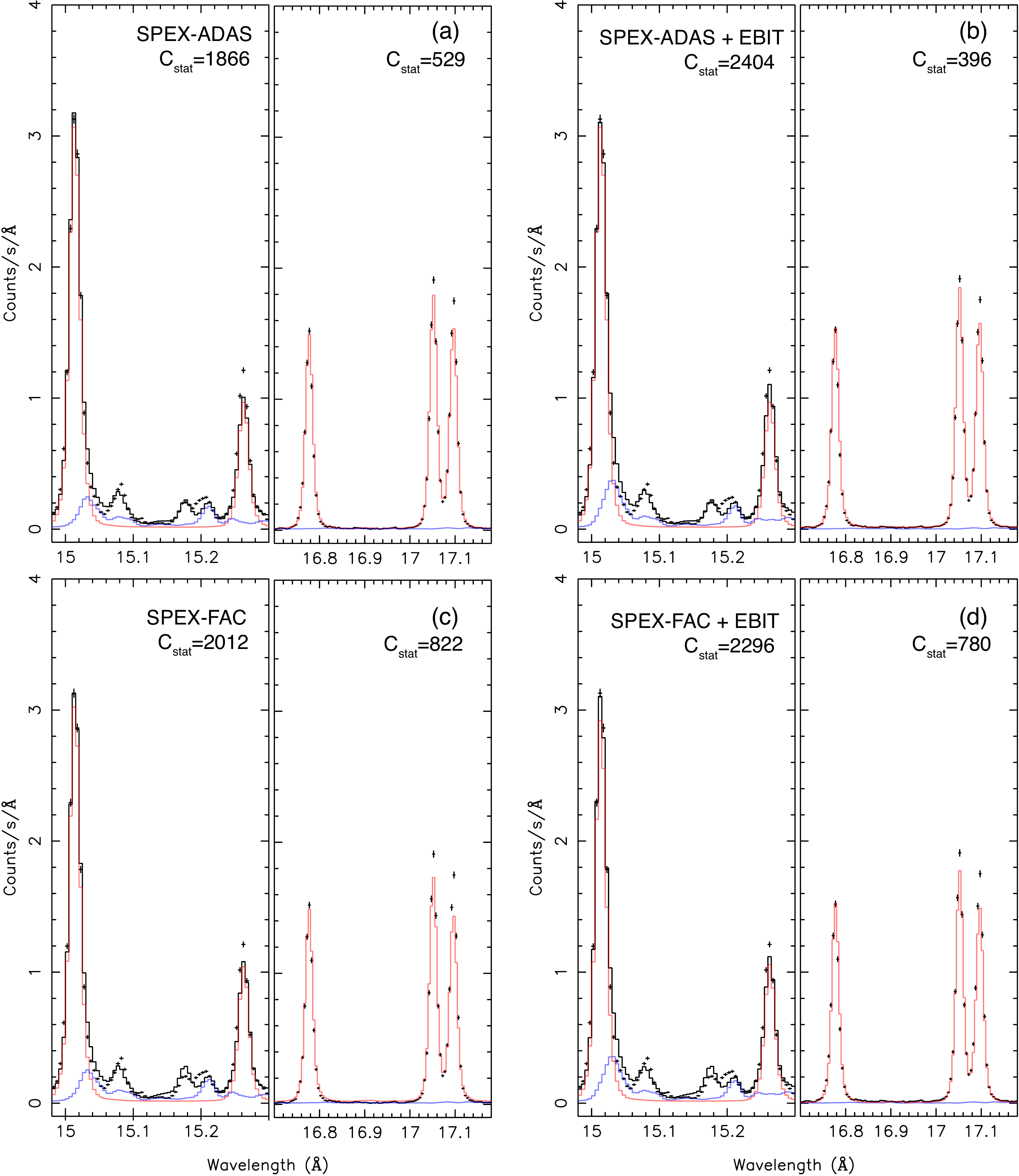}}
\caption{Advanced fits shown in the wavelength ranges of $14.98-15.3$~{\AA} and $16.7-17.18$~{\AA}. The fits with SPEX-ADAS and SPEX-FAC calculations
are plotted in panels (a) and (c), and those with EBIT-calibrated rates are shown in panels (b) and (d). Total model, \ion{Fe}{XVII} transitions,
and \ion{Fe}{XVI} transitions are shown in black, red, and blue. Local C-stat values are given at the top. The total model is omitted in the 17~{\AA} plot, 
as it would overlap with the \ion{Fe}{XVII} lines. }
\label{fig:withebit}
\end{figure*}

As shown in Fig.~\ref{fig:fepro}, the Fe abundances change significantly as a function of temperature. The abundances of the $0.6-1.0$~keV and 
$<0.3$~keV components are found to be about one solar or higher, while the abundances of other components are clearly sub-solar. On the other hand, 
advanced modeling is also used to constrain the mean electron density of the source. The upper limits on the density are obtained to be 
$0.6\times10^{9}$ cm$^{-3}$ (SPEX-ADAS) and $1.4\times10^{9}$ cm$^{-3}$ (SPEX-FAC). Our values agree with the previous results, e.g., 
$<2.4\times10^{9}$ cm$^{-3}$ by \citet{ness2001} and $<7\times10^{9}$ cm$^{-3}$ by \citet{mewe2001}. These results were derived from the \ion{O}{VII} 
triplet line ratios measured with the {\it Chandra} Low Energy Transmission Grating Spectrometer (LETGS) data.

In Appendix B, we present a semi-quantitative discussion on the fit quality with the advanced model. The SPEX-ADAS, SPEX-FAC, and APEC models are
compared systematically, identifying key differences between calculations, as well as problem areas in the fit with each code. It reveals several
issues on the wavelengths and line fluxes of the present calculations, which can be directly fed into the prioritization of future laboratory experiments. 
Both APEC and SPEX codes are in tension with the Capella data on wavelengths for about 10\% of the observed lines. By fixing the apparent wavelength errors in the SPEX code, fixing the central energies based on APEC or Chianti version 9.0 when these provide a better match to the data, we further improve the advanced model to its ``ultimate'' form shown in Table~\ref{tab:parameters},
Figures~\ref{fig:dem} and \ref{fig:fepro}. Details can be found in Appendix B and specifically Table~\ref{tab:wav_ult}.

\subsection{Results with EBIT-calibrated rates}
\label{sec:ebitfit}

The \ion{Fe}{XVII} and \ion{Fe}{XVI} rates corrected through the EBIT data (\S~\ref{sec:ebit}) are applied in the advanced fits.
As shown in Table~\ref{tab:parameters}, the EBIT-corrected models yield slightly poorer fits than the models with theoretical rates,
though the final differences in C-stat are small. Table~\ref{tab:parameters} also tabulates the changes of the non-Fe abundances, mostly $<10$\%, by
incorporating the EBIT rates. The emission measure distributions using the EBIT rates remain nearly the same as using theoretical calculations 
presented earlier.

In Fig.~\ref{fig:withebit}, we take a detailed look at the \ion{Fe}{XVI} and \ion{Fe}{XVII} lines in the 15~{\AA} and 17~{\AA} regions, where
the EBIT rates should affect the fit directly. The advanced models with EBIT rates and theoretical rates give nearly the same fit quality to the 
\ion{Fe}{XVII} 3$C$ and 3$F$ lines at 15.01~{\AA} and 16.78~{\AA}, and the models with EBIT rates improve, though marginally, the fits to the 
3$D$ (15.26~{\AA}), 3$G$ (17.05~{\AA}), and $M$2 (17.10~{\AA}) lines. Note that these three lines are affected indirectly, as the EBIT rate is
set only for the 3$C$ transition (\S~\ref{sec:ebitde}). Overall, the models with EBIT rates can provide a reasonable fit to the \ion{Fe}{XVII} lines. 

As shown in Fig.~\ref{fig:withebit}, the EBIT models appear to overestimate the \ion{Fe}{XVI} dielectronic recombination lines near 15.03~{\AA}. 
It seems that the \ion{Fe}{XVI} lines from the original calculations already a bit exceed the observed level, and the EBIT rates make the discrepancy
further larger. The \ion{Fe}{XVI} lines likely contribute the main C-stat differences between the fits with the EBIT and theoretical models.
However, considering that the \ion{Fe}{XVI} line is only a weak transition with a much lower flux than its close neighbour, the fit of the line could
have been affected by many systematic factors: a small error in the line spread function at the wing, non-Gaussianity, or an error in wavelength, might indirectly 
cause the poor fit. The line strength is also influenced by the determination of the ionization concentration and abundance of \ion{Fe}{XVI}, for which
there is a trade-off
with the global fit of the broadband spectrum. Therefore, we cannot conclude based on the current fit that there is a real discrepancy between 
the EBIT and Capella observations on the \ion{Fe}{XVI} lines.

We note that the current model seems to slightly underestimate the \ion{Fe}{XVII} lines at 17~{\AA}, in particular the $M$2 line, even though the direct 
and resonant excitation cross sections of the $M$2 and 3$G$ lines are already verified with the EBIT data (\S~\ref{sec:ebitde}), and the photon resonant scattering
has been modeled with the {\it hot} component. A possible explanation is that these two lines
are also significantly populated by the cascades following dielectronic recombination from \ion{Fe}{XVIII} ions, which are missing in the current experiment (S19, \S~\ref{sec:ebitde}).
%Paper I reported that the dielectronic recombination becomes the dominant indirect excitation channel at $\geq 1$~keV, contributing to $\sim 35$\% of
%the total cross section for the $M$2 line, and $\sim 10$\% for the 3$G$ line. 
This calls for follow-up works on the dielectronic recombination contribution
to the Ne-like lines at 17~{\AA}.

\section{Needs for benchmarks with laboratory data}
\label{sec:need}

Our fits to the {\it Chandra} grating spectrum of Capella not only revealed limits of the best available plasma codes but also showed the requirements for further testing of codes using high-accuracy laboratory measurements. 
In fact, a truly accurate fit to the Capella data requires a huge number of laboratory benchmarks, including transition energies and reaction rates of various processes, for both the L- and K- shells of all astrophysically relevant ions. 
%
%While in practice a complete test is not feasible, a few necessary tests, apart from the study of the \ion{Fe}{XVII} excitation and recombination (\S~\ref{sec:ebit}), can be mentioned. 
%
While in practice such a complete test is not feasible, a few necessary tests, apart from the present study of the \ion{Fe}{XVII} excitation and recombination (\S~\ref{sec:ebit}), should be prioritized.

First, the absolute cross section benchmarks for electron-impact/resonant excitation and dielectronic recombination followed by cascades, in particular for high-$n$ states, for the remaining Fe-L ions, are probably of a high priority because they determine the intensities of most lines seen in the Capella spectrum. 
The accuracy of the cross section measurements is required to be better than 10\%.
These future measurements should be arranged with a full awareness of what is available at present: the IRON project with a series of theoretical and experimental works (following the first paper \citealt{hummer1993}); individual EBIT works on excitation of \ion{Fe}{XVIII} $-$ \ion{Fe}{XXIV} 
\citep{chen2006}, \ion{Fe}{XXI} $-$ \ion{Fe}{XXIV} \citep{chen2005}, \ion{Fe}{XXIV} \citep{chen2002a, gu1999}, and \ion{Fe}{XXI} $-$ \ion{Fe}{XXIV} \citep{gu2001} lines; storage ring measurements of dielectronic recombination forming \ion{Fe}{XVIII} $-$ \ion{Fe}{XXII} \citep{savin2002b, savin2002, savin2003, savin2006}.
Typical uncertainties on the measured cross section data are $\sim 10-25$\%. 
This can be improved for an example by implementing new measurement techniques at the EBIT, such as simultaneous X-ray observations with wide-band high-resolution transition-edge sensor microcalorimeters~\citep{durkin2019,szypryt2019} and polarization measurements~\citep{shah2015,shah2016,shah2018} using dedicated X-ray polarimeters~\citep{weber2015,beiersdorfer2016}.
Further improvements can also be achieved by measuring the electron beam and ion cloud overlap in the EBIT~\citep{liang2009,arthanayaka2020}.

Second, the wavelengths of strong lines need to be calibrated to an accuracy of the order of 0.01~{\AA}. A list of tentative candidates, where wavelength mismatches between spectral codes and the Capella data are detected in the present work, can be found in Appendix B. 
Existing EBIT measurements using crystal spectrometers provided Fe-L wavelengths precise enough to allow reliable line identification~\citep{brown1998,brown2002}. 
However, line energies can be significantly improved (to parts-per-million accuracy) by the application of resonant photoexcitation of highly charged ions using the EBIT at synchrotrons and free-electron lasers~\citep{epp2007,simon2010,bernitt2012,rudolph2013,shah2020}.
Besides line energies, these dedicated experiments can also provide accurate information on the radiative and auger decay rates~\citep{steinbrugge2015,togawa2020}.

In addition, similar benchmarks using the laboratory data on the total ionization and recombination cross sections should be conducted, as they form the basis to determine the ionization balance as a function of electron temperature and density. 
These benchmarks will be utilized for further optimizing the quality of fits for the archival grating data from a variety of celestial sources. 
They are even more needed for \textit{XRISM} and \textit{Athena} with their superb sensitivity and large bandwidths. 

%Usual issues:
%Calibration with RR (accurate up to 5 \%)
%Polarization correction
%Detector efficiency overall
%background model crystal spectrometer
%Total error: 10 - 25 \% 

%M-shell Fe stuff
%Inner-Shell Absorption Lines of Fe VI-Fe XVI: A Many-Body Perturbation Theory Approach
%-- Ming F. Gu et al 2006 ApJ 641 1227 -- https://doi.org/10.1086/500640

%The Iron Unresolved Transition Array in Active Galactic Nuclei
%-- Hagai Netzer 2004 ApJ 604 551 --
%https://doi.org/10.1086/382038
%and by a book by A. M\"uller
%https://doi.org/10.1016/S1049-250X(07)55006-8

%\textbf{Atomic data need and Laboratory astrophysics reviews
%}

%A review by Kallman
%https://doi.org/10.1103/RevModPhys.79.79

%A review by Beiersdorfer
%https://doi.org/10.1146/annurev.astro.41.011802.094825

\section{Conclusion}
\label{sec:con}

We calibrate theoretical calculations of the Fe-L cross sections through EBIT measurements and {\it Chandra} grating observations
of Capella. By utilizing a two-dimensional component analysis, the EBIT cross sections of dielectronic recombination from Na-like Fe, the resonant excitation,
and the direct excitation of the Ne-like Fe are independently determined. We find reasonable agreement with the theoretical calculation for the excitation of the  
3$s$ - 2$p$ transitions, while the known discrepancies in the 3$d$ - 2$p$ dielectronic recombination and direct excitation rates found in earlier works are confirmed with the new experimental data. 
The updated theoretical calculation and the EBIT results are then fed into global modeling of the Capella spectrum. 
The inclusion of the new atomic calculation improves significantly the fit to the observed spectrum, while the effect of the EBIT calibration remains inconclusive, in particular for the 3$d$ - 2$p$ transitions,
as it deeply couples with the astrophysical source modeling and instrumental calibration. However, the present work shows for the first time that the EBIT experimental data can be directly applied to benchmark and improve existing hot plasma models. Furthermore, future targeted EBIT experiments with the high-resolution photon detectors would certainly improve experimental accuracy that eventually will make models better. 
We conclude that the present Fe-L atomic calculation is almost ready to be  
delivered to the community, except for a few issues on wavelengths and rates, which are to be addressed with follow-up calculations and dedicated laboratory measurements.

\begin{acknowledgements}
L. Gu is supported by the RIKEN Special Postdoctoral Researcher Program.
SRON is supported financially by NWO, the Netherlands Organization for
Scientific Research. 
Work by C. Shah was supported by the Max-Planck-Gesellschaft (MPG), the Deutsche Forschungsgemeinschaft (DFG) Project No. 266229290, and by an appointment to the NASA Postdoctoral Program at the NASA Goddard Space Flight Center, administered by Universities Space Research Association under contract with NASA.
P. Amaro acknowledges the support from Funda\c{c}\~{a}o para a
Ci\^{e}ncia e a Tecnologia (FCT), Portugal, under Grant No. UID/FIS/04559/2020(LIBPhys).
\end{acknowledgements}

\bibliographystyle{aa}
\bibliography{main}

\newpage

\begin{appendix}

\onecolumn

\section{EBIT line list}
The energies and fluxes of the dielectronic recombination and resonant excitation peaks measured
at the EBIT experiment (\S~\ref{sec:ebit}), as well as their statistical uncertainties determined
from the two-dimensional fits, are listed in the following tables. The systematic uncertainties
are given in \S~\ref{sec:uncert}.

%\LTcapwidth=\textwidth
\begin{longtable}{ccccccccccc}
\caption{\label{tab:ebit_de} EBIT dielectronic recombination line list } \\
\hline\hline
beam energy & beam energy error & X-ray energy & X-ray energy error & total counts$^{\rm a}$ & stat. error on counts \\
(eV)        & (eV)              & (eV)         & (eV)               &              &                  \\
\hline
\endfirsthead
\caption{continued.}\\
\hline\hline
beam energy & beam energy error & X-ray energy & X-ray energy error & total counts$^{\rm a}$ & stat. error on counts \\
(eV)        & (eV)              & (eV)         & (eV)               &              &                  \\
\hline
\endhead
\hline
\endfoot
      261.1 &        0.6 &      690.4 &        4.6 &      4238 &       173 \\
      272.1 &        0.2 &      745.2 &        4.5 &      4239 &       171 \\
      284.2 &        0.6 &      721.0 &        3.5 &      5912 &       199 \\
      309.9 &        0.3 &      726.7 &        3.9 &     10852 &       424 \\
      317.8 &        0.2 &      828.1 &        1.3 &     83678 &       677 \\
      326.9 &        0.1 &      814.6 &        1.4 &     38884 &       469 \\
      334.7 &        0.1 &      831.6 &        1.2 &     44765 &       393 \\
      349.3 &        0.3 &      799.9 &        1.4 &     46251 &       918 \\
      357.6 &        0.2 &      810.8 &        1.0 &    107773 &      1328 \\
      361.6 &        0.1 &      820.7 &        0.9 &    109859 &      1414 \\
      367.6 &        0.1 &      821.3 &        1.0 &     86465 &      1098 \\
      372.8 &        0.3 &      834.0 &        2.4 &     26917 &      1112 \\
      382.0 &        0.4 &      822.1 &        1.9 &     31731 &       454 \\
      395.1 &        0.2 &      804.5 &        1.4 &    113537 &      3575 \\
      398.1 &        0.1 &      857.5 &        3.5 &     89076 &      3199 \\
      404.8 &        0.1 &      808.9 &        0.7 &    158898 &       994 \\
      411.5 &        0.1 &      817.0 &        0.6 &    159839 &       840 \\
      418.3 &        0.2 &      817.5 &        2.1 &     43450 &       550 \\
      428.3 &        0.1 &      879.8 &        1.7 &     45157 &       427 \\
      436.5 &        0.3 &      803.3 &        3.6 &     10306 &       345 \\
      441.7 &        0.3 &      855.6 &        5.7 &      9246 &       373 \\
      452.6 &        0.5 &      849.3 &        4.7 &     17105 &       317 \\
      465.0 &        0.2 &      768.8 &        3.9 &     14286 &       272 \\
      475.3 &        0.2 &      896.5 &        3.2 &     11735 &       276 \\
      482.4 &        0.2 &      900.7 &        3.4 &     11121 &       279 \\
      482.7 &        0.3 &      694.0 &        3.0 &     12673 &       289 \\
      488.4 &        0.2 &      915.8 &        5.1 &      6710 &       218 \\
      497.4 &        0.2 &      708.5 &        2.7 &     12796 &       238 \\
      499.2 &        0.2 &      982.9 &        3.5 &      5544 &       174 \\
      508.0 &        0.1 &      713.8 &        1.8 &     19371 &       273 \\
      510.8 &        0.3 &      992.3 &        6.6 &      2505 &       143 \\
      520.0 &        0.2 &      725.1 &        2.6 &      9642 &       207 \\
      529.9 &        0.4 &      695.4 &        3.8 &     11449 &       273 \\
      534.6 &        0.1 &     1003.4 &        1.2 &     46724 &       470 \\
      542.2 &        0.2 &      785.7 &        3.4 &     24337 &       437 \\
      545.7 &        0.3 &     1012.8 &        3.0 &     14158 &      2346 \\
      546.8 &        0.3 &     1019.9 &        10.0 &     32491 &      2592 \\
      558.2 &        0.2 &     1018.6 &        3.0 &      9166 &       226 \\
      559.8 &        0.2 &      791.6 &        3.3 &     23142 &       357 \\
      568.8 &        0.1 &      806.0 &        1.4 &     55846 &       471 \\
      578.9 &        0.2 &     1020.3 &        1.7 &     44620 &       844 \\
      581.9 &        0.1 &      806.0 &        0.7 &    191284 &       993 \\
      589.6 &        1.0 &     1004.8 &        4.0 &     19494 &       773 \\
      590.7 &        0.1 &      808.1 &        1.9 &     83944 &       931 \\
      592.7 &        0.2 &     1031.5 &        1.6 &     25033 &      1432 \\
      598.0 &        0.1 &      819.9 &        0.5 &    278070 &      1135 \\
      607.2 &        0.1 &      821.4 &        0.6 &    164666 &       770 \\
      615.2 &        0.3 &      806.9 &        3.1 &     27252 &       382 \\
      617.7 &        0.1 &     1094.5 &        1.7 &     18627 &       259 \\
      622.5 &        0.2 &      796.0 &        2.5 &     29807 &       455 \\
      629.2 &        0.1 &     1100.7 &        2.0 &     13151 &       228 \\
      630.8 &        0.2 &      711.4 &        2.3 &     31272 &       475 \\
      636.5 &        0.2 &      848.2 &        5.2 &     25441 &       535 \\
      643.4 &        0.5 &     1107.0 &        7.5 &      1444 &       116 \\
      644.8 &        0.2 &      726.8 &        3.5 &     15481 &       262 \\
      656.5 &        0.3 &      788.5 &        3.0 &     57530 &      1112 \\
      662.2 &        0.1 &     1104.7 &        1.2 &     37882 &       371 \\
      666.8 &        0.4 &      809.7 &        1.8 &    104616 &      2711 \\
      673.3 &        0.1 &      824.2 &        1.1 &    178482 &      3000 \\
      675.0 &        0.1 &     1117.3 &        2.1 &     16467 &       270 \\
      679.4 &        0.3 &      903.0 &        5.5 &     14475 &       849 \\
      684.0 &        0.1 &      826.9 &        0.5 &    383093 &      1388 \\
      687.2 &        0.5 &     1144.7 &        3.7 &      7386 &       268 \\
      695.3 &        0.3 &     1128.4 &        3.5 &      8165 &       267 \\
      695.6 &        0.3 &      752.8 &        2.5 &     50833 &       569 \\
      700.8 &        0.7 &     1151.1 &        5.3 &      6573 &       271 \\
      706.0 &        0.1 &      775.2 &        1.5 &     95860 &       643 \\
      706.8 &        0.3 &     1141.9 &        2.3 &     15463 &       368 \\
      713.6 &        0.8 &     1164.9 &        7.2 &      3711 &       238 \\
      714.7 &        0.1 &      815.4 &        1.1 &    119787 &       747 \\
      719.9 &        0.2 &     1158.5 &        3.1 &      8617 &       255 \\
      722.2 &        0.1 &      855.1 &        1.1 &    142046 &      1045 \\
      728.7 &        0.1 &      825.1 &        0.7 &    319260 &      1459 \\
      730.9 &        0.7 &     1169.4 &        5.9 &      3300 &       156 \\
      747.9 &        0.7 &     1184.3 &        7.4 &      2381 &       149 \\
      754.9 &        0.1 &      830.7 &        1.1 &    290141 &      2630 \\
      759.6 &        1.0 &     1063.7 &       29.0 &      5071 &       429 \\
      770.3 &        0.2 &      817.9 &        2.4 &    132865 &      3142 \\
      771.2 &        0.7 &     1041.5 &        2.6 &      1749 &       216 \\
      772.2 &        0.1 &      845.0 &        4.1 &     55476 &      2359 \\
      779.9 &        0.2 &     1025.0 &        2.4 &      4345 &       336 \\
      783.8 &        0.1 &      815.5 &        1.8 &    160797 &      2636 \\
      785.1 &        0.7 &     1220.8 &        7.6 &      1837 &       121 \\
      792.1 &        0.1 &      816.7 &        2.3 &     64168 &      1911 \\
      792.6 &        0.1 &     1036.2 &        2.3 &      5019 &       496 \\
      798.3 &        0.1 &      807.1 &        2.5 &     88459 &      1718 \\
      804.2 &        0.1 &      813.3 &        1.6 &    137045 &      1979 \\
      810.6 &        0.3 &      809.3 &        1.2 &    193403 &      2576 \\
      823.5 &        0.1 &      800.5 &        1.0 &    132303 &      1227 \\
\end{longtable}
%\begin{tablenotes}
\begin{itemize}
\item[$a$] Count numbers of resonant peaks corrected for the filter transmission, detector response, and polarization.
\end{itemize}
%\end{tablenotes}

\begin{longtable}{ccccccccccc}
\caption{\label{tab:ebit_re} EBIT resonant excitation line list } \\
\hline\hline
beam energy & beam energy error & X-ray energy & X-ray energy error & total counts$^{\rm a}$ & stat. error on counts \\
(eV)        & (eV)              & (eV)         & (eV)               &              &                  \\
\hline
\endfirsthead
\caption{continued.}\\
\hline\hline
beam energy & beam energy error & X-ray energy & X-ray energy error & total counts$^{\rm a}$ & stat. error on counts \\
(eV)        & (eV)              & (eV)         & (eV)               &              &                  \\
\hline
\endhead
\hline
\endfoot
      734.4 &        0.1 &      705.9 &        0.7 &    138875 &       928 \\
      738.4 &        0.6 &      717.1 &        1.4 &    145685 &      3083 \\
      744.8 &        0.2 &      726.2 &        1.2 &    219839 &      6304 \\
      751.3 &        0.2 &      751.0 &        1.7 &    144253 &      3687 \\
      759.3 &        0.7 &      715.1 &        2.6 &    112798 &      4674 \\
      763.7 &        0.1 &      737.5 &        1.1 &    335881 &      4487 \\
      771.5 &        0.3 &      727.5 &        1.2 &    201733 &      3548 \\
      778.5 &        0.1 &      722.7 &        0.7 &    328825 &      3210 \\
      786.6 &        0.2 &      720.7 &        1.1 &    185150 &      3631 \\
      792.2 &        0.2 &      727.2 &        1.4 &    234471 &      4222 \\
      799.2 &        0.2 &      725.5 &        1.5 &    190307 &      3652 \\
      803.3 &        0.6 &      723.9 &        2.5 &     79544 &      2869 \\
      807.9 &        1.3 &      741.6 &        9.3 &      6235 &      3932 \\
      832.4 &        0.1 &      724.7 &        1.5 &     37720 &       450 \\
      839.4 &        0.1 &      740.6 &        0.7 &    151615 &       889 \\
      846.0 &        0.1 &      758.0 &        2.3 &     49355 &      1223 \\
      852.5 &        0.2 &      726.1 &        0.9 &    145702 &      1222 \\
      862.1 &        0.2 &      754.0 &        0.8 &    211282 &      1783 \\
      868.8 &        0.1 &      797.1 &        1.2 &    127459 &      2109 \\
      877.7 &        0.2 &      764.5 &        0.7 &    227553 &      1673 \\
      888.4 &        0.3 &      723.3 &        1.1 &     85808 &      1017 \\
      889.2 &        0.7 &      769.9 &        1.7 &     41852 &      1118 \\
      906.1 &        0.3 &      715.1 &        0.9 &    103042 &      1940 \\
      909.6 &        0.1 &      783.3 &        6.1 &     34507 &      1157 \\
      918.7 &        0.2 &      773.1 &        3.0 &     49833 &      1346 \\
      923.3 &        0.8 &      702.1 &        1.1 &    100434 &      1346 \\
      936.7 &        0.1 &      773.7 &        2.1 &     90548 &      1252 \\
      948.3 &        0.2 &      796.3 &        2.0 &     79209 &      1482 \\
      953.5 &        0.2 &      696.6 &        1.3 &     94457 &      1490 \\
      963.7 &        0.1 &      767.3 &        4.6 &     90833 &      2099 \\
      967.5 &        0.3 &      718.2 &        1.6 &     83968 &      1978 \\
      982.0 &        0.2 &      577.6 &        3.6 &      7605 &       131 \\
      982.4 &        0.2 &      695.9 &        1.1 &     75812 &       565 \\
      987.4 &        1.1 &      723.3 &        2.7 &     42598 &      1531 \\
      997.2 &        0.2 &      778.9 &        4.3 &      5248 &       359 \\
      998.4 &        0.4 &      708.9 &        1.3 &     91980 &      2032 \\
     1007.0 &        0.3 &      707.5 &        1.8 &     41704 &      1906 \\
     1007.5 &        0.4 &      790.8 &        5.9 &     11790 &       705 \\
     1045.2 &        0.2 &      721.2 &        1.3 &     23229 &       313 \\
     1057.3 &        0.6 &      727.1 &        9.8 &      6690 &       296 \\
     1075.9 &        0.2 &      711.7 &        1.1 &     42605 &       413 \\
     1087.9 &        0.1 &      718.8 &        2.5 &      8190 &       199 \\
     1095.2 &        0.2 &      768.0 &        2.7 &      6702 &       201 \\
\end{longtable}
%\begin{tablenotes}
\begin{itemize}
\item[$a$] Count numbers of resonant peaks corrected for the filter transmission, detector response, and polarization.
\end{itemize}
%\end{tablenotes}

\section{Quality of the Capella fits}

Figures~B.1-B.3 show the full band Capella spectrum fit with advanced model using SPEX-ADAS (with the EBIT correction), and the relative differences
of the fits from various atomic calculations. In table~\ref{tab:allline}, we further evaluate the quality of the fits to the strong transitions 
(i.e., lines detected $>5\sigma$) in the data.

As a short summary, we find
\begin{itemize}
\item[$(1)$] Table~\ref{tab:allline} presents a detailed comparison of the present plasma code calculations for each strong line, reflecting
the current knowledge and systematic uncertainties in the underlying atomic constants. It provides
a list of problematic regions that would prioritize the dedicated laboratory measurements. Furthermore, the line-by-line evaluation of the 
model quality can be used as a reference in the future spectroscopic analysis with the XRISM and Athena data.
\item[$(2)$] More than 60\% of the observed transitions listed in Table~\ref{tab:allline} are reasonably reproduced, both in wavelengths
and line fluxes, by the SPEX-ADAS and SPEX-FAC calculations. This includes the main Fe-L transitions, such as the 15~{\AA} and 17~{\AA} \ion{Fe}{XVII} lines.
For 30\% of the lines, the model fluxes (therefore the transition cross sections) are off by $>20$\% from the values indicated by the data. 
\item[$(3)$] Both APEC and SPEX codes are in tension with the Capella data on wavelengths for about 10\% of the observed lines. The SPEX code cannot fit
the central energies of the \ion{Fe}{XVI} line at 16.62~{\AA}, \ion{Fe}{XVII} lines at 11.03~{\AA}, 13.12~{\AA}, 13.82~{\AA} and 16.24~{\AA}, \ion{Fe}{XVIII} lines 
at 10.36~{\AA}, 10.54~{\AA}, 11.33~{\AA}, 11.43~{\AA}, 14.53~{\AA}, 15.49~{\AA}, 15.63~{\AA}, 16.07~{\AA}, 17.36~{\AA}, and 17.61~{\AA}, \ion{Fe}{XIX} lines
at 9.84~{\AA}, 10.81~{\AA}, 13.49~{\AA}, 13.63~{\AA}, 13.93~{\AA}, 16.11~{\AA}, and 16.28~{\AA}, \ion{Fe}{XX} lines at 9.06~{\AA}
and 12.59~{\AA}, \ion{Fe}{XXI} line at 12.29~{\AA}, \ion{Mg}{XI} line at 7.85~{\AA}, as well as a few Ni and Ca L-shell lines. The APEC code
seems to match better for a few Fe lines, but it is off in several other transitions. As seen in the residual plots (Figs.~B.1-B.3),
these wavelength errors contribute substantially to the C-stat of the fits. It is therefore crucial to correct the 
wavelength database using data acquired in laboratory experiments, as well as through advanced theoretical calculations.  
\item[$(4)$] There are potential line features in the spectrum, such as those at 9.62~{\AA}, 9.79~{\AA}, 9.89~{\AA}, 14.41~{\AA}, 17.8~{\AA}, and 18.1~{\AA}, which 
cannot be (fully) interpreted by the current model. A further study is required to identify the origin of these transitions. 
\item[$(5)$] Capella data provide an excellent test to a 0.6~keV plasma by resolving the \ion{Fe}{XVII} transitions up to $n=8$,
and \ion{Fe}{XVIII} transitions up to $n=5$ (Table~\ref{tab:allline}). However, the L-shell emission of more ionized species, such as \ion{Fe}{XXI} - \ion{Fe}{XXIV}, are much
weaker in the Capella data due to the drop in ionization concentration. Other astrophysical and/or laboratory sources are needed to calibrate 
these lines.
\end{itemize}

By fixing the apparent wavelength errors in the SPEX code, we are able to obtain a final fit of the source spectrum. The central energies of the 
lines listed in item (3) above are set to the corresponding values in the APEC code, when APEC makes a better match. We further look into the
Chianti database version 9.0 \footnote{https://www.chiantidatabase.org/chianti_direct_data.html} for the transitions where both APEC and SPEX wavelengths are off 
(see detail in Table~\ref{tab:wav_ult}).
Most of the wavelength errors are fixed by using the APEC/Chianti data, and the quality of the fit has been significantly improved (Table~\ref{tab:parameters}). Although
the C-stat is formally still far from acceptable, this ``ultimate'' fit does represent the state-of-the-art understanding of the
emission spectrum from the Capella corona. The obtained emission measure distribution (Fig.~\ref{fig:dem}) and elemental abundances (Table~\ref{tab:parameters})
differ significantly from the advanced fits, suggesting that the accuracy of wavelengths does matter for the high-resolution spectroscopy.

\begin{figure*}[!htbp]
\centering
\resizebox{0.6\hsize}{!}{\includegraphics[angle=0]{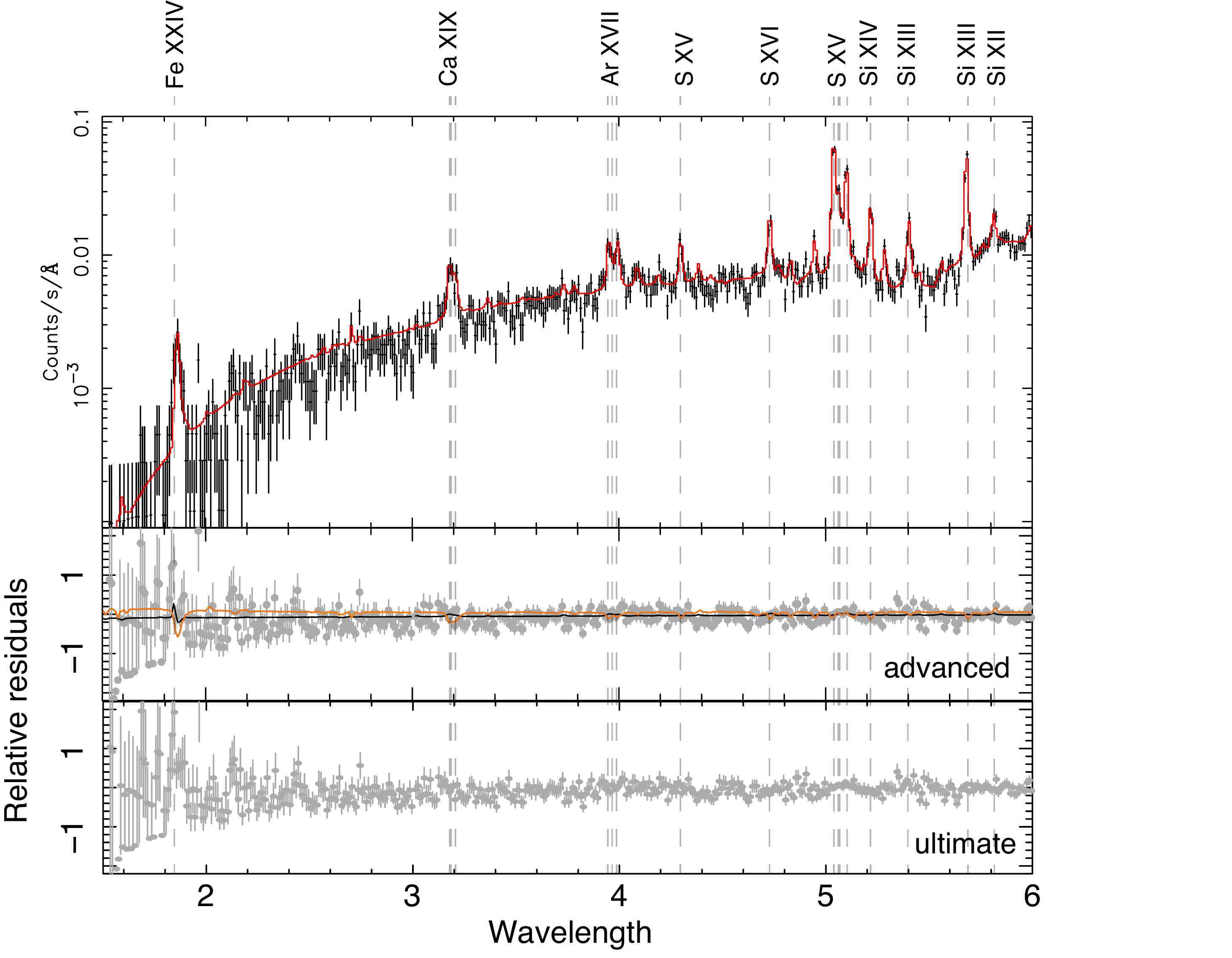}}
\resizebox{0.9\hsize}{!}{\includegraphics[angle=0]{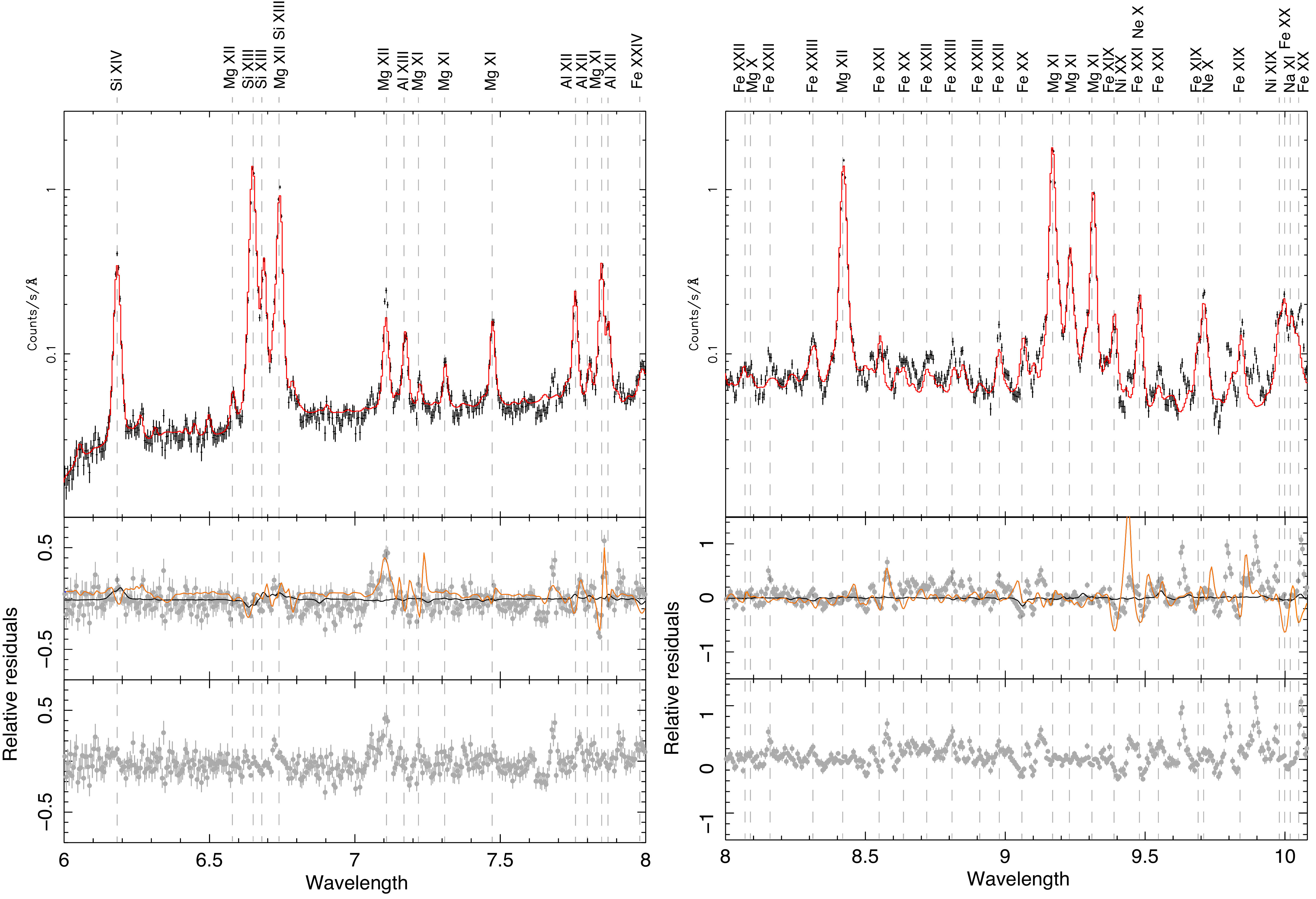}}
\caption{Chandra grating spectrum of Capella in $1.5-10.08$~{\AA} fit with different models. Upper panels in each subfigure 
show fit with the advanced SPEX-ADAS model with EBIT calibration (red). Middle panels show the fit residual with the advanced models (grey points), as well as
the ratio between the SPEX-FAC
fit and the SPEX-ADAS fit in black, and the APEC-to-SPEX-ADAS ratio in orange. Lower panels show the fit residual with the ultimate model.  }
\label{fig:allspec1}
\end{figure*}

\begin{figure*}[!htbp]
\centering
\resizebox{0.9\hsize}{!}{\includegraphics[angle=0]{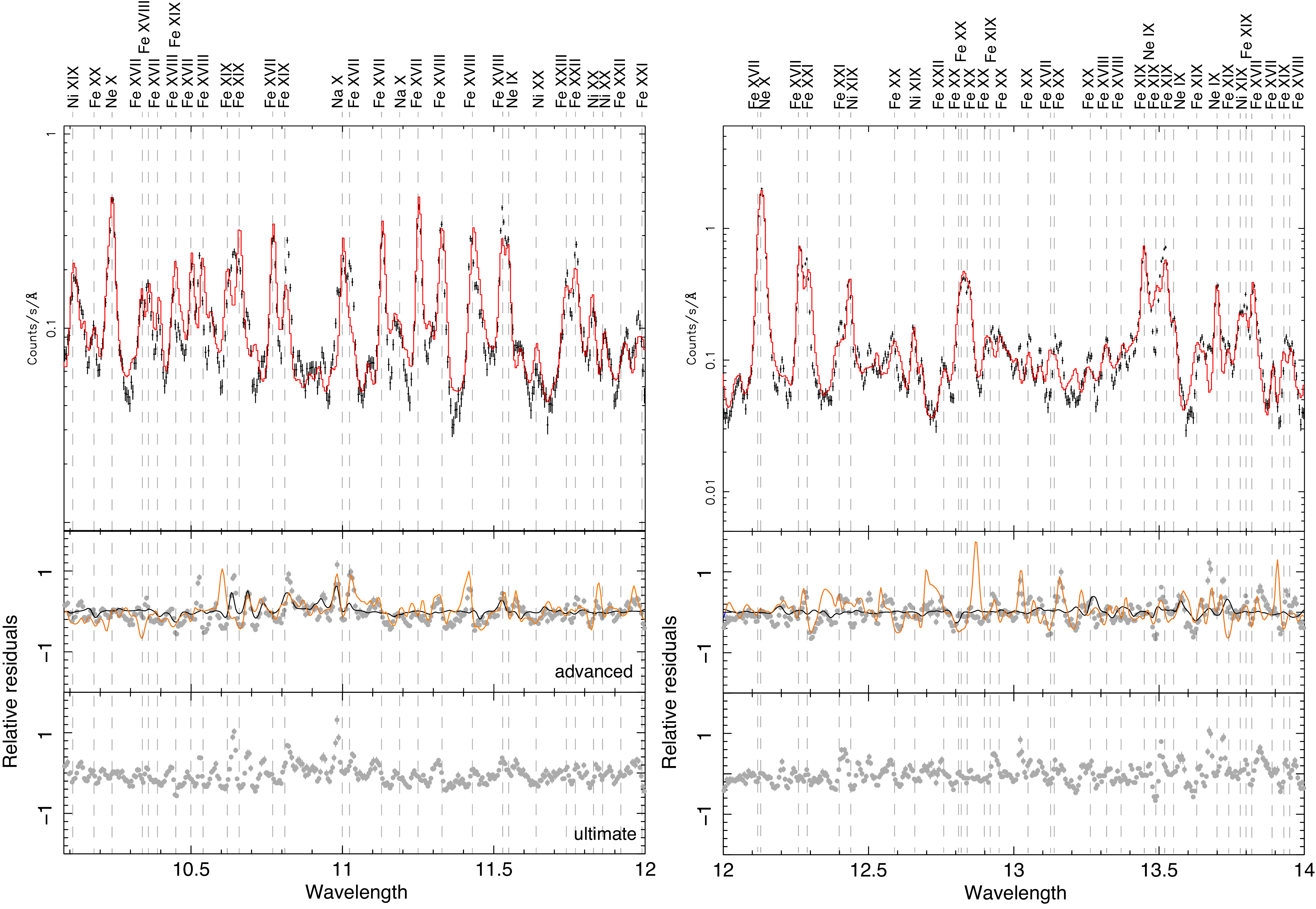}}
\resizebox{0.9\hsize}{!}{\includegraphics[angle=0]{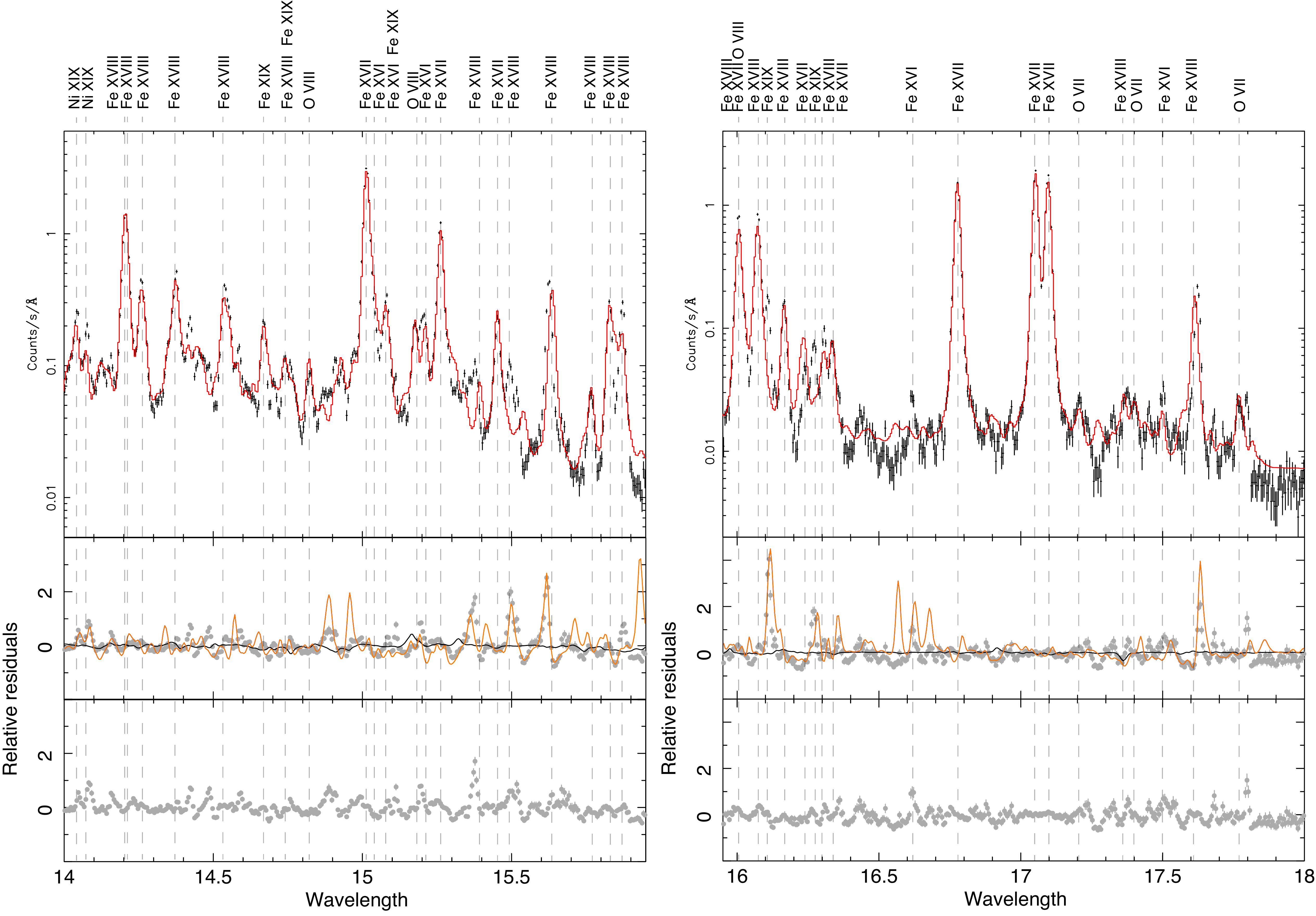}}
\caption{Same as Fig.~\ref{fig:allspec1}, but in $10.08-18$~{\AA}.}
\label{fig:allspec2}
\end{figure*}

\begin{figure*}[!htbp]
\centering
\resizebox{0.9\hsize}{!}{\includegraphics[angle=0]{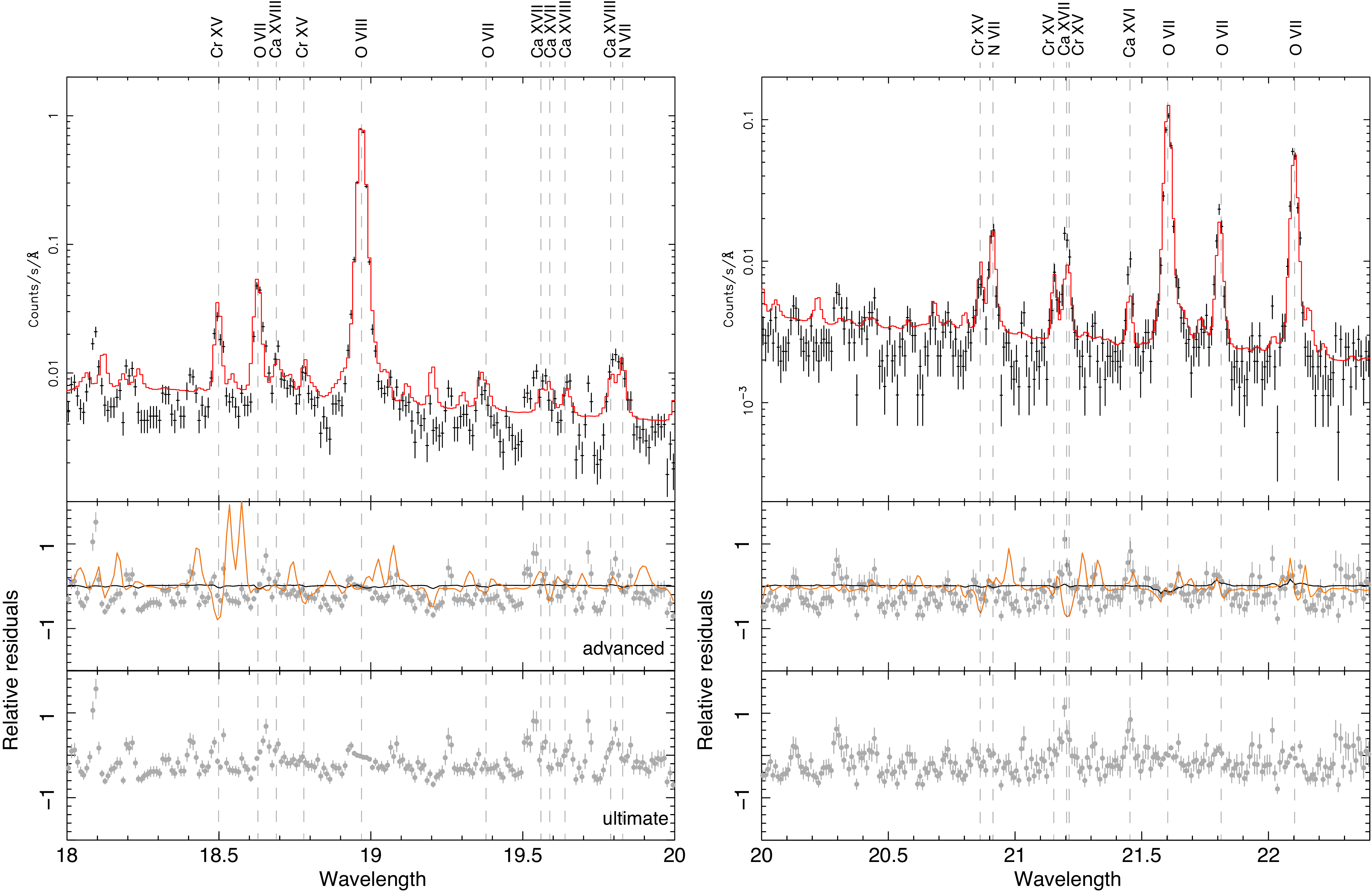}}
\resizebox{0.9\hsize}{!}{\includegraphics[angle=0]{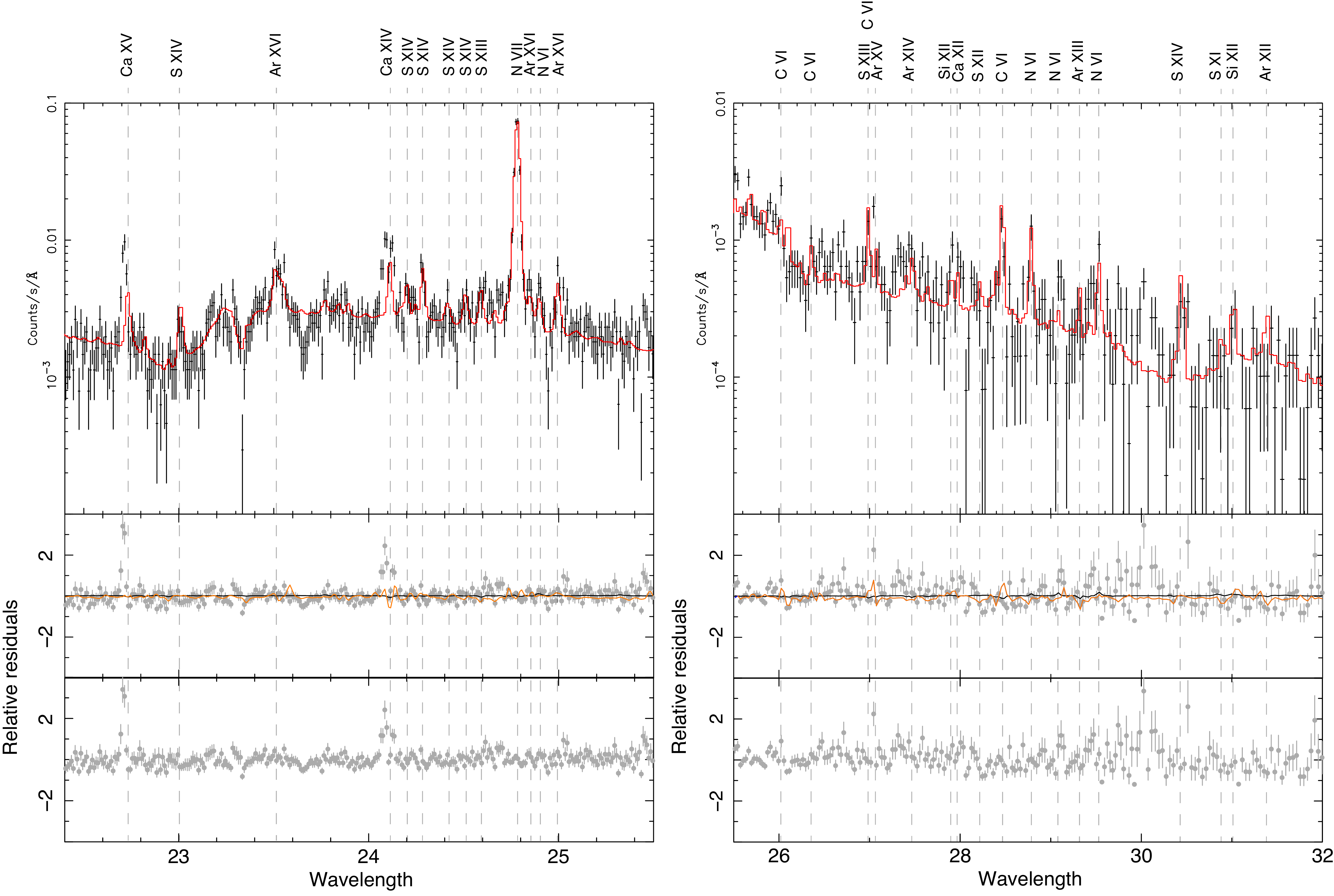}}
\caption{Same as Fig.~\ref{fig:allspec1}, but in $18-32$~{\AA}.}
\label{fig:allspec3}
\end{figure*}

\begin{longtable}{c@{\hskip 0.01in}c@{\hskip 0.01in}c@{\hskip 0.01in}c@{\hskip 0.05in}c@{\hskip 0.05in}c@{\hskip 0.01in}c}
\caption{\label{tab:allline} Capella line list$^{a}$ } \\
\hline\hline
Name & transition & wavelength$^{b}$ & \multicolumn{3}{c}{quality of the fit$^{c}$} & note \\
     &            & ({\AA})    & \scriptsize{SPEX-ADAS}$^{d}$ & \scriptsize{SPEX-FAC}$^{d}$ & \scriptsize{APEC} & \\
\hline     
\endfirsthead
\caption{continued.}\\
\hline\hline
Name & transition & wavelength$^{b}$ & \multicolumn{3}{c}{quality of the fit$^{c}$} & note \\
     &            & ({\AA})    & \scriptsize{SPEX-ADAS}$^{d}$ & \scriptsize{SPEX-FAC}$^{d}$ & \scriptsize{APEC} & \\
\hline
\endhead
\hline
\endfoot
\ion{N}{VII} Ly$\alpha1$   & 1s ($^2$S$_{1/2}$) - 2p ($^2$P$_{3/2}$)                                & 24.78 & A & A & A & \\
\ion{N}{VII} Ly$\alpha2$   & 1s ($^2$S$_{1/2}$) - 2p ($^2$P$_{1/2}$)                                & 24.78 & A & A & A & \\
\ion{Ca}{XIV}              & 2p$^3$ ($^4$S$_{3/2}$) - 2p$^2$3d ($^4$P$_{5/2}$)                      & 24.11 & D & D & D & \scriptsize{SPEX v2 calculation}\\
\ion{Ar}{XVI}              & 2s ($^2$S$_{1/2}$) - 3p ($^2$P$_{3/2}$)                                & 23.51 & A & A & A & \\
\ion{Ca}{XV}               & 2s$^2$2p$^2$ ($^3$P$_0$) - 2s$^2$2p3d ($^3$D$_1$)                      & 22.73 & D & D & D & \\
\ion{O}{VII} He$\alpha$    & 1s$^2$ ($^1$S$_0$) - 1s2s ($^3$S$_1$)                                  & 22.10 & A & A & A & \\
\ion{O}{VII} He$\alpha$    & 1s$^2$ ($^1$S$_0$) - 1s2p ($^3$P$_1$)                                  & 21.81 & A & A & A  & \\
\ion{O}{VII} He$\alpha$    & 1s$^2$ ($^1$S$_0$) - 1s2p ($^1$P$_1$)                                  & 21.60 & A & A & A & \\
\ion{Ca}{XVI}              & 2s$^2$2p ($^2$P$_{1/2}$) - 2s$^2$3d ($^2$D$_{3/2}$)                    & 21.45 & B & B & B & \\
\ion{Cr}{XV}               & 2s$^2$2p$^6$ ($^1$S$_0$) - 2s$^2$2p$^5$3s ($^3$P$_2$)                  & 21.21 & B & B & $-$ & \scriptsize{line not in APEC} \\
\ion{Ca}{XVII}             & 2s2p ($^1$P$_1$) - 2s3d ($^1$D$_2$)                                    & 21.20 & B & B & $-$ & \scriptsize{line not in APEC} \\
\ion{Cr}{XV}               & 2s$^2$2p$^6$ ($^1$S$_0$) - 2s$^2$2p$^5$3s ($^1$P$_1$)                  & 21.15 & A & A & A & \\
\ion{N}{VII} Ly$\beta1$    & 1s ($^2$S$_{1/2}$) - 3p ($^2$P$_{3/2}$)                                & 20.91 & A & A & A & \\
\ion{N}{VII} Ly$\beta2$    & 1s ($^2$S$_{1/2}$) - 3p ($^2$P$_{1/2}$)                                & 20.91 & A & A & A & \\
\ion{Cr}{XV}               & 2s$^2$2p$^6$ ($^1$S$_0$) - 2s$^2$2p$^5$3s ($^3$P$_1$)                  & 20.86 & A & A & $-$ & \scriptsize{line not in APEC} \\
\ion{N}{VII} Ly$\gamma1$   & 1s ($^2$S$_{1/2}$) - 4p ($^2$P$_{3/2}$)                                & 19.83 & A & A & A & \\
\ion{N}{VII} Ly$\gamma2$   & 1s ($^2$S$_{1/2}$) - 4p ($^2$P$_{1/2}$)                                & 19.83 & A & A & A & \\
\ion{Ca}{XVIII}            & 2p ($^2$P$_{3/2}$) - 3d ($^2$D$_{5/2}$)                                & 19.79 & B & B & A & \\
\ion{O}{VIII} Ly$\alpha1$  & 1s ($^2$S$_{1/2}$) - 2p ($^2$P$_{3/2}$)                                & 18.97 & A & A & A & \\
\ion{O}{VIII} Ly$\alpha2$  & 1s ($^2$S$_{1/2}$) - 2p ($^2$P$_{1/2}$)                                & 18.97 & A & A & A & \\
\ion{O}{VII} He$\beta$     & 1s$^2$ ($^1$S$_0$) - 1s3p ($^1$P$_1$)                                  & 18.63 & C & C & C & \\
\ion{Cr}{XV}               & 2s$^2$2p$^6$ ($^1$S$_0$) - 2s$^2$2p$^5$3d ($^1$P$_1$)                  & 18.50 & B & B & $-$ & \scriptsize{line not in APEC} \\
\ion{O}{VII} He$\gamma$    & 1s$^2$ ($^1$S$_0$) - 1s4p ($^1$P$_1$)                                  & 17.77 & A & A & A & \scriptsize{miss lines at 17.8~{\AA} and 18.1~{\AA}} \\
\ion{Fe}{XVIII}            & 2s2p$^6$ ($^2$S$_{1/2}$) - 2s$^2$2p$^4$3p ($^2$P$_{3/2}$)              & 17.61 & C & C & B & \\
\ion{Fe}{XVI}              & 2s$^2$2p$^6$3p ($^2$P$_{3/2}$) - 2p$^5$3s3p ($^4$P$_{5/2}$)            & 17.50 & B & B & C & \\
\ion{O}{VII} He$\delta$    & 1s$^2$ ($^1$S$_0$) - 1s5p ($^1$P$_1$)                                  & 17.40 & A & A & C & \\
\ion{Fe}{XVIII}            & 2s2p$^6$ ($^2$S$_{1/2}$) - 2s$^2$2p$^4$3p ($^2$P$_{3/2}$)              & 17.36 & C & D & D & \\
\ion{Fe}{XVI}              & 2s$^2$2p$^6$3p ($^2$P$_{3/2}$) - 2p$^5$3s3p ($^2$D$_{5/2}$)            & 17.21 & B & B & B & \\
\ion{O}{VII} He$\epsilon$  & 1s$^2$ ($^1$S$_0$) - 1s6p ($^1$P$_1$)                                  & 17.20 & B & B & B & \\
\ion{Fe}{XVII}             & 2s$^2$2p$^6$ ($^1$S$_0$) - 2s$^2$2p$^5$3s ($^3$P$_2$)                  & 17.10 & A & B & B & \\
\ion{Fe}{XVII}             & 2s$^2$2p$^6$ ($^1$S$_0$) - 2s$^2$2p$^5$3s ($^1$P$_1$)                  & 17.05 & A & A & A & \\\
\ion{Fe}{XVII}             & 2s$^2$2p$^6$ ($^1$S$_0$) - 2s$^2$2p$^5$3s ($^3$P$_1$)                  & 16.78 & A & A & B & \\
\ion{Fe}{XVI}              & 2s$^2$2p$^6$3p ($^2$P$_{3/2}$) - 2p$^5$3s3p ($^2$S$_{1/2}$)            & 16.62 & $-$ & $-$ & B & \scriptsize{line not in SPEX} \\
\ion{Fe}{XVII}             & 2s$^2$2p$^6$ ($^1$S$_0$) - 2s$^2$2p$^5$3p ($^3$D$_2$)                  & 16.34 & A & A & C & \\
\ion{Fe}{XVIII}            & 2s2p$^6$ ($^2$S$_{1/2}$) - 2s2p$^5$3s ($^4$P$_{3/2}$)                  & 16.30 & B & B & C & \\
\ion{Fe}{XIX}              & 2s2p$^5$ ($^3$P$_1$) - 2s$^2$2p$^3$3p ($^3$P$_2$)                      & 16.28 & $-$ & $-$ & C & \scriptsize{line not in SPEX} \\
\ion{Fe}{XVII}             & 2s$^2$2p$^6$ ($^1$S$_0$) - 2s$^2$2p$^5$3p ($^3$P$_2$)                  & 16.24 & C & C & C & \\
\ion{Fe}{XVIII}            & 2s2p$^6$ ($^2$S$_{1/2}$) - 2s2p$^5$3s ($^2$P$_{3/2}$)                  & 16.17 & A & A & A & \\
\ion{Fe}{XIX}              & 2s2p$^5$ ($^3$P$_2$) - 2s$^2$2p$^3$3p ($^3$P$_2$)                      & 16.11 & $-$ & $-$ & C & \scriptsize{line not in SPEX} \\
\ion{Fe}{XVIII}            & 2s$^2$2p$^5$ ($^2$P$_{3/2}$) - 2s$^2$2p$^4$3s ($^4$P$_{5/2}$)          & 16.07 & C & C & C & \\
\ion{Fe}{XVIII}            & 2s$^2$2p$^5$ ($^2$P$_{3/2}$) - 2s$^2$2p$^4$3s ($^2$P$_{3/2}$)          & 16.00 & B & B & B & \\
\ion{Fe}{XVII}             & 2s$^2$2p$^6$ ($^1$S$_0$) - 2s$^2$2p$^5$3p ($^1$D$_2$)                  & 16.00 & B & B & B & \\
\ion{O}{VIII} Ly$\beta1$   & 1s ($^2$S$_{1/2}$) - 3p ($^2$P$_{3/2}$)                                & 16.00 & B & B & B & \\
\ion{O}{VIII} Ly$\beta2$   & 1s ($^2$S$_{1/2}$) - 3p ($^2$P$_{1/2}$)                                & 16.00 & B & B & B & \\
\ion{Fe}{XVIII}            & 2s$^2$2p$^5$ ($^2$P$_{1/2}$) - 2s$^2$2p$^4$3s ($^2$D$_{3/2}$)          & 15.87 & B & B & B & \\
\ion{Fe}{XVIII}            & 2s$^2$2p$^5$ ($^2$P$_{3/2}$) - 2s$^2$2p$^4$3s ($^4$P$_{3/2}$)          & 15.83 & A & A & C & \\
\ion{Fe}{XVIII}            & 2s$^2$2p$^5$ ($^2$P$_{3/2}$) - 2s$^2$2p$^4$3s ($^2$P$_{1/2}$)          & 15.77 & A & A & B & \\
\ion{Fe}{XVIII}            & 2s$^2$2p$^5$ ($^2$P$_{3/2}$) - 2s$^2$2p$^4$3s ($^2$D$_{5/2}$)          & 15.63 & C & C & A & \\
\ion{Fe}{XVIII}            & 2s2p$^6$ ($^2$S$_{1/2}$) - 2s2p$^5$3s ($^2$P$_{3/2}$)                  & 15.49 & $-$ & $-$ & B & \scriptsize{line not in SPEX} \\
\ion{Fe}{XVII}             & 2s$^2$2p$^6$ ($^1$S$_0$) - 2s$^2$2p$^5$3d ($^3$P$_1$)                  & 15.45 & A & A & A & \\
\ion{Fe}{XVIII}            & 2s$^2$2p$^5$ ($^2$P$_{1/2}$) - 2s$^2$2p$^4$3s ($^2$S$_{1/2}$)          & 15.45 & A & A & A & \\
\ion{Fe}{XVIII}            & 2s$^2$2p$^5$ ($^2$P$_{3/2}$) - 2s$^2$2p$^4$3p ($^2$D$_{5/2}$)          & 15.39 & D & D & D & \\
\ion{Fe}{XVII}             & 2s$^2$2p$^6$ ($^1$S$_0$) - 2s$^2$2p$^5$3d ($^3$D$_1$)                  & 15.26 & B & A & B & \\
\ion{Fe}{XVI}              & 2s$^2$2p$^6$3s ($^2$S$_{1/2}$) - 2p$^5$3s3d ($^2$P$_{1/2}$)            & 15.21 & B & B & $-$ & \scriptsize{line not in APEC} \\
\ion{O}{VIII} Ly$\gamma1$  & 1s ($^2$S$_{1/2}$) - 4p ($^2$P$_{3/2}$)                                & 15.18 & A & B & A & \\
\ion{O}{VIII} Ly$\gamma2$  & 1s ($^2$S$_{1/2}$) - 4p ($^2$P$_{1/2}$)                                & 15.18 & A & B & A & \\
\ion{Fe}{XVI}              & 2s$^2$2p$^6$4d ($^2$D$_{5/2}$) - 2p$^5$3d4d ($^4$D$_{7/2}$)            & 15.08 & A & A & B & \\
\ion{Fe}{XIX}              & 2s$^2$2p$^4$ ($^3$P$_2$) - 2s$^2$2p$^3$3s ($^5$S$_2$)                  & 15.08 & A & A & B & \\
\ion{Fe}{XVI}              & 2s$^2$2p$^6$4f ($^2$F$_{7/2}$) - 2p$^5$3d4f ($^4$G$_{9/2}$)            & 15.03 & B & B & B & \\
\ion{Fe}{XVII}             & 2s$^2$2p$^6$ ($^1$S$_0$) - 2s$^2$2p$^5$3d ($^1$P$_1$)                  & 15.01 & A & A & A & \\
\ion{O}{VIII} Ly$\delta1$  & 1s ($^2$S$_{1/2}$) - 5p ($^2$P$_{3/2}$)                                & 14.82 & B & A & A & \\
\ion{O}{VIII} Ly$\delta2$  & 1s ($^2$S$_{1/2}$) - 5p ($^2$P$_{1/2}$)                                & 14.82 & B & A & A & \\
\ion{Fe}{XIX}              & 2s$^2$2p$^4$ ($^3$P$_2$) - 2s$^2$2p$^3$3s ($^3$D$_3$)                  & 14.67 & A & B & D & \\
\ion{Fe}{XVIII}            & 2s$^2$2p$^5$ ($^2$P$_{3/2}$) - 2s$^2$2p$^4$3d ($^2$F$_{5/2}$)          & 14.53 & C & C & D & \\
\ion{Fe}{XVIII}            & 2s$^2$2p$^5$ ($^2$P$_{3/2}$) - 2s$^2$2p$^4$3d ($^2$D$_{5/2}$)          & 14.37 & B & B & B & \scriptsize{miss a line at 14.41~{\AA}} \\
\ion{Fe}{XVIII}            & 2s$^2$2p$^5$ ($^2$P$_{3/2}$) - 2s$^2$2p$^4$3d ($^2$S$_{1/2}$)          & 14.26 & A & A & B & \\
\ion{Fe}{XVIII}            & 2s$^2$2p$^5$ ($^2$P$_{3/2}$) - 2s$^2$2p$^4$3d ($^2$P$_{3/2}$)          & 14.21 & A & A & C & \\
\ion{Fe}{XVIII}            & 2s$^2$2p$^5$ ($^2$P$_{3/2}$) - 2s$^2$2p$^4$3d ($^2$D$_{5/2}$)          & 14.20 & A & A & C & \\
\ion{Ni}{XIX}              & 2s$^2$2p$^6$ ($^1$S$_0$) - 2s$^2$2p$^5$3s ($^3$P$_2$)                  & 14.07 & B & B & A & \\
\ion{Ni}{XIX}              & 2s$^2$2p$^6$ ($^1$S$_0$) - 2s$^2$2p$^5$3s ($^1$P$_1$)                  & 14.04 & B & B & A & \\
\ion{Fe}{XVIII}            & 2s$^2$2p$^5$ ($^2$P$_{3/2}$) - 2s$^2$2p$^4$3d ($^2$D$_{5/2}$)          & 13.95 & B & B & B & \\
\ion{Fe}{XIX}              & 2s$^2$2p$^4$ ($^3$P$_2$) - 2s$^2$2p$^3$3d ($^5$D$_3$)                  & 13.93 & D & D & $-$ & \scriptsize{line not in APEC} \\
\ion{Fe}{XVII}             & 2s$^2$2p$^6$ ($^1$S$_0$) - 2s2p$^6$3p ($^3$P$_1$)                      & 13.89 & A & A & D & \\
\ion{Fe}{XVII}             & 2s$^2$2p$^6$ ($^1$S$_0$) - 2s2p$^6$3p ($^1$P$_1$)                      & 13.82 & C & C & B & \\
\ion{Fe}{XIX}              & 2s$^2$2p$^4$ ($^3$P$_2$) - 2s$^2$2p$^3$3d ($^3$D$_3$)                  & 13.80 & B & B & A & \\
\ion{Ni}{XIX}              & 2s$^2$2p$^6$ ($^1$S$_0$) - 2s$^2$2p$^5$3s ($^3$P$_1$)                  & 13.78 & B & B & A & \\
\ion{Fe}{XIX}              & 2s$^2$2p$^4$ ($^1$D$_2$) - 2s$^2$2p$^3$3d ($^1$F$_3$)                  & 13.74 & A & A & $-$ & \scriptsize{line not in APEC} \\
\ion{Ne}{IX} He$\alpha$    & 1s$^2$ ($^1$S$_0$) - 1s2s ($^3$S$_1$)                                  & 13.70 & A & B & A & \scriptsize{\ion{Ne}{VIII} blend} \\
\ion{Fe}{XIX}              & 2s$^2$2p$^4$ ($^3$P$_2$) - 2s$^2$2p$^3$3d ($^3$F$_3$)                  & 13.63 & D & D & C & \\
\ion{Ne}{IX} He$\alpha$    & 1s$^2$ ($^1$S$_0$) - 1s2p ($^3$P$_1$)                                  & 13.55 & A & A & B & \scriptsize{\ion{Ne}{VIII} blend} \\
\ion{Fe}{XIX}              & 2s$^2$2p$^4$ ($^3$P$_2$) - 2s$^2$2p$^3$3d ($^3$D$_3$)                  & 13.52 & B & B & B & \\
\ion{Fe}{XIX}              & 2s$^2$2p$^4$ ($^3$P$_2$) - 2s$^2$2p$^3$3d ($^3$P$_2$)                  & 13.49 & C & C & D & \\
\ion{Fe}{XIX}              & 2s$^2$2p$^4$ ($^3$P$_2$) - 2s$^2$2p$^3$3d ($^3$S$_1$)                  & 13.45 & A & A & A & \\
\ion{Ne}{IX} He$\alpha$    & 1s$^2$ ($^1$S$_0$) - 1s2p ($^1$P$_1$)                                  & 13.45 & A & A & A & \scriptsize{\ion{Ne}{VIII} blend} \\
\ion{Fe}{XVIII}            & 2s$^2$2p$^5$ ($^2$P$_{3/2}$) - 2s2p$^5$3p ($^2$D$_{5/2}$)              & 13.37 & A & B & D & \\
\ion{Fe}{XVIII}            & 2s$^2$2p$^5$ ($^2$P$_{3/2}$) - 2s2p$^5$3p ($^4$P$_{5/2}$)              & 13.32 & A & A & D & \\
\ion{Fe}{XX}               & 2s$^2$2p$^3$ ($^2$D$_{5/2}$) - 2s$^2$2p$^2$3d ($^4$F$_{7/2}$)          & 13.27 & B & B & D & \\
\ion{Fe}{XX}               & 2s$^2$2p$^3$ ($^2$D$_{5/2}$) - 2s$^2$2p$^2$3d ($^4$P$_{5/2}$)          & 13.14 & A & A & A & \\
\ion{Fe}{XVII}             & 2s$^2$2p$^6$ ($^1$S$_0$) - 2s2p$^6$3d ($^1$D$_2$)                      & 13.12 & C & C & A & \\  
\ion{Fe}{XX}               & 2s$^2$2p$^3$ ($^4$S$_{3/2}$) - 2s$^2$2p$^2$3d ($^4$D$_{5/2}$)          & 13.05 & B & B & D & \\
\ion{Fe}{XX}               & 2s$^2$2p$^3$ ($^4$S$_{3/2}$) - 2s$^2$2p$^2$3d ($^4$F$_{5/2}$)          & 12.95 & A & A & B & \\
\ion{Fe}{XIX}              & 2s$^2$2p$^4$ ($^3$P$_2$) - 2s2p$^4$3p ($^5$D$_3$)                      & 12.92 & A & A & A & \\
\ion{Fe}{XX}               & 2s$^2$2p$^3$ ($^4$S$_{3/2}$) - 2s$^2$2p$^2$3d ($^2$F$_{5/2}$)          & 12.90 & B/C & B/C & A & \\
\ion{Fe}{XX}               & 2s$^2$2p$^3$ ($^4$S$_{3/2}$) - 2s$^2$2p$^2$3d ($^4$P$_{5/2}$)          & 12.84 & A & A & D & \\
\ion{Fe}{XX}               & 2s$^2$2p$^3$ ($^4$S$_{3/2}$) - 2s$^2$2p$^2$3d ($^4$P$_{3/2}$)          & 12.82 & A & A & D & \\
\ion{Fe}{XX}               & 2s$^2$2p$^3$ ($^4$S$_{3/2}$) - 2s$^2$2p$^2$3d ($^4$P$_{1/2}$)          & 12.81 & A & A & D & \\
\ion{Fe}{XXII}             & 2s2p$^2$ ($^2$D$_{3/2}$) - 2s2p3s ($^2$P$_{1/2}$)                      & 12.76 & A & A & A & \\
\ion{Ni}{XIX}              & 2s$^2$2p$^6$ ($^1$S$_0$) - 2s$^2$2p$^5$3d ($^3$D$_1$)                  & 12.66 & A & A & A & \\
\ion{Fe}{XX}               & 2s$^2$2p$^3$ ($^4$S$_{3/2}$) - 2s2p$^3$3p ($^4$P$_{5/2}$)              & 12.59 & C & C & B & \\
\ion{Ni}{XIX}              & 2s$^2$2p$^6$ ($^1$S$_0$) - 2s$^2$2p$^5$3d ($^1$P$_1$)                  & 12.44 & A & A & B & \\
\ion{Fe}{XXI}              & 2s$^2$2p$^2$ ($^3$P$_1$) - 2s$^2$2p3d ($^3$D$_1$)                      & 12.40 & B & B & B & \\
\ion{Fe}{XXI}              & 2s$^2$2p$^2$ ($^3$P$_0$) - 2s$^2$2p3d ($^3$D$_1$)                      & 12.29 & C & C & B & \\
\ion{Fe}{XVII}             & 2s$^2$2p$^6$ ($^1$S$_0$) - 2s$^2$2p$^5$4d ($^3$D$_1$)                  & 12.26 & A & A & A & \\
\ion{Ne}{X} Ly$\alpha2$    & 1s ($^2$S$_{1/2}$) - 2p ($^2$P$_{1/2}$)                                & 12.14 & A & A & A & \\
\ion{Ne}{X} Ly$\alpha1$    & 1s ($^2$S$_{1/2}$) - 2p ($^2$P$_{3/2}$)                                & 12.13 & A & A & A & \\
\ion{Fe}{XVII}             & 2s$^2$2p$^6$ ($^1$S$_0$) - 2s$^2$2p$^5$4d ($^1$P$_1$)                  & 12.12 & A & A & A & \\
\ion{Fe}{XXI}              & 2s$^2$2p$^2$ ($^3$P$_0$) - 2s2p$^2$3p ($^5$P$_1$)                      & 11.99 & B & B & A & \\
\ion{Fe}{XXII}             & 2s$^2$2p ($^2$P$_{3/2}$) - 2s$^2$3d ($^2$D$_{5/2}$)                    & 11.92 & A & A & A & \\
\ion{Ni}{XX}               & 2s$^2$2p$^5$ ($^2$P$_{3/2}$) - 2s$^2$2p$^4$3d ($^2$D$_{3/2}$)          & 11.86 & A & A & B & \scriptsize{SPEX v2 calculation}\\
\ion{Ni}{XX}               & 2s$^2$2p$^5$ ($^2$P$_{3/2}$) - 2s$^2$2p$^4$3d ($^2$D$_{5/2}$)          & 11.83 & C & C & C & \scriptsize{SPEX v2 calculation}\\
\ion{Fe}{XXII}             & 2s$^2$2p ($^2$P$_{1/2}$) - 2s$^2$3d ($^2$D$_{3/2}$)                    & 11.77 & B & B & B & \\
\ion{Fe}{XXIII}            & 2s2p ($^1$P$_1$) - 2s3d ($^1$D$_2$)                                    & 11.74 & A & A & B & \\
\ion{Ni}{XX}               & 2s$^2$2p$^5$ ($^2$P$_{3/2}$) - 2s$^2$2p$^4$3d ($^2$D$_{5/2}$)          & 11.64 & B & B & A & \scriptsize{SPEX v2 calculation}\\
\ion{Ne}{IX} He$\beta$     & 1s$^2$ ($^1$S$_0$) - 1s3p ($^1$P$_1$)                                  & 11.55 & A & A & A & \\
\ion{Fe}{XVIII}            & 2s$^2$2p$^5$ ($^2$P$_{3/2}$) - 2s$^2$2p$^4$4d ($^2$F$_{5/2}$)          & 11.53 & B & A & A & \\
\ion{Fe}{XVIII}            & 2s$^2$2p$^5$ ($^2$P$_{3/2}$) - 2s$^2$2p$^4$4d ($^2$D$_{5/2}$)          & 11.43 & C & C & C & \\
\ion{Fe}{XVIII}            & 2s$^2$2p$^5$ ($^2$P$_{3/2}$) - 2s$^2$2p$^4$4d ($^2$F$_{5/2}$)          & 11.33 & C & C & B & \\
\ion{Fe}{XVII}             & 2s$^2$2p$^6$ ($^1$S$_0$) - 2s$^2$2p$^5$5d ($^1$P$_1$)                  & 11.25 & A & A & A & \\
\ion{Na}{X} He$\alpha$     & 1s$^2$ ($^1$S$_0$) - 1s2s ($^3$S$_1$)                                  & 11.19 & B & A & A & \\
\ion{Fe}{XVII}             & 2s$^2$2p$^6$ ($^1$S$_0$) - 2s$^2$2p$^5$5d ($^3$D$_1$)                  & 11.13 & A & A & A & \\
\ion{Fe}{XVII}             & 2s$^2$2p$^6$ ($^1$S$_0$) - 2s2p$^6$4p ($^1$P$_1$)                      & 11.03 & $-$ & $-$ & A & \scriptsize{line not in SPEX} \\
\ion{Na}{X} He$\alpha$     & 1s$^2$ ($^1$S$_0$) - 1s2p ($^1$P$_1$)                                  & 11.00 & B & B & A & \\
\ion{Fe}{XIX}              & 2s$^2$2p$^4$ ($^3$P$_2$) - 2s$^2$2p$^3$4d ($^3$D$_3$)                  & 10.81 & D & D & D & \\
\ion{Fe}{XVII}             & 2s$^2$2p$^6$ ($^1$S$_0$) - 2s$^2$2p$^5$6d ($^1$P$_1$)                  & 10.77 & A & A & A & \\
\ion{Fe}{XIX}              & 2s$^2$2p$^4$ ($^3$P$_2$) - 2s$^2$2p$^3$4d ($^3$D$_3$)                  & 10.66 & B/C & B/C & B & \\
\ion{Fe}{XIX}              & 2s$^2$2p$^4$ ($^3$P$_2$) - 2s$^2$2p$^3$4d ($^3$S$_1$)                  & 10.62 & B & B & C & \\
\ion{Fe}{XVIII}            & 2s$^2$2p$^5$ ($^2$P$_{3/2}$) - 2s$^2$2p$^4$5d ($^2$D$_{5/2}$)          & 10.54 & C & C & C & \\
\ion{Fe}{XVII}             & 2s$^2$2p$^6$ ($^1$S$_0$) - 2s$^2$2p$^5$7d ($^1$P$_1$)                  & 10.50 & A & A & A & \\
\ion{Fe}{XVIII}            & 2s$^2$2p$^5$ ($^2$P$_{3/2}$) - 2s$^2$2p$^4$5d ($^2$F$_{5/2}$)          & 10.45 & B & B & C & \\
\ion{Fe}{XIX}              & 2s$^2$2p$^4$ ($^3$P$_2$) - 2s$^2$2p$^3$4d ($^3$D$_3$)                  & 10.45 & B & B & C & \\
\ion{Fe}{XVII}             & 2s$^2$2p$^6$ ($^1$S$_0$) - 2s$^2$2p$^5$7d ($^3$D$_1$)                  & 10.39 & B & A & B & \\
\ion{Fe}{XVIII}            & 2s$^2$2p$^5$ ($^2$P$_{3/2}$) - 2s$^2$2p$^4$5d ($^2$F$_{5/2}$)          & 10.36 & C & C & C & \\
\ion{Fe}{XVII}             & 2s$^2$2p$^6$ ($^1$S$_0$) - 2s$^2$2p$^5$8d ($^1$P$_1$)                  & 10.34 & A & A & $-$ & \scriptsize{line not in APEC} \\
\ion{Ne}{X} Ly$\beta1$     & 1s ($^2$S$_{1/2}$) - 3p ($^2$P$_{3/2}$)                                & 10.24 & A & A & A & \\
\ion{Ne}{X} Ly$\beta2$     & 1s ($^2$S$_{1/2}$) - 3p ($^2$P$_{1/2}$)                                & 10.24 & A & A & A & \\
\ion{Fe}{XX}               & 2s$^2$2p$^3$ ($^2$P$_{3/2}$) - 2s$^2$2p$^2$4d ($^2$P$_{3/2}$)          & 10.18 & A & B & B & \\
\ion{Ni}{XIX}              & 2s$^2$2p$^6$ ($^1$S$_0$) - 2s$^2$2p$^5$4d ($^3$D$_1$)                  & 10.11 & B & B & B & \\
\ion{Fe}{XX}               & 2s$^2$2p$^3$ ($^4$S$_{3/2}$) - 2s$^2$2p$^2$4d ($^4$F$_{5/2}$)          & 10.05 & B/C & B/C & $-$ & \scriptsize{line not in APEC} \\
\ion{Na}{XI} Ly$\alpha2$   & 1s ($^2$S$_{1/2}$) - 2p ($^2$P$_{1/2}$)                                & 10.02 & A & A & A & \\
\ion{Na}{XI} Ly$\alpha1$   & 1s ($^2$S$_{1/2}$) - 2p ($^2$P$_{3/2}$)                                & 10.02 & A & A & A & \\
\ion{Fe}{XX}               & 2s$^2$2p$^3$ ($^4$P$_{3/2}$) - 2s$^2$2p$^2$4d ($^4$P$_{3/2}$)          & 10.00 & A & A & $-$ & \scriptsize{line not in APEC} \\
\ion{Ni}{XIX}              & 2s$^2$2p$^6$ ($^1$S$_0$) - 2s$^2$2p$^5$4d ($^1$P$_1$)                  &  9.98 & A & A & B & \\
\ion{Fe}{XIX}              & 2s$^2$2p$^4$ ($^3$P$_2$) - 2s$^2$2p$^3$5d ($^3$D$_3$)                  &  9.84 & C & C & C & \scriptsize{miss lines at 9.89~{\AA} and 9.79~{\AA}} \\
\ion{Ne}{X} Ly$\gamma1$    & 1s ($^2$S$_{1/2}$) - 4p ($^2$P$_{3/2}$)                                &  9.71 & A & A & A & \\
\ion{Ne}{X} Ly$\gamma2$    & 1s ($^2$S$_{1/2}$) - 4p ($^2$P$_{1/2}$)                                &  9.71 & A & A & A & \\
\ion{Fe}{XIX}              & 2s$^2$2p$^4$ ($^3$P$_2$) - 2s$^2$2p$^3$5d ($^3$P$_2$)                  &  9.69 & A & A & A & \\
\ion{Fe}{XXI}              & 2s$^2$2p$^2$ ($^3$P$_1$) - 2s$^2$2p4d ($^3$D$_1$)                      &  9.55 & B & B & C & \scriptsize{miss a line at 9.62~{\AA}}\\
\ion{Fe}{XXI}              & 2s$^2$2p$^2$ ($^3$P$_0$) - 2s$^2$2p4d ($^3$D$_1$)                      &  9.48 & A & A & B & \\
\ion{Ne}{X} Ly$\delta1$    & 1s ($^2$S$_{1/2}$) - 5p ($^2$P$_{3/2}$)                                &  9.48 & A & A & B & \\
\ion{Ne}{X} Ly$\delta2$    & 1s ($^2$S$_{1/2}$) - 5p ($^2$P$_{1/2}$)                                &  9.48 & A & A & B & \\
\ion{Fe}{XIX}              & 2s$^2$2p$^4$ ($^3$P$_2$) - 2s$^2$2p$^3$6d ($^3$D$_3$)                  &  9.39 & B & B & $-$ & \scriptsize{line not in APEC} \\
\ion{Ni}{XX}               & 2s$^2$2p$^5$ ($^2$P$_{3/2}$) - 2s$^2$2p$^4$4d ($^2$P$_{5/2}$)          &  9.39 & B & B & $-$ & \scriptsize{SPEX v2; line not in APEC}  \\
\ion{Mg}{XI} He$\alpha$    & 1s$^2$ ($^1$S$_0$) - 1s2s ($^3$S$_1$)                                  &  9.31 & A & A & A & \scriptsize{\ion{Mg}{X} blend} \\
\ion{Mg}{XI} He$\alpha$    & 1s$^2$ ($^1$S$_0$) - 1s2p ($^3$P$_1$)                                  &  9.23 & A & A & A & \scriptsize{\ion{Mg}{X} blend} \\
\ion{Mg}{XI} He$\alpha$    & 1s$^2$ ($^1$S$_0$) - 1s2p ($^1$P$_1$)                                  &  9.17 & A & A & A & \scriptsize{\ion{Mg}{X} blend} \\
\ion{Fe}{XX}               & 2s$^2$2p$^3$ ($^4$S$_{3/2}$) - 2s$^2$2p$^2$5d ($^4$P$_{3/2}$)          &  9.06 & C & D & C & \\
\ion{Fe}{XXII}             & 2s$^2$2p ($^2$P$_{1/2}$) - 2s$^2$4d ($^2$D$_{3/2}$)                    &  8.98 & B & B & D & \\
\ion{Fe}{XXIII}            & 2s2p ($^1$P$_1$) - 2s4s ($^1$S$_0$)                                    &  8.91 & A & A & A & \\
\ion{Fe}{XXIII}            & 2s2p ($^1$P$_1$) - 2s4d ($^1$D$_2$)                                    &  8.81 & B & B & D & \\
\ion{Fe}{XXII}             & 2s$^2$2p ($^2$P$_{1/2}$) - 2s2p4d ($^2$D$_{3/2}$)                      &  8.72 & B/C & B/C & B/C & \\
\ion{Fe}{XX}               & 2s$^2$2p$^3$ ($^4$P$_{3/2}$) - 2s$^2$2p$^2$6d ($^4$P$_{3/2}$)          &  8.63 & B & B & D & \\
\ion{Fe}{XXI}              & 2s$^2$2p$^2$ ($^3$P$_0$) - 2s$^2$2p5d ($^3$D$_1$)                      &  8.55 & B/C & B/C & A & \\
\ion{Mg}{XII} Ly$\alpha1$  & 1s ($^2$S$_{1/2}$) - 2p ($^2$P$_{3/2}$)                                &  8.42 & A & A & A & \\
\ion{Mg}{XII} Ly$\alpha2$  & 1s ($^2$S$_{1/2}$) - 2p ($^2$P$_{1/2}$)                                &  8.42 & A & A & A & \\
\ion{Fe}{XXIII}            & 2s2p ($^1$P$_1$) - 2s4p ($^1$P$_1$)                                    &  8.30 & A & A & B & \\
\ion{Fe}{XXII}             & 2s$^2$2p ($^2$P$_{3/2}$) - 2s$^2$5d ($^2$D$_{5/2}$)                    &  8.16 & B & B & D & \\
\ion{Fe}{XXII}             & 2s$^2$2p ($^2$P$_{1/2}$) - 2s$^2$5d ($^2$D$_{3/2}$)                    &  8.09 & A & A & A & \\
\ion{Mg}{X}                & 2p ($^2$P$_{3/2}$) - 1s2p3p ($^2$D$_{5/2}$)                            &  8.07 & A & A & A & \\
\ion{Fe}{XXIV}             & 2s ($^2$S$_{1/2}$) - 4p ($^2$P$_{3/2}$)                                &  7.98 & A & A & B & \\
\ion{Al}{XII} He$\alpha$   & 1s$^2$ ($^1$S$_0$) - 1s2s ($^3$S$_1$)                                  &  7.87 & A & A & A & \scriptsize{\ion{Al}{XI} blend} \\
\ion{Mg}{XI} He$\beta$     & 1s$^2$ ($^1$S$_0$) - 1s3p ($^1$P$_1$)                                  &  7.85 & C & C & A & \\
\ion{Al}{XII} He$\alpha$   & 1s$^2$ ($^1$S$_0$) - 1s2p ($^3$P$_1$)                                  &  7.80 & A & A & A & \scriptsize{\ion{Al}{XI} blend} \\
\ion{Al}{XII} He$\alpha$   & 1s$^2$ ($^1$S$_0$) - 1s2p ($^1$P$_1$)                                  &  7.76 & A & A & A & \scriptsize{\ion{Al}{XI} blend} \\
\ion{Mg}{XI} He$\gamma$    & 1s$^2$ ($^1$S$_0$) - 1s4p ($^1$P$_1$)                                  &  7.47 & A & A & A & \\
\ion{Mg}{XI} He$\delta$    & 1s$^2$ ($^1$S$_0$) - 1s5p ($^1$P$_1$)                                  &  7.31 & A & A & A & \\
\ion{Mg}{XI} He$\epsilon$  & 1s$^2$ ($^1$S$_0$) - 1s6p ($^1$P$_1$)                                  &  7.22 & A & A & C & \\
\ion{Al}{XIII} Ly$\alpha1$ & 1s ($^2$S$_{1/2}$) - 2p ($^2$P$_{3/2}$)                                &  7.17 & A & A & A & \\
\ion{Al}{XIII} Ly$\alpha2$ & 1s ($^2$S$_{1/2}$) - 2p ($^2$P$_{1/2}$)                                &  7.17 & A & A & A & \\
\ion{Mg}{XII} Ly$\beta1$   & 1s ($^2$S$_{1/2}$) - 3p ($^2$P$_{3/2}$)                                &  7.11 & A & A & A & \\
\ion{Mg}{XII} Ly$\beta2$   & 1s ($^2$S$_{1/2}$) - 3p ($^2$P$_{1/2}$)                                &  7.11 & B & B & A & \\
\ion{Si}{XIII} He$\alpha$  & 1s$^2$ ($^1$S$_0$) - 1s2s ($^3$S$_1$)                                  &  6.74 & A & A & A & \scriptsize{\ion{Si}{XII} blend} \\
\ion{Mg}{XII} Ly$\gamma1$  & 1s ($^2$S$_{1/2}$) - 4p ($^2$P$_{3/2}$)                                &  6.74 & A & A & A & \\
\ion{Mg}{XII} Ly$\gamma2$  & 1s ($^2$S$_{1/2}$) - 4p ($^2$P$_{1/2}$)                                &  6.74 & A & A & A & \\
\ion{Si}{XIII} He$\alpha$  & 1s$^2$ ($^1$S$_0$) - 1s2p ($^3$P$_1$)                                  &  6.68 & A & A & A & \scriptsize{\ion{Si}{XII} blend} \\
\ion{Si}{XIII} He$\alpha$  & 1s$^2$ ($^1$S$_0$) - 1s2p ($^1$P$_1$)                                  &  6.65 & A & A & A & \scriptsize{\ion{Si}{XII} blend} \\
\ion{Mg}{XII} Ly$\delta1$  & 1s ($^2$S$_{1/2}$) - 5p ($^2$P$_{3/2}$)                                &  6.58 & A & A & A & \\
\ion{Mg}{XII} Ly$\delta2$  & 1s ($^2$S$_{1/2}$) - 5p ($^2$P$_{1/2}$)                                &  6.58 & A & A & A & \\
\ion{Si}{XIV} Ly$\alpha1$  & 1s ($^2$S$_{1/2}$) - 2p ($^2$P$_{3/2}$)                                &  6.18 & A & A & A & \\
\ion{Si}{XIV} Ly$\alpha2$  & 1s ($^2$S$_{1/2}$) - 2p ($^2$P$_{1/2}$)                                &  6.18 & A & A & A & \\
\ion{Si}{XII}              & 2p ($^2$P$_{3/2}$) - 1s2p3p ($^2$D$_{5/2}$)                            &  5.82 & A & A & A & \\
\ion{Si}{XIII} He$\beta$   & 1s$^2$ ($^1$S$_0$) - 1s3p ($^1$P$_1$)                                  &  5.68 & A & A & A & \\
\ion{Si}{XIII} He$\gamma$  & 1s$^2$ ($^1$S$_0$) - 1s4p ($^1$P$_1$)                                  &  5.40 & A & A & A & \\
\ion{Si}{XIV} Ly$\beta1$   & 1s ($^2$S$_{1/2}$) - 3p ($^2$P$_{3/2}$)                                &  5.22 & A & A & A & \\
\ion{Si}{XIV} Ly$\beta2$   & 1s ($^2$S$_{1/2}$) - 3p ($^2$P$_{1/2}$)                                &  5.22 & A & A & A & \\
\ion{S}{XV} He$\alpha$     & 1s$^2$ ($^1$S$_0$) - 1s2s ($^3$S$_1$)                                  &  5.10 & A & A & A & \scriptsize{\ion{S}{XIV} blend} \\
\ion{S}{XV} He$\alpha$     & 1s$^2$ ($^1$S$_0$) - 1s2p ($^3$P$_1$)                                  &  5.07 & A & A & A & \scriptsize{\ion{S}{XIV} blend} \\
\ion{S}{XV} He$\alpha$     & 1s$^2$ ($^1$S$_0$) - 1s2p ($^3$P$_2$)                                  &  5.06 & A & A & A & \scriptsize{\ion{S}{XIV} blend} \\
\ion{S}{XV} He$\alpha$     & 1s$^2$ ($^1$S$_0$) - 1s2p ($^1$P$_1$)                                  &  5.04 & A & A & A & \scriptsize{\ion{S}{XIV} blend} \\
\ion{S}{XVI} Ly$\alpha1$   & 1s ($^2$S$_{1/2}$) - 2p ($^2$P$_{3/2}$)                                &  4.73 & A & A & A & \\
\ion{S}{XVI} Ly$\alpha2$   & 1s ($^2$S$_{1/2}$) - 2p ($^2$P$_{1/2}$)                                &  4.73 & A & A & A & \\
\ion{S}{XV} He$\beta$      & 1s$^2$ ($^1$S$_0$) - 1s3p ($^1$P$_1$)                                  &  4.30 & A & A & A & \\
\ion{Ar}{XVII} He$\alpha$  & 1s$^2$ ($^1$S$_0$) - 1s2s ($^3$S$_1$)                                  &  3.99 & A & A & A & \scriptsize{\ion{Ar}{XVI} blend}  \\
\ion{Ar}{XVII} He$\alpha$  & 1s$^2$ ($^1$S$_0$) - 1s2p ($^3$P$_1$)                                  &  3.97 & A & A & A & \scriptsize{\ion{Ar}{XVI} blend} \\
\ion{Ar}{XVII} He$\alpha$  & 1s$^2$ ($^1$S$_0$) - 1s2p ($^3$P$_2$)                                  &  3.97 & A & A & A & \scriptsize{\ion{Ar}{XVI} blend} \\
\ion{Ar}{XVII} He$\alpha$  & 1s$^2$ ($^1$S$_0$) - 1s2p ($^1$P$_1$)                                  &  3.95 & A & A & A & \scriptsize{\ion{Ar}{XVI} blend} \\
\ion{Ca}{XIX} He$\alpha$   & 1s$^2$ ($^1$S$_0$) - 1s2s ($^3$S$_1$)                                  &  3.21 & A & A & A & \scriptsize{\ion{Ca}{XVIII} blend} \\
\ion{Ca}{XIX} He$\alpha$   & 1s$^2$ ($^1$S$_0$) - 1s2p ($^3$P$_1$)                                  &  3.19 & A & A & A & \scriptsize{\ion{Ca}{XVIII} blend} \\
\ion{Ca}{XIX} He$\alpha$   & 1s$^2$ ($^1$S$_0$) - 1s2p ($^3$P$_2$)                                  &  3.19 & A & A & A & \scriptsize{\ion{Ca}{XVIII} blend} \\
\ion{Ca}{XIX} He$\alpha$   & 1s$^2$ ($^1$S$_0$) - 1s2p ($^1$P$_1$)                                  &  3.18 & A & A & A & \scriptsize{\ion{Ca}{XVIII} blend} \\
\ion{Fe}{XXV} He$\alpha$   & 1s$^2$ ($^1$S$_0$) - 1s2s ($^3$S$_1$)                                  &  1.87 & A & A & B & \scriptsize{\ion{Fe}{XXIV} blend} \\
\ion{Fe}{XXV} He$\alpha$   & 1s$^2$ ($^1$S$_0$) - 1s2p ($^3$P$_1$)                                  &  1.86 & A & A & B & \scriptsize{\ion{Fe}{XXIV} blend} \\
\ion{Fe}{XXV} He$\alpha$   & 1s$^2$ ($^1$S$_0$) - 1s2p ($^3$P$_2$)                                  &  1.86 & A & A & B & \scriptsize{\ion{Fe}{XXIV} blend} \\
\ion{Fe}{XXV} He$\alpha$   & 1s$^2$ ($^1$S$_0$) - 1s2p ($^1$P$_1$)                                  &  1.85 & A & A & B & \scriptsize{\ion{Fe}{XXIV} blend} \\
\hline
\end{longtable}
%\begin{tablenotes}
\begin{itemize}
\item[$a:$] Lists of lines detected at $> 5\sigma$ in the Capella spectrum.
\item[$b:$] Rest-frame wavelengths in SPEX database.
\item[$c:$] A: Model line wavelength and flux match with the data. B: Wavelength matches, but the model line flux is off by $> 20$\%. C: Flux matches within 20\%,
but the line central energy differs by $> 3\sigma$. D: Though the line is present, neither its wavelength nor the flux matches with the data.
\item[$d:$] EBIT rates (\S~\ref{sec:ebit}) incorporated.
\end{itemize}
%\end{tablenotes}

\newpage

\begin{table}[!htbp]
\centering
\caption{\label{tab:wav_ult} Wavelength correction for the ultimate modeling}
\begin{threeparttable}
\begin{tabular}{ccccccccccc}
\hline
Name & atomic state & SPEX energy & corr. energy & source  \\
     &            & (eV)          & (eV)           &  \\
\hline
\ion{Fe}{XVII}  & 2s2p$^6$3d ($^1$D$_{2}$)        & 944.79  & 942.17  & APEC \\
\ion{Fe}{XVII}  & 2s2p$^6$4p ($^1$P$_{1}$)        & 1126.86 & 1124.78 & APEC \\
\ion{Fe}{XVIII} & 2s$^2$2p$^4$3s ($^4$P$_{1/2}$)  & 781.25  & 782.37  & APEC \\
\ion{Fe}{XVIII} & 2s$^2$2p$^4$3s ($^2$D$_{5/2}$)  & 793.00  & 793.65  & APEC \\
\ion{Fe}{XVIII} & 2s$^2$2p$^4$3p ($^2$D$_{5/2}$)  & 805.48  & 807.70  & Chianti \\
\ion{Fe}{XVIII} & 2s$^2$2p$^4$3p ($^2$P$_{3/2}$)  & 835.97  & 835.58  & Chianti \\
\ion{Fe}{XVIII} & 2s$^2$2p$^4$3d ($^2$F$_{5/2}$)  & 853.06  & 852.89  & Chiant\\
\ion{Fe}{XVIII} & 2s$^2$2p$^4$4d ($^2$D$_{5/2}$)  & 1084.39 & 1085.68 & Chianti \\
\ion{Fe}{XVIII} & 2s$^2$2p$^4$4d ($^2$F$_{5/2}$)  & 1094.64 & 1094.68 & APEC \\
\ion{Fe}{XVIII} & 2s$^2$2p$^4$5d ($^2$D$_{5/2}$)  & 1176.73 & 1178.23 & Chianti \\
\ion{Fe}{XVIII} & 2s$^2$2p$^4$5d ($^2$F$_{5/2}$)  & 1196.73 & 1197.86 & Chianti \\
\ion{Fe}{XIX}   & 2s$^2$2p$^3$3p ($^3$P$_2$)      & 886.17  & 884.04  & Chianti \\
\ion{Fe}{XIX}   & 2s$^2$2p$^3$3d ($^5$D$_3$)      & 890.23  & 889.54  & Chianti \\
\ion{Fe}{XIX}   & 2s$^2$2p$^3$3d ($^3$F$_3$)      & 909.33  & 908.44  & Chianti \\
\ion{Fe}{XIX}   & 2s$^2$2p$^3$3d ($^3$P$_2$)      & 918.97  & 917.97  & Chianti \\
\ion{Fe}{XIX}   & 2s$^2$2p$^3$4d ($^3$D$_3$)      & 1146.61 & 1146.30 & Chianti \\
\ion{Fe}{XIX}   & 2s$^2$2p$^3$5d ($^3$D$_3$)      & 1259.74 & 1258.98 & Chianti \\
\ion{Fe}{XXI}   & 2s$^2$2p3d ($^3$D$_3$)          & 1008.68 & 1009.45 & APEC \\
\ion{Mg}{XI}    & 1s3p ($^1$P$_1$)                & 1580.00 & 1579.31 & APEC \\
\hline
\end{tabular}
\end{threeparttable}
\end{table}

\end{appendix}

\end{document}